\documentclass{JHEP3}
\pdfoutput=1
\usepackage[english]{babel}
\usepackage{amsmath,amssymb}
\usepackage[pdftex]{graphicx}
\usepackage{amsfonts}
\usepackage{bbm}
\usepackage{cite}

\newcommand{\bc}{\begin{center}}
\newcommand{\ec}{\end{center}}
\newcommand{\bt}{\begin{tabular}}
\newcommand{\et}{\end{tabular}}
\newcommand{\be}{\begin{equation}}
\newcommand{\ee}{\end{equation}}
\newcommand{\bea}{\begin{eqnarray}}
\newcommand{\eea}{\end{eqnarray}}
\newcommand{\bfig}{\begin{figure}}
\newcommand{\efig}{\end{figure}}
\newcommand{\ie} {{\it i.e.}}
\newcommand\fverb{\setbox\fverbbox=\hbox\bgroup\verb}
\newcommand\fverbdo{\egroup\medskip\noindent%
			\fbox{\unhbox\fverbbox}\ }
\newcommand\fverbit{\egroup\item[\fbox{\unhbox\fverbbox}]}
\newbox\fverbbox

\title{Dark Matter Direct Detection Signals inferred from a Cosmological N-body
Simulation with Baryons}

\author{F.-S. Ling$^1$, E. Nezri$^2$, E. Athanassoula$^2$, R. Teyssier$^{3,4}$\\
$^1${\it
Service de Physique Th\'eorique, Universit\'e Libre de Bruxelles,\\
CP225, Bld du Triomphe, 1050 Brussels, Belgium\\}
\vspace{0.35cm} \\
$^2${\it Laboratoire d'Astrophysique de Marseille,\\ 
Observatoire Astronomique de Marseille Provence\\
CNRS/Universit\'e de Provence\\38 rue Joliot Curie, 13388 Marseille, France\\}
\vspace{0.35cm} \\
$^3${\it IRFU, CEA Saclay \\
L'Orme des merisiers 91191 Gif-sur-Yvette, France\\}
\vspace{0.35cm} \\
$^4${\it Institute f\"ur Theoretische Physik, Universit\"at Z\"urich,\\
Winterthurerstrasse 190, CH-8057 Z\"urich, Switzerland} \\
\vspace{0.35cm} \\
E-mails: \email{fling@ulb.ac.be}, \email{Emmanuel.Nezri@oamp.fr},
\email{lia@oamp.fr}, \email{romain.teyssier@gmail.com}
}

\preprint{ULB-TH/09-30}

\abstract{
We extract at redshift $z=0$ a Milky Way sized object including gas, stars and dark matter (DM) from a recent, 
high-resolution cosmological N-body simulation with baryons. 
Its resolution is sufficient to witness the formation of a rotating disk and bulge at the center of the halo potential,
therefore providing a realistic description of the birth and the evolution of galactic structures 
in the $\Lambda$CDM cosmology paradigm.

The phase-space structure of the central galaxy reveals that, throughout a thick region, the dark halo 
is co-rotating on average with the stellar disk.
At the Earth's location, the rotating component, sometimes called dark disk in the literature,
is characterized by a minimum lag velocity $v_{lag} \simeq 75$~km/s,
in which case it contributes to around 25\% of the total DM local density, whose value is $\rho_{DM} \simeq 0.37~{\rm GeV/cm^3}$.
The velocity distributions also show strong deviations from pure Gaussian and Maxwellian distributions, 
with a sharper drop of the high velocity tail.

We give a detailed study of the impact of these features on the predictions for DM signals in direct detection experiments.
In particular, the question of whether the modulation signal observed by DAMA is or is not excluded by limits set by other
experiments (CDMS, XENON and CRESST...) is re-analyzed and compared to the case of a standard Maxwellian halo.
We consider spin-independent interactions for both the elastic and the inelastic scattering scenarios.
For the first time, we calculate the allowed regions for DAMA and the exclusion limits of other null experiments
\emph{directly} from the velocity distributions found in the simulation. We then compare these results with 
the predictions of various analytical distributions.

We find that the compatibility between DAMA and the other experiments is improved.
In the elastic scenario, the DAMA modulation signal is slightly enhanced in the so-called channeling region,
as a result of several effects that include a departure from a Maxwellian distribution and anisotropies
in the velocity dispersions due to the dark disk.
For the inelastic scenario, the improvement of the fit is mainly attributable to the departure from a Maxwellian distribution at high velocity.
It is correctly modeled by a generalized Maxwellian distribution with a parameter $\alpha \simeq 1.95$, or by a Tsallis distribution
with $q \simeq 0.75$.
}


\begin{document}

\section{Introduction}
\label{sec:intro}

Nowadays, the hypothesis of dark matter (DM) is considered as the most plausible explanation for various observations from galactic to 
cosmological scales.
Flat rotation curves in spiral galaxies~\cite{Sofue:2000jx,Bosma:2003yv}, 
velocity dispersions in galaxy clusters~\cite{Zwicky:1933gu,Lewis:2002mfa},
gravitational lensing mass reconstructions~\cite{Refregier:2003ct,Clowe:2003tk,Gavazzi:2008aq},
hierarchical patterns in large scale structures~\cite{Cole:2005sx,Seljak:2004xh}
and cosmic microwave background anisotropies~\cite{Spergel:2006hy}
all point towards the existence of a new neutral, non baryonic, cold ({\it i.e.} with low thermal velocities) and weakly interacting 
massive particle (WIMP) in the universe.

Theories beyond the Standard Model of particle physics~\cite{Amsler:2008zzb,Iliopoulos:2008fc,Baer:2009nt} provide many candidates for DM 
particles (see for example Refs.~\cite{Peccei:1977ur, Ellis:1983ew, Cheng:2002ej, Cirelli:2005uq}).
However, its existence will not be proved and its properties will not be determined until a clear identification is made
through signals in indirect or direct detection experiments or in particle accelerators.
Recently, several astrophysical excesses have been reported~\cite{Knodlseder:2005yq, Hunger:1997we, Adriani:2008zr, Chang:2008zzr},
and associated with a DM signal in more or less exotic scenarios (see, among many others, Refs.~\cite{Boehm:2003bt,deBoer:2004ab,Frere:2006hp,Fox:2008kb,Cirelli:2008pk,Ibarra:2009dr}).
However, up to now, none of them requires DM as an irrefutable explanation.
The INTEGRAL 511~keV line $\gamma$-ray signal~\cite{Knodlseder:2005yq} can be explained by the emission of low mass X-ray binaries~\cite{Weidenspointner:2008zz}.
The EGRET diffuse $\gamma$-ray GeV anomaly~\cite{Hunger:1997we} was most likely due to an error in the energy calibration~\cite{Stecker:2007xp}, 
and has since indeed been ruled out by the FERMI instrument~\cite{Porter:2009sg}.
The positron and electron peak seen by ATIC~\cite{Chang:2008zzr} has been flattened by HESS~\cite{Aharonian:2009ah} and FERMI~\cite{Abdo:2009zk} measurements.
The positron fraction rise observed by PAMELA~\cite{Adriani:2008zr}, although striking, can be accommodated in
standard astrophysical scenarios with pulsars~\cite{Hooper:2008kg,Malyshev:2009tw}, or with an inhomogeneous distribution of supernovae remnants~\cite{Piran:2009tx}.

Direct detection provides a unique opportunity to disentangle DM signals from other astrophysical processes.
The idea is to directly measure the energy deposited in a low noise detector during the collision between a DM particle and a detector nucleus. 
Several experiments have been designed for this aim~\cite{Bernabei:2000qi, Ahmed:2008eu, Angle:2007uj, Angle:2008we,Angloher:2008jj}. 
Up to now, all of them are compatible with null results, except the DAMA/LIBRA collaboration which claims to have observed a
signal with an annual modulation characteristic of an DM signal~\cite{Bernabei:2008yi}.
This modulation is induced by the Earth rotation around the Sun, and is presented as a signature hard to reproduce by background events.
The DAMA result is still quite controversial because it leads to tensions with other experiment measurements
(for a recent status, see Refs.~\cite{Petriello:2008jj,Fairbairn:2008gz, Savage:2009mk, Arina:2009um}).

Predicting direct detection signals in any particle physics model requires astrophysical assumptions in the form of a value for the local DM density 
as well as velocity distributions in the solar neighborhood. 
Usually, the local DM density is taken as $\rho_{DM} = 0.3~{\rm GeV/cm^3} = 0.079~{\rm M_{Sun}/pc^3}$, 
a value which is supported by the Milky-Way rotation curve~\cite{Flores:1987qi,Sofue:2008wt},
and stars spectrophotometric data~\cite{Chen:2003rh}.
A recent determination which takes numerous dynamical observables into account gives a larger value, $\rho_{DM} = 0.39~{\rm GeV/cm^3}$~\cite{Catena:2009mf}.
However, it should be mentioned that determinations based on mass models only give average values.
A local peak in the DM density, although improbable without a dynamical reason~\cite{Lavalle:1900wn}, is in principle not excluded
by very local bounds based on planetary data~\cite{Sereno:2006mw,Frere:2007pi}.
Another common over-simplified assumption is to take the velocity distribution as isotropic and Maxwellian.
Incomplete virialization, streams and interaction with the stellar disk could contribute to create anisotropies or a non Maxwellian spectrum.  
Moreover, even in a completely virialized structure at equilibrium, anisotropy is expected as a result of
the inhomogeneous gravitational potential with a power law density profile~\cite{Hansen:2004qs,Hansen:2005yj}.
The consequences for direct detection of some of these departures from the standard halo case have been analyzed 
by several authors~\cite{Green:2000jg,Green:2002ht,Green:2003yh,Savage:2006qr,Vergados:2007nc,Vergados:2008ez,Kamionkowski:2008vw,MarchRussell:2008dy}. 

Numerical simulations have been used for decades as a powerful virtual laboratory to explore complex dynamical systems.
On cosmological scales, early studies enabled to assess the role of DM in the hierarchical structure formation sheme~\cite{White:1987yr,Tormen:1996fc}.
Recent techniques~\cite{Teyssier:2001cp,Springel:2005mi} as well as improving computing capabilities have brought the possibility to extract
galactic DM halos with tremendous resolution.
The so-called \emph{Via Lactea} simulation~\cite{Diemand:2008in} or the \emph{Aquarius} project~\cite{Springel:2008cc} have resolved
a Milky-Way sized galactic halo with more than a billion particles. 
It appears that numerous sub-halos are present, anisotropies and deviations from standard Maxwellian velocity distributions are also found
and lead to somewhat different signatures in direct detection 
experiments~\cite{Hansen:2004qs,Hansen:2005yj,Vogelsberger:2008qb,Fairbairn:2008gz,MarchRussell:2008dy}. 
Despite their impressive resolution, these DM only simulations fail as a realistic description of a galactic halo for the simple reason that
they totally neglect the baryonic components (stars and gas in the galactic disk and bulge), although these dominate are known to dominate 
the dynamics near the galactic center.
This shortcoming is being fixed by very recent cosmological simulations which include baryons~\cite{Gibson:2008ev,Read:2009iv}.
It appears that the DM spatial distribution is neither spherical, or triaxial, but has a thick oblate shape around the galactic disk,
in what some authors have called a \emph{dark disk}~\cite{Read:2008fh,Bruch:2008rx}, that is co-rotating with the galactic stellar disk.
These particular characteristics lead to a potential enhancement of the direct detection signal at low recoil energies~\cite{Bruch:2008rx},
although there is an important spread in the predictions, depending on the particular merger history for each galaxy.

In this paper, a recent and advanced cosmological simulation with baryons is used as a realistic framework to extract detailed predictions
for direct detection. The paper is organized as follows.
In Sec.~\ref{sec:simu}, the simulation is presented and described. Velocity distributions in the solar neighborhood are extracted,
analyzed and compared with standard simplified assumptions. The presence of a co-rotating dark disk is also discussed.
In Sec.~\ref{sec:dde}, after a brief overview of the event rate formalism and the experimental status, the direct detection predictions are given
for both the elastic and the inelastic scenarios. Finally, our results are summarized in Sec.~\ref{sec:conclu}.

\section{A cosmological simulation with baryons }
\label{sec:simu}

Although galaxy formation is far from being completely understood, both theory and observation
have made significant progresses in the recent years. Based on first principles, it is now possible 
to perform simulations of Milky-Way--like galaxies in reasonable agreement with observed galaxies of 
similar circular velocities~\cite{Mayer:2008mr,Governato:2006cq,Governato:2008qp}. 
Some long-lasting problems seem however to remain, especially for massive galaxies like our own: 
the simulated bulge seems systematically too massive (sometimes referred to as the angular momentum problem), 
and the total baryon fraction too large (the overcooling problem). 
Following previous work initiated by Bruch {\it et al.}~\cite{Bruch:2008rx}, 
we nevertheless would like to use a Milky--Way--like galaxy simulation
including both dark matter and baryons dynamics to infer the expected signal in dark matter direct detection experiments.

\subsection{Features}

\begin{figure}[t]
\begin{center}
\includegraphics[width=0.6\textwidth]{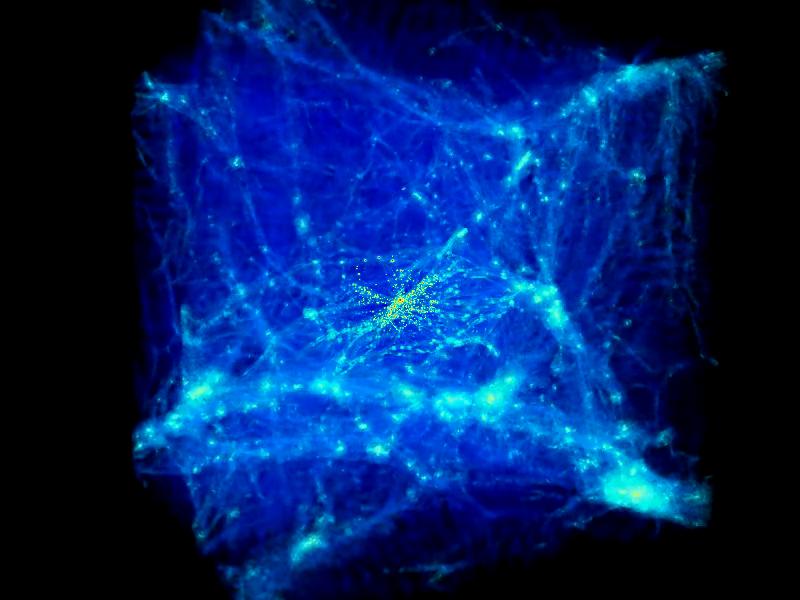}\\
\includegraphics[width=0.6\textwidth]{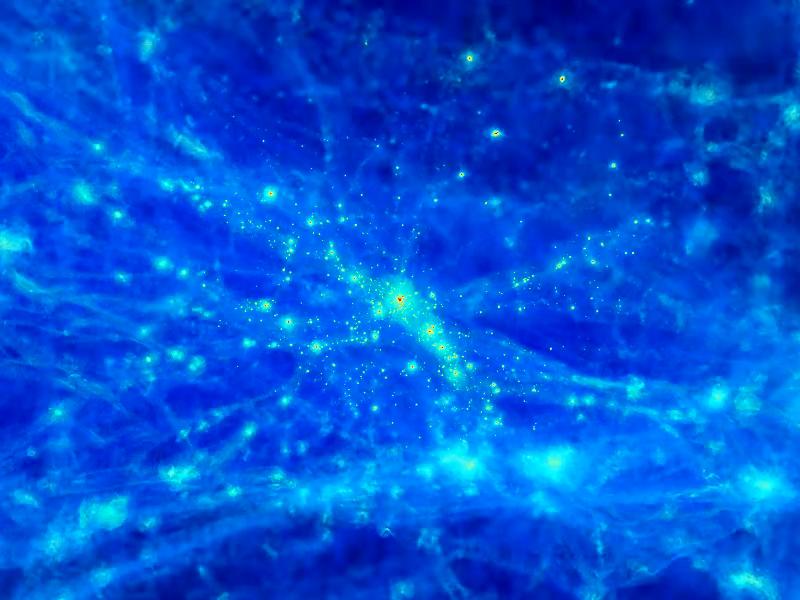}
\caption{\small \it
View of the simulation box (top), with a zoom on the central and better resolved region (bottom).
The size of the simulation box is 20 Mpc$/h$. The central galaxy is embedded in a large scale filamentary structure.}
\label{fig:simubox}
\end{center}
\end{figure}
\begin{figure}[t]
\begin{center}
\includegraphics[width=0.6\textwidth]{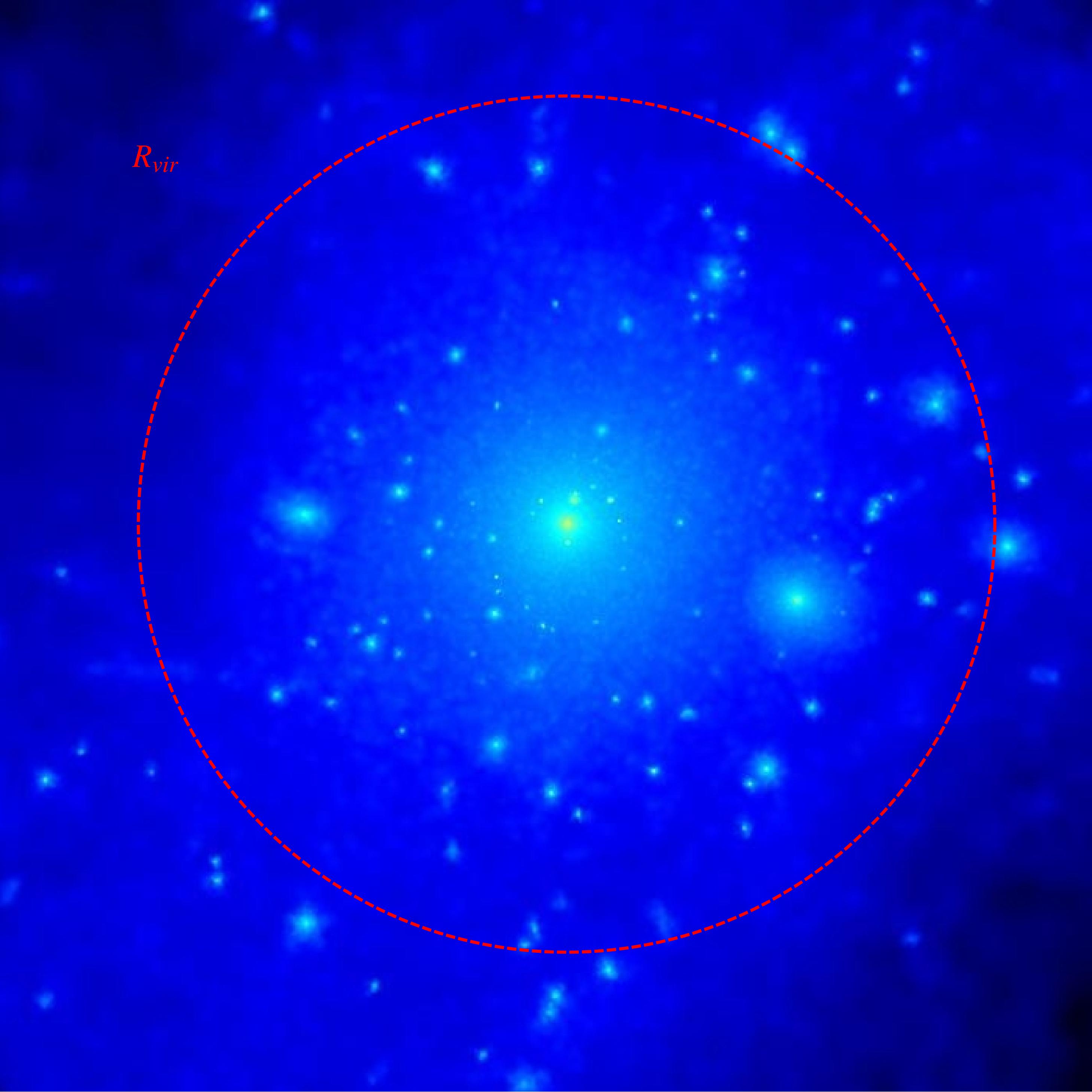}
\caption{\small \it 
View of the dark matter halo of the central galaxy of Fig.~\ref{fig:simubox}. This
halo will be used in all analyses in this paper. The radius of the red circle
is equal to its virial radius, i.e. 264 kpc. As expected, this halo
is centrally concentrated and has a large number of subhalos. 
}
\label{fig:halo}
\end{center}
\end{figure}
\begin{figure}[t]
\begin{center}
\begin{tabular}{cc}
\includegraphics[width=0.45\textwidth]{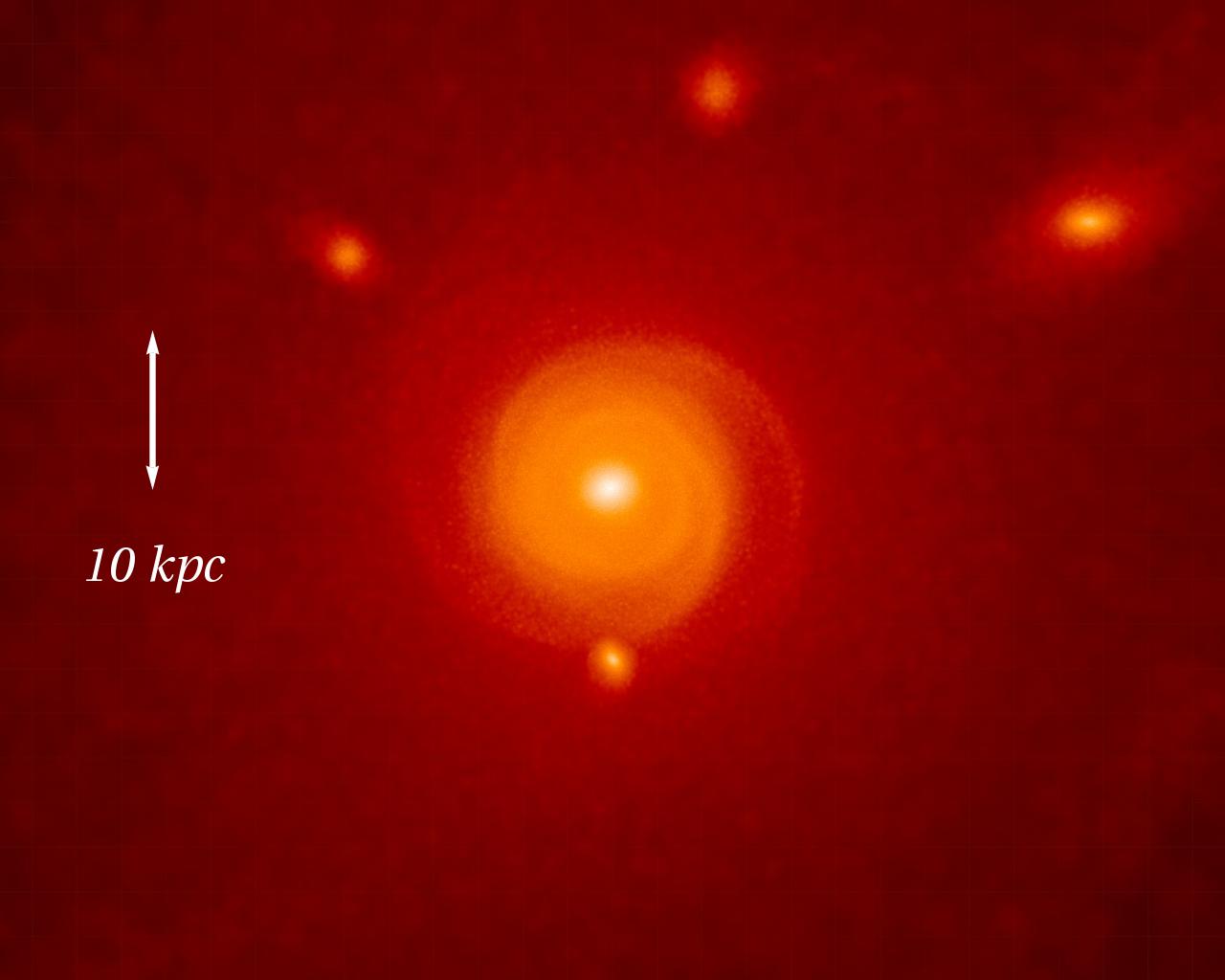}&
\includegraphics[width=0.45\textwidth]{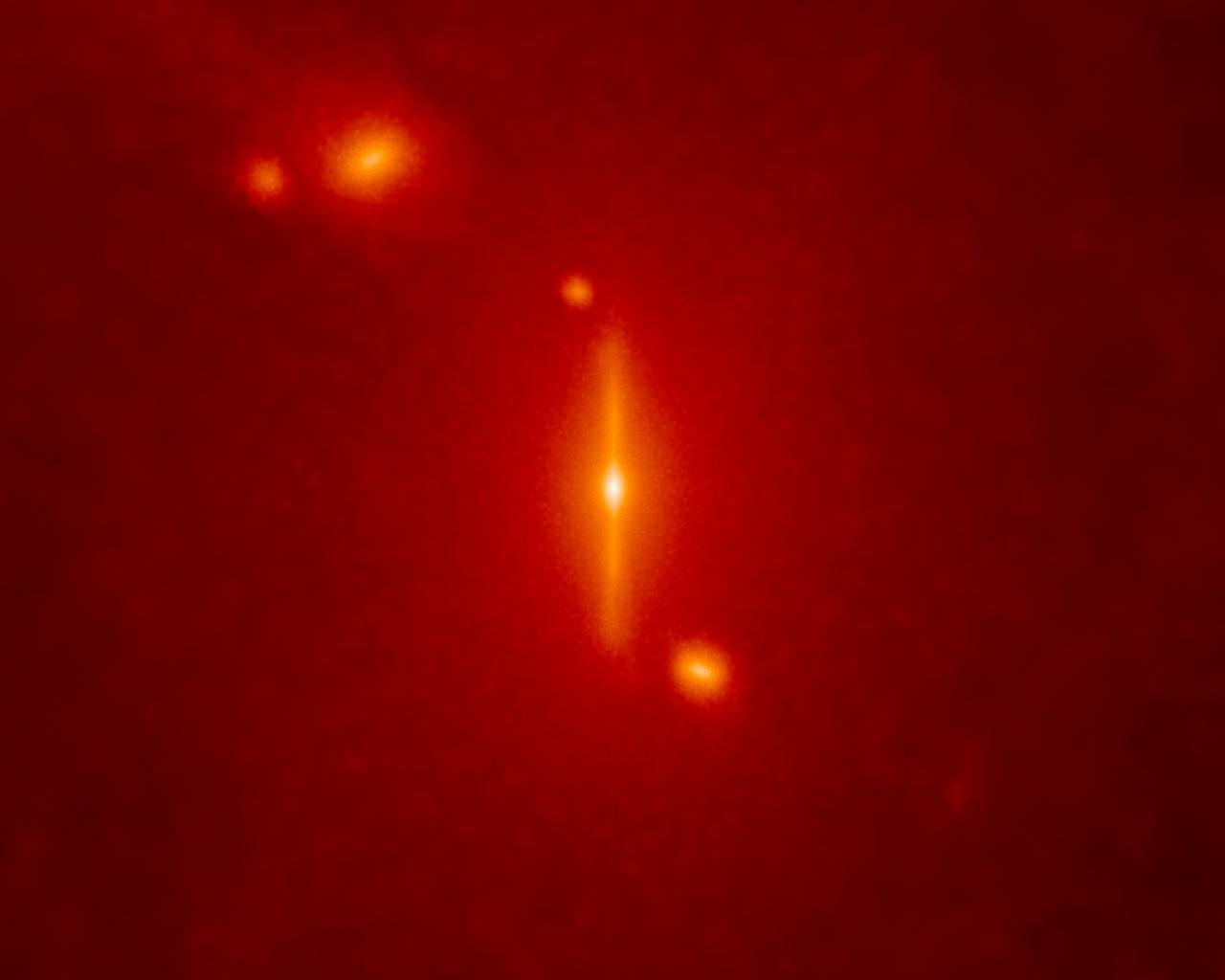}\\
a) Stars face on & b) Stars edge on\\
\includegraphics[width=0.45\textwidth]{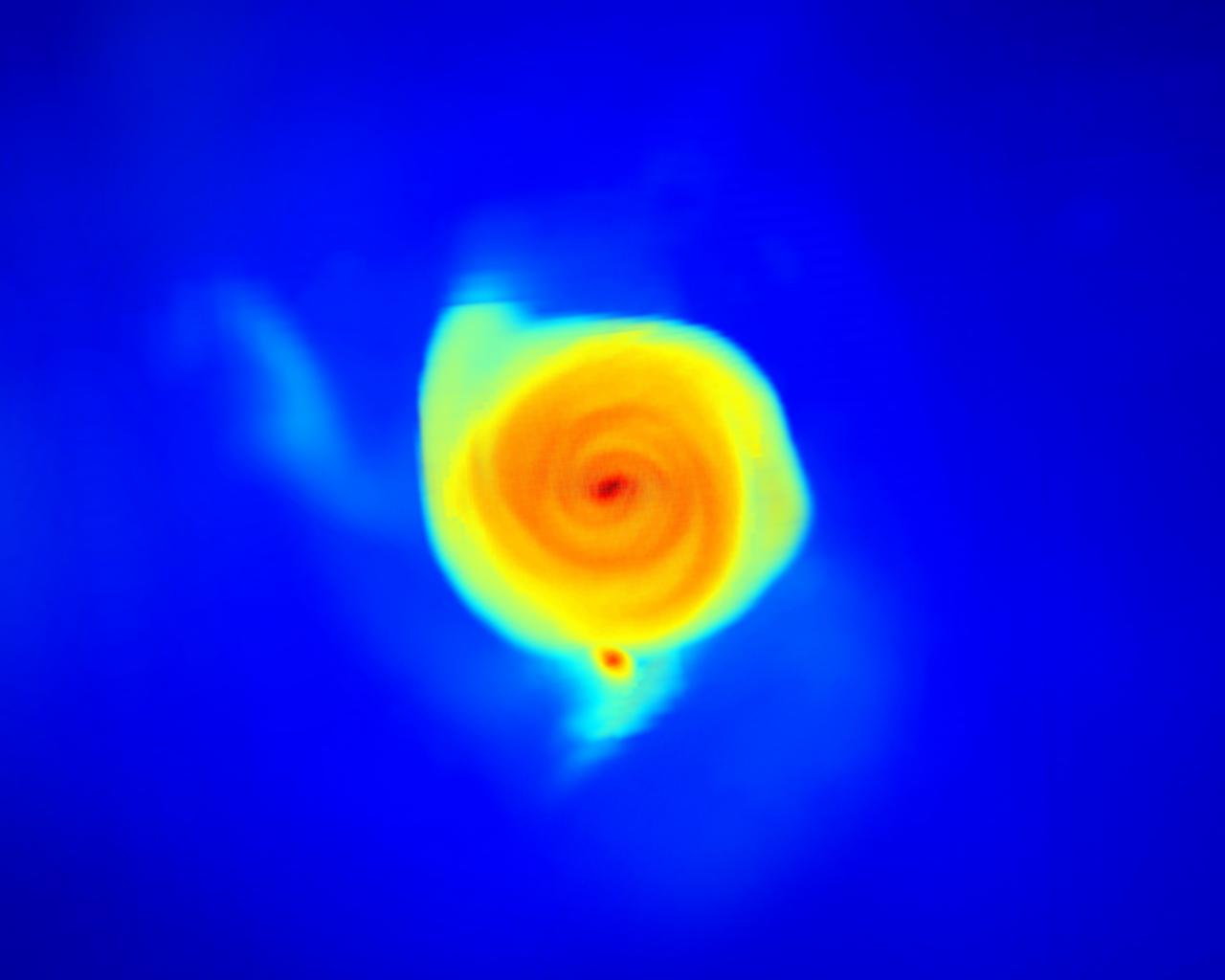}&
\includegraphics[width=0.45\textwidth]{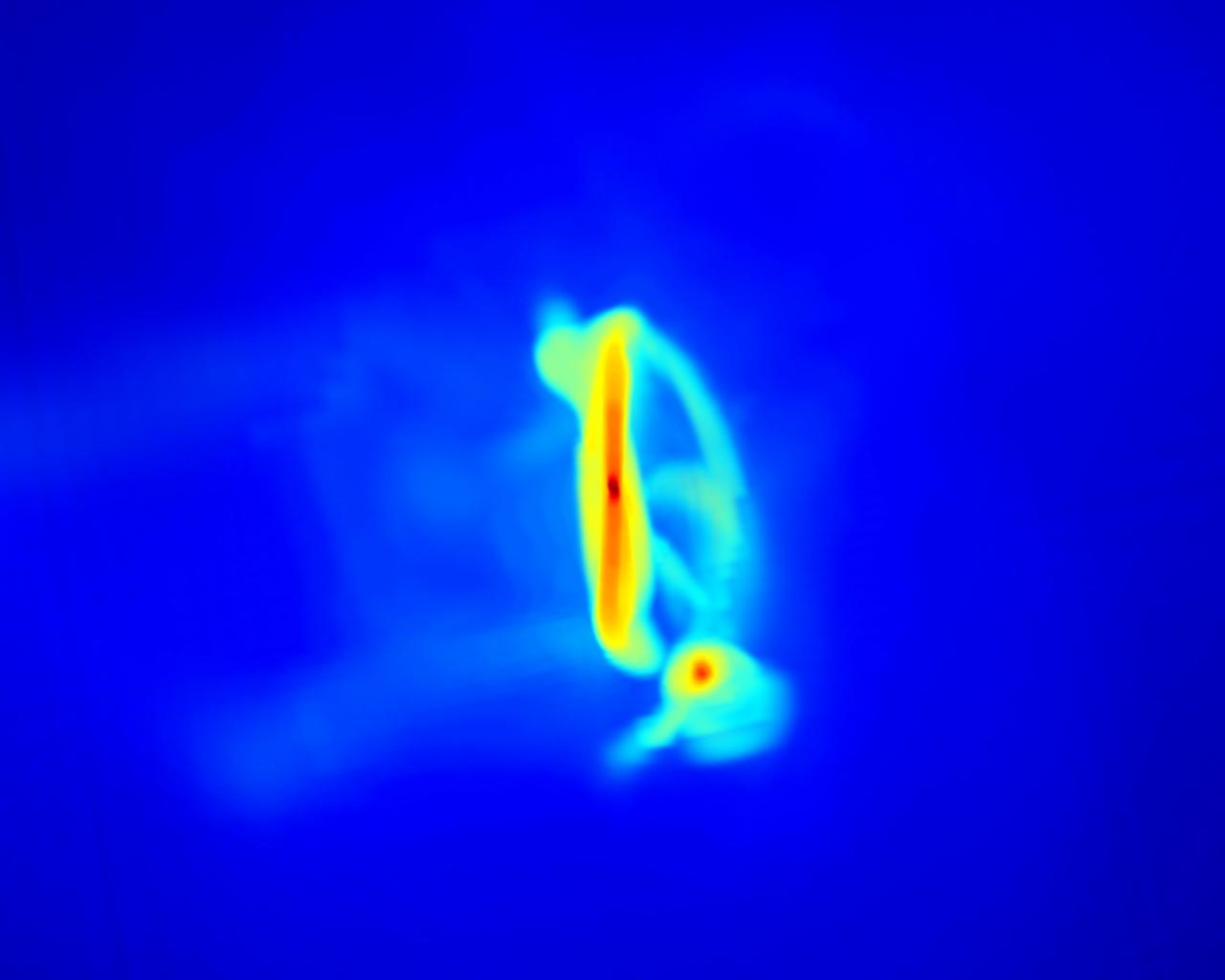}\\
c) Gas face on & d) Gas edge on
\end{tabular}
\caption{\small \it
Face-on and edge-on distributions of the baryonic material in our
galaxy. The white vertical line in the upper left panel has a length of
10 kpc. Note that both
the stars (upper panels) and the gas (lower panels) have formed a thin
disk, in good agreement with observations. A number of satellites can
also be seen, more numerous in the stars than in the gas, the gas having been
expelled from a large fraction of the substructures. The filamentary structures in the
gas edge-on view outlines past satellite trajectories.}
\label{fig:stars-gas}
\end{center}
\end{figure}
The simulation was performed using the cosmological Adaptive Mesh Refinement (AMR) code 
RAMSES~\cite{RAMSES}. We follow the evolution of 2 collisionless fluids, namely dark matter 
and stars, coupled through gravity to the dissipative baryonic component. Gas dynamics is 
described using a second-order unsplit Godunov scheme. Gas is allowed to cool radiatively, based
on a standard equilibrium photo-ionised mixture of Hydrogen and Helium~\cite{Katz:1992db}, taken into account
the excess cooling due to metals~\cite{Sutherland:1993ec}. Star formation is implemented using a standard Schmidt law,
spawning stars as a Poisson random process~\cite{Rasera:2005gq}. Our rather standard recipe for 
supernovae feedback and the associated chemical enrichment is described in Ref.~\cite{Dubois:2008mz}. 

We used the so-called ``zoom-in'' simulation technique, for which we identify a candidate halo using a low resolution,
dark matter only simulation, and then re-simulate it at higher resolution, degrading both spatial and mass
resolutions as we move further away from the central region. We checked that our final halo is not contaminated by
high mass, low resolution dark matter particles. Our effective resolution in the initial conditions corresponds
to a Cartesian grid of $1024^3$ elements covering our $20$~Mpc/$h$ periodic box. The corresponding dark matter particle
mass is therefore $7.46 \times 10^5$ solar masses, given we used the following cosmological model $\Omega_m=0.3,
\Omega_{\Lambda}=0.7,\Omega_b=0.045,H_0=70$~km/s/Mpc. The initial Gaussian random field was generated according to the 
Eisenstein and Hu (1999) $\Lambda$CDM model~\cite{Hu:1998tj} for the transfer function.

At redshift $z=0$, the complete simulation volume contains 8,376,152 dark matter particles of various masses, 
11,702,370 star particles and 19,677,225 AMR cells. We have fixed the spatial resolution to be $200$~pc
in {\it physical units}, which ensures that the vertical scale height of the stellar disc is barely resolved.
The final halo Virial radius $R_{vir}$\footnote{The Virial radius $R_{vir}$ is defined as a function of the 
cosmological matter density $\rho_m$ by $\frac{M(R<R_{vir})}{4/3 \pi R^3_{vir}}=200\rho_m$}
is found to be 264 kpc. It contains $N_{DM}=842,768$ dark matter particles corresponding to a halo mass $M_h=6.3 \ 10^{11} {\rm
  M_{sun}}$. Within the central 60~kpc, we have $N_{*}=5,416,865$ stars particles (within 15~kpc, this number
goes down to $N_*=$4,927,325 particles).
We have carefully selected our halo so that its mass accretion history is typical
of an late-type galaxy, with no major merger and a steady accretion rate in the last 8 Gyr.
Moreover, our halo mass is small enough ($6\times 10^{11}$ M$_{\odot}$) so that its
initial Lagrangian volume, with a comoving radius of 2 Mpc/h, is not affected by spurious effects
due to the periodic replication of the 20 Mpc/h box.
Our simulated halo, as can be seen in Figure 1, is embedded in a large
scale filament, which is quite typical for a $L_*$ galaxy.

As discussed above, our central galaxy is slightly too concentrated, with a bulge mass of the order of 4$\times 10^{10} {\rm
  M_{sun}} $ and a similar disc mass, while the most recent observational constraints in the Milky Way suggest
a bulge mass of 2$\times 10^{10} {\rm M_{sun}}$ for a disc mass of 6$\times 10^{10} {\rm M_{sun}}$~\cite{Sofue:2008wt}. 
Although the total mass in stars of our simulated galaxy agrees quite well with observations, we stress that 
we have considered here a rather low mass halo model ($M_{vir} \simeq 6 \times 10^{11} {\rm M_{sun}}$), resulting in a total 
baryon fraction in the galaxy close to the universal value of $15\%$. 
One should bear in mind that the larger bulge mass of our simulated
galaxy will have the effect of concentrating the dark matter distribution further in. 
This process, usually referred to as ``Adiabatic Contraction" (AC), results in a decrease 
of the dark matter characteristic scale radius, and in an increase of the dark matter 
circular velocity~\cite{Blumenthal:1985qy,Gnedin:2004cx,Abadi:2009ve}.
When the effects of AC are combined with a small total halo mass
($6\times 10^{11}$ M$_{\odot}$), we obtain a dark matter density in the solar neighbourhood close to
0.37 GeV/cm$^3$. Reducing the bulge mass by a factor of 2 would result in a 10\% decrease of the
circular velocity and in a 20\% decrease of the dark matter density in the solar neighbourhood (see
Ref.~\cite{Gnedin:2004cx} for a quantitative estimate).

The central dark halo, stars and gas are shown respectively on figure \ref{fig:halo} and \ref{fig:stars-gas}.
Figure~\ref{fig:halo} shows the dark matter distribution. As expected, it is
centrally concentrated and has a considerable number of
subhaloes. Using Adaptahop to extract the substructure information, 
we find that the central galaxy contains 92 subhaloes, of which 79 are inside the virial radius.
The subhalo mass fraction is 4.8\%, a rather small value due to the limited mass resolution in the simulation. 
The closest clump to the galactic center is at a radial distance of 12.7 kpc. 
Therefore, as no clump lies in the the solar neighborhood region ($7<R<9$~kpc and $-1<z<1$~kpc),
the influence of substructures on DM direct detection signals cannot be inferred from this simulation.

Figure~\ref{fig:stars-gas} shows the distribution
of the stars and the gas. Both display a very thin disk of a radial
extent of roughly 10 kpc and a clear spiral structure. For the stellar
population, this is mainly seen in the outer parts of the disc, while for
the gas it can be traced to a region very near the center. In both
cases it is two-armed, but has clear asymmetries. Secondary branchings
of the arms, coupled to the main arms result in a ring-like shape in
the gas distribution.

Note also that there is an important central concentration in both
distributions. This is of short extent, both in the equatorial plane
and perpendicular to it. The latter property, as well its rich gaseous
content which has roughly the same extent as the stellar distribution,
exclude the possibility that it is a classical bulge and argue
strongly that is a discy bulge~\cite{Athanassoula:2005xh}. Further
investigations of this point is nevertheless necessary.  

In the stellar distribution the main galaxy is
surrounded by a number of small condensations, big enough to be
considered as small satellite galaxies. In the gaseous distribution,
however, only one of these condensations can be seen. This means that
the remaining condensations have no gaseous counterpart; the gas
having been expelled either by ram pressure stripping, or by a central
supernova, or a combination of both. The clear filamentary structures
in the gaseous edge-on plot outline past satellite trajectories.

\subsection{ Velocity distributions}
\label{sec:velocity}

\begin{figure}[t]
\begin{center}
\begin{tabular}{cc}
\includegraphics[width=0.45\textwidth]{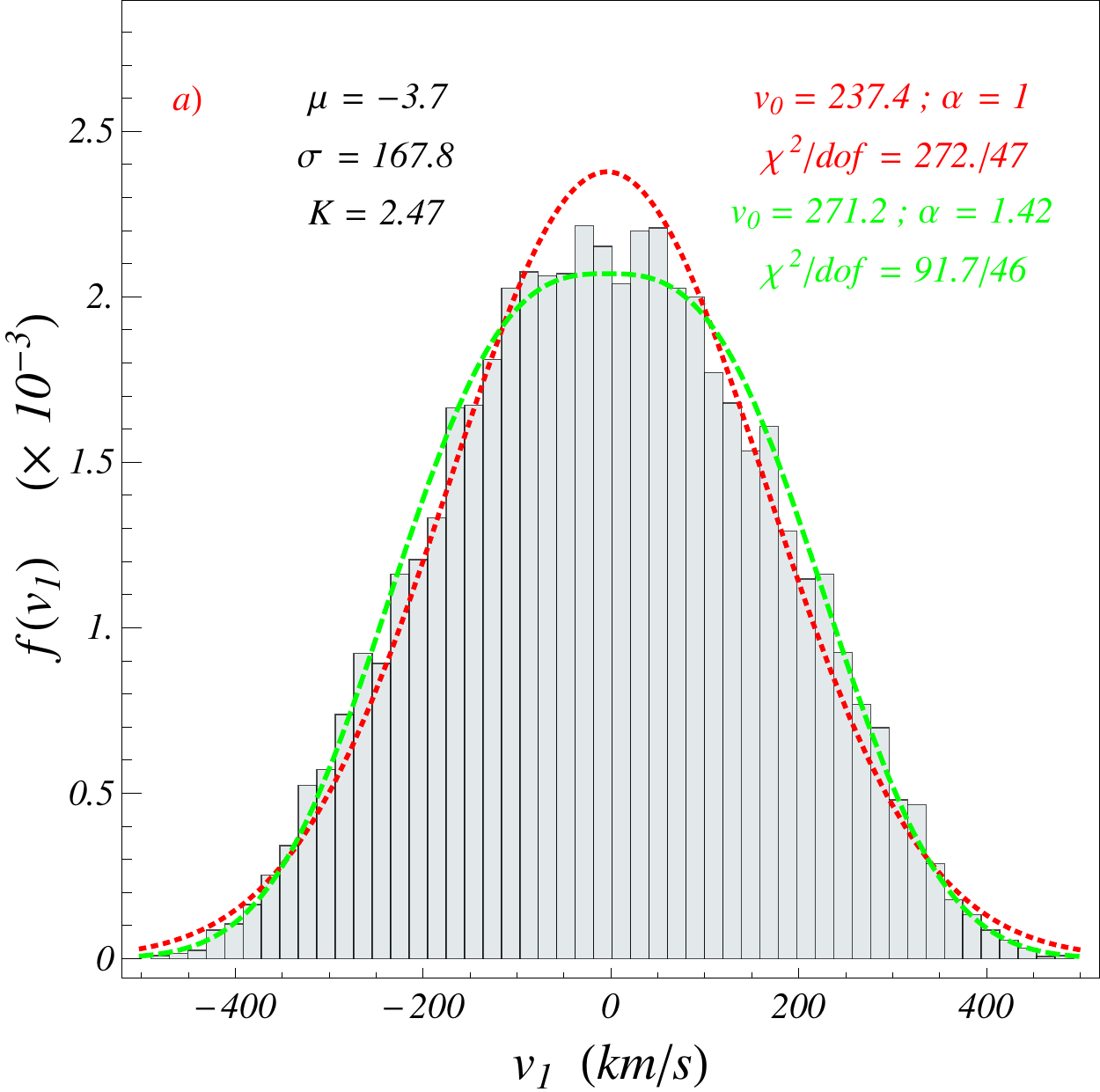}&
\includegraphics[width=0.45\textwidth]{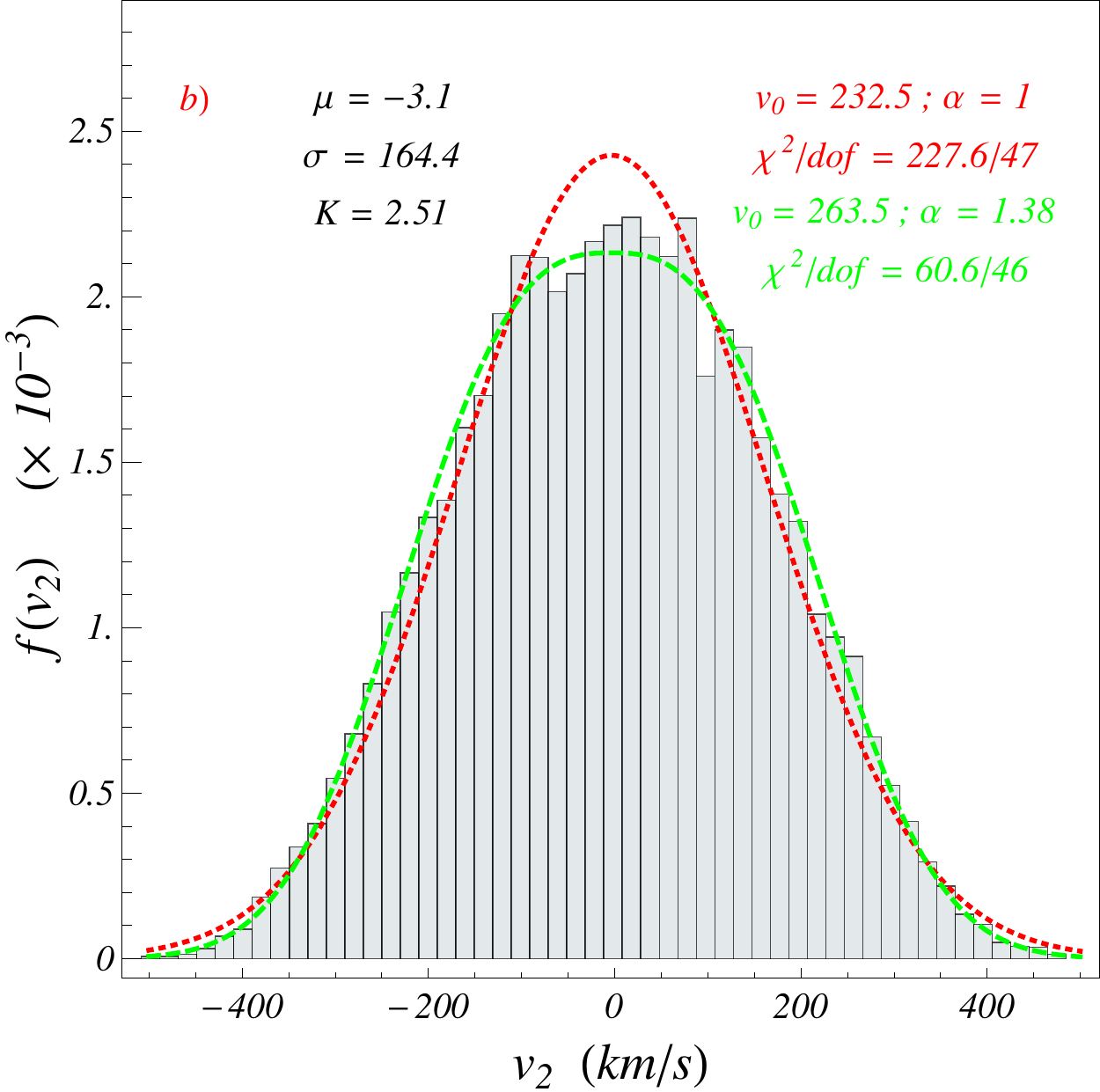}\\
\includegraphics[width=0.45\textwidth]{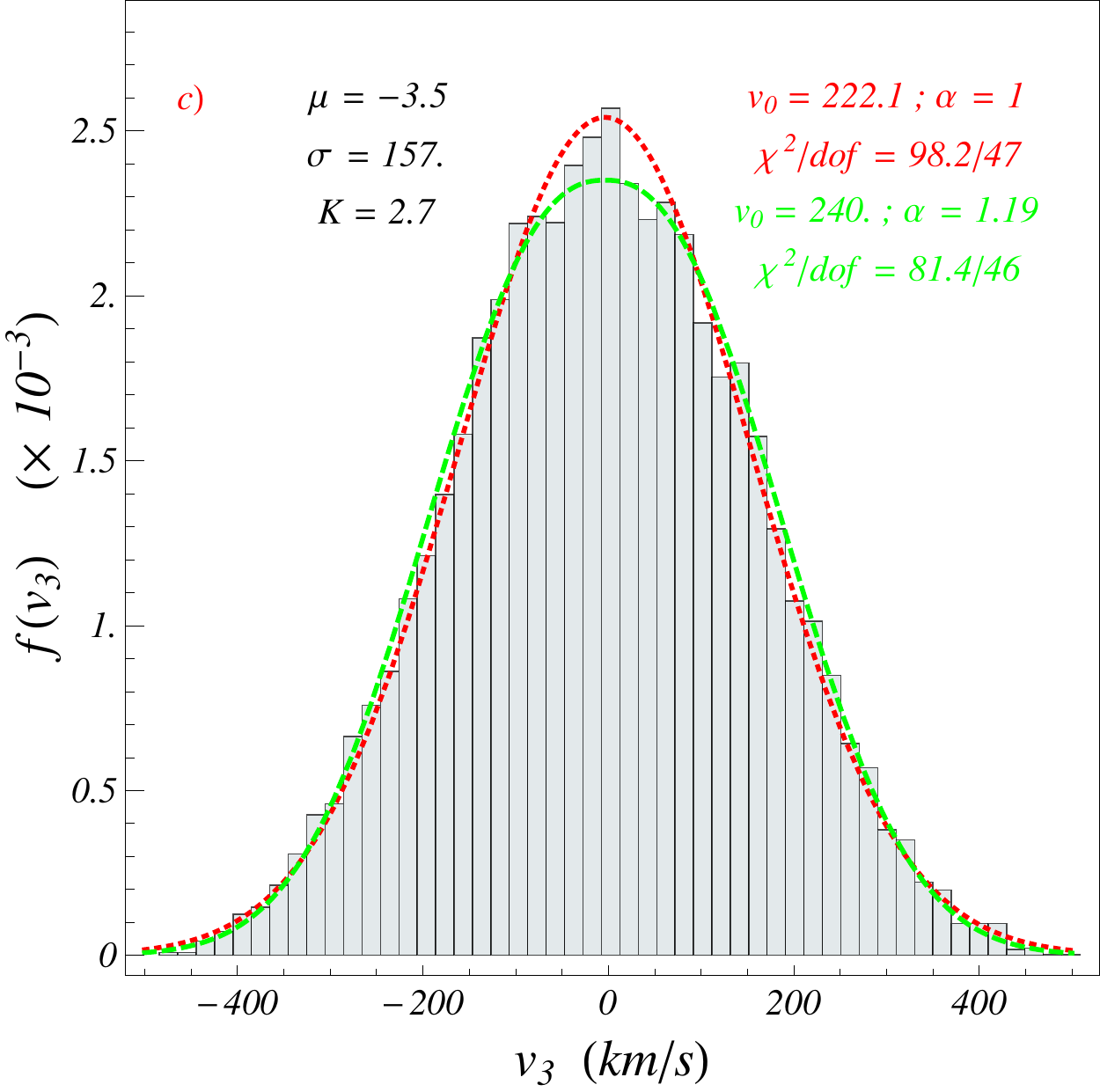}&
\includegraphics[width=0.45\textwidth]{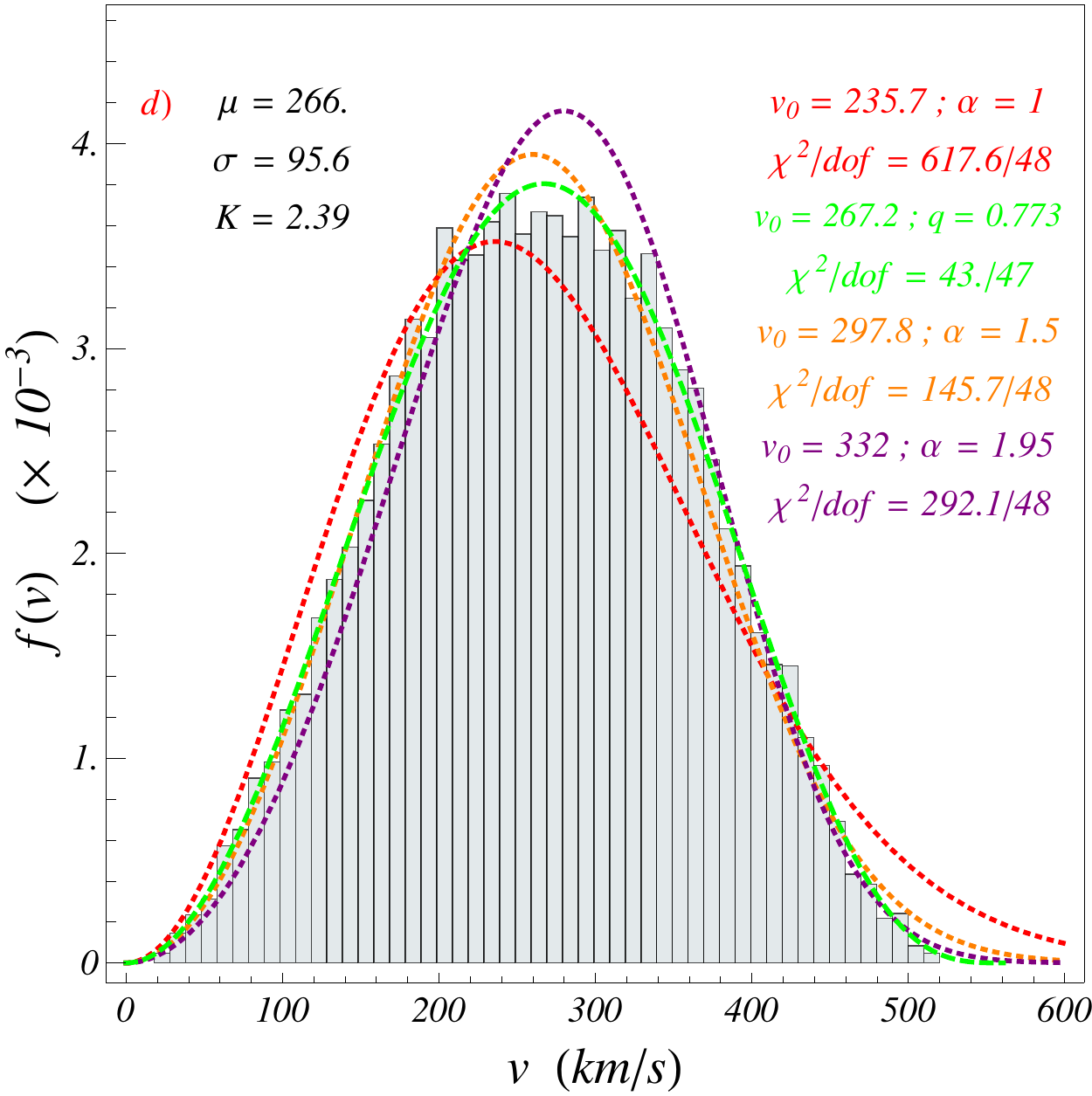}\\
\end{tabular}
\caption{\small \it 
Velocity distributions of dark matter particles ($N_{shell}=16,545$) in a spherical shell $7<R<9$ kpc around the galactic center.
\newline
a), b) and c) Velocity components along the principal axes of the velocity dispersion tensor,
together with the Gaussian (red) and a generalized Gaussian (green) distribution fits (cfr. Eq.~(\ref{eq:ggauss})). 
\newline
d) Velocity module, with Maxwellian (red), Tsallis (green) and generalized Maxwellian (orange and purple) fits 
(cfr. Eqs.~(\ref{eq:gmaxw},\ref{eq:tsallis})).
\newline
$\mu$, $\sigma$ (both in km/s) and $K$ stand for the mean, the standard deviation and the Kurtosis parameter of the distribution.
The goodness of fit is indicated by the value of the $\chi^2$ vs. the number of degrees of freedom (dof).
}
\label{fig:histovAquariuslike}
\end{center}
\end{figure}
\begin{figure}[t]
\begin{center}
\begin{tabular}{cc}
\includegraphics[width=0.45\textwidth]{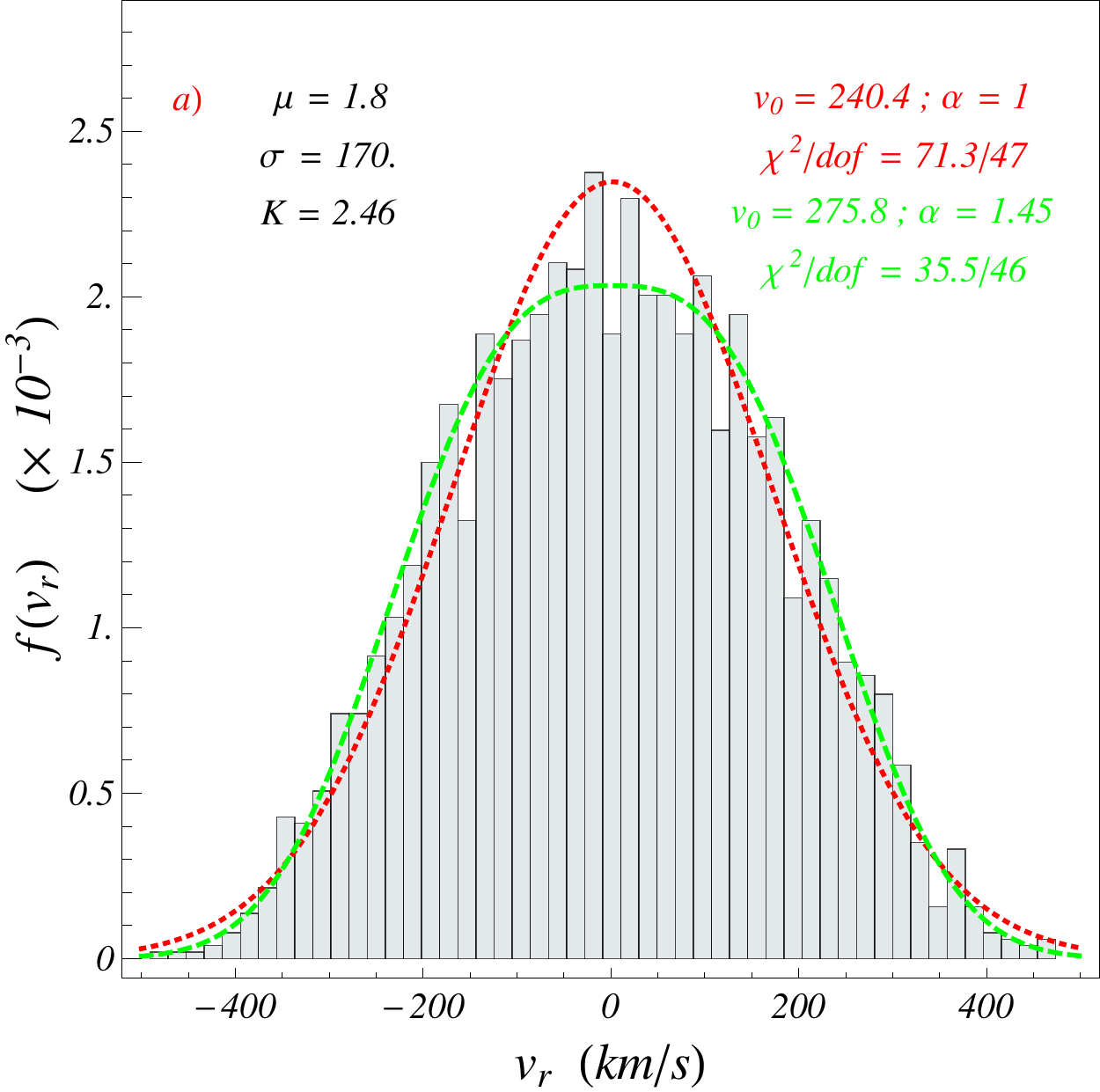}&
\includegraphics[width=0.45\textwidth]{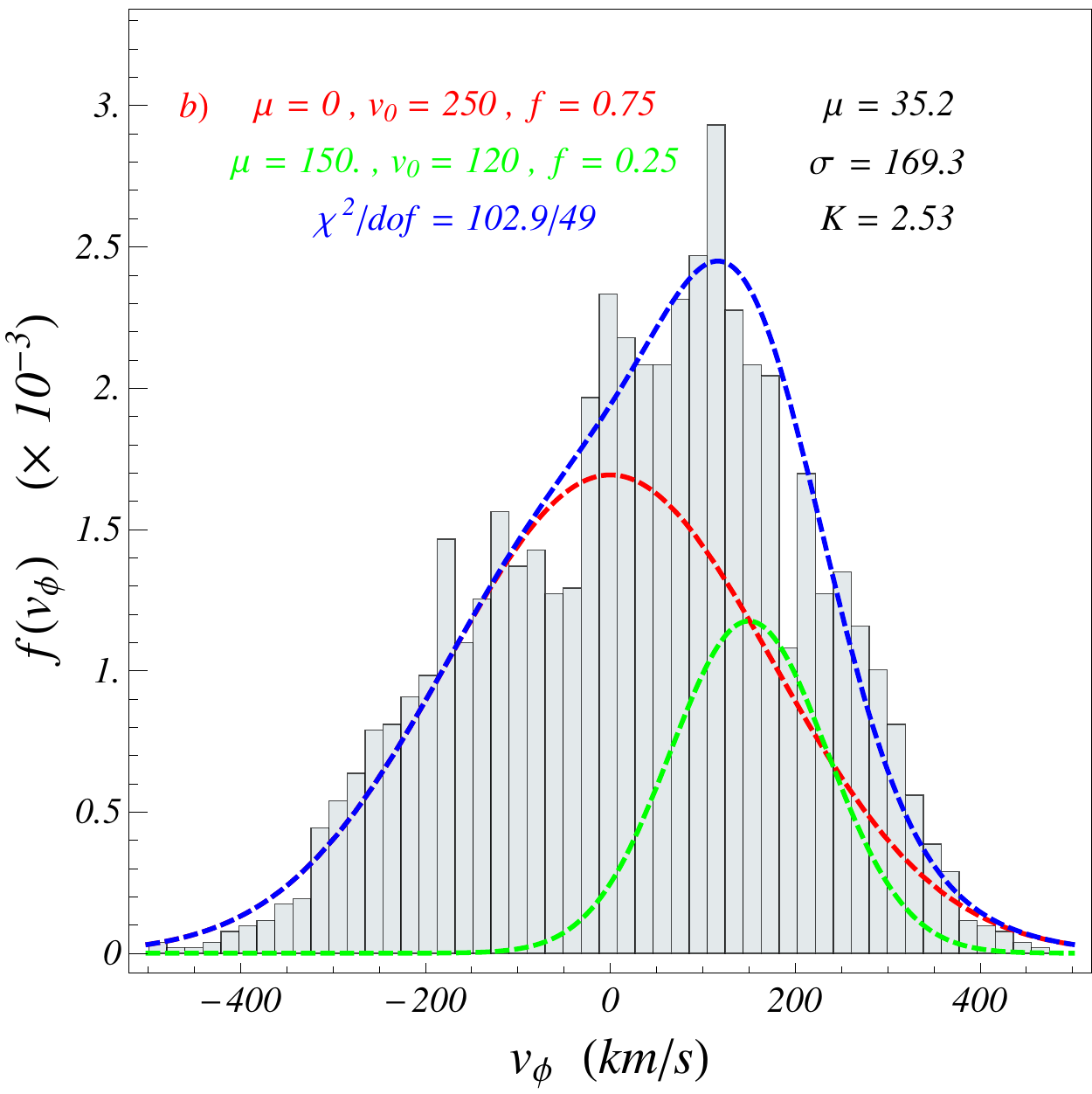}\\
\includegraphics[width=0.45\textwidth]{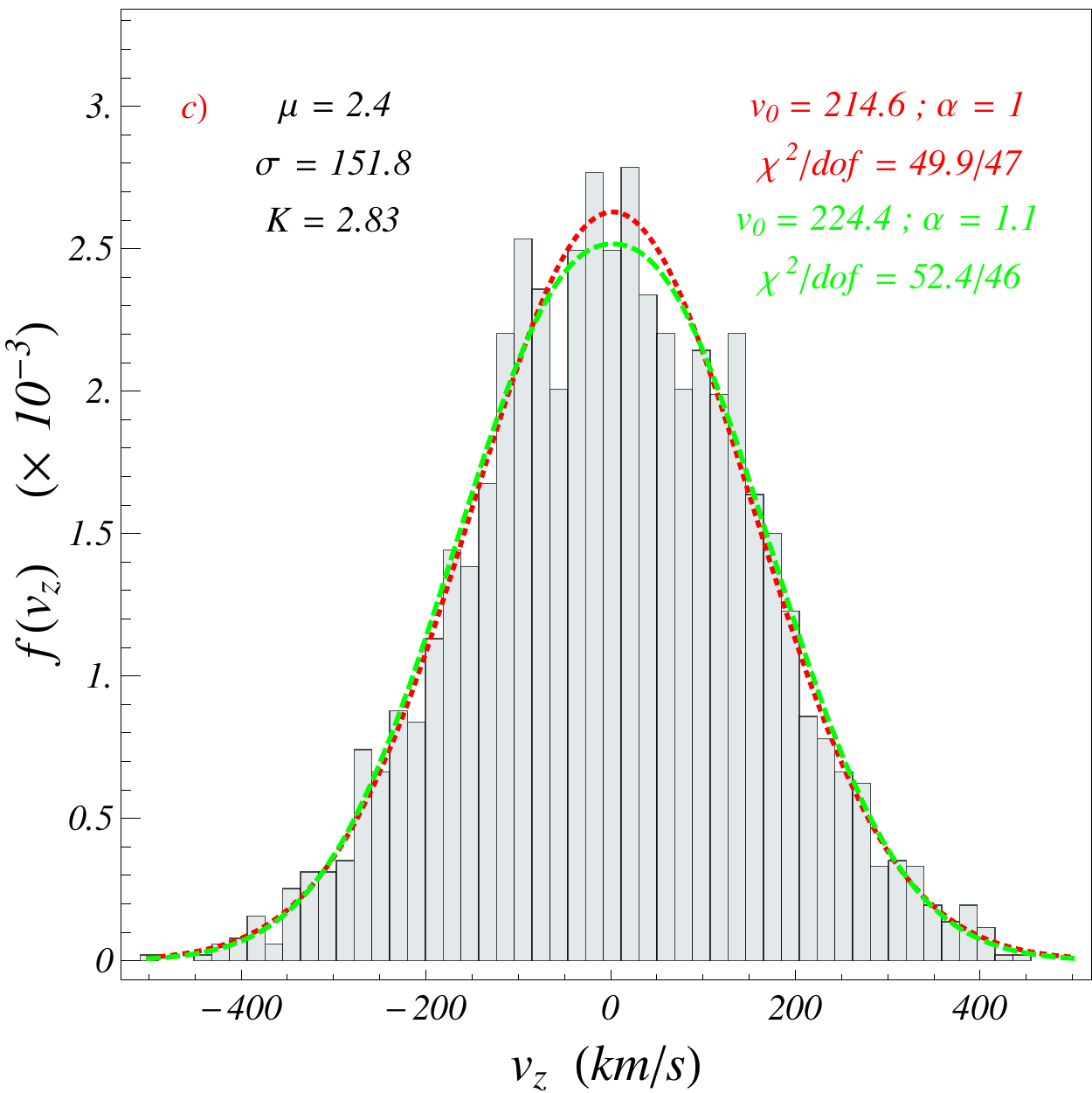}&
\includegraphics[width=0.45\textwidth]{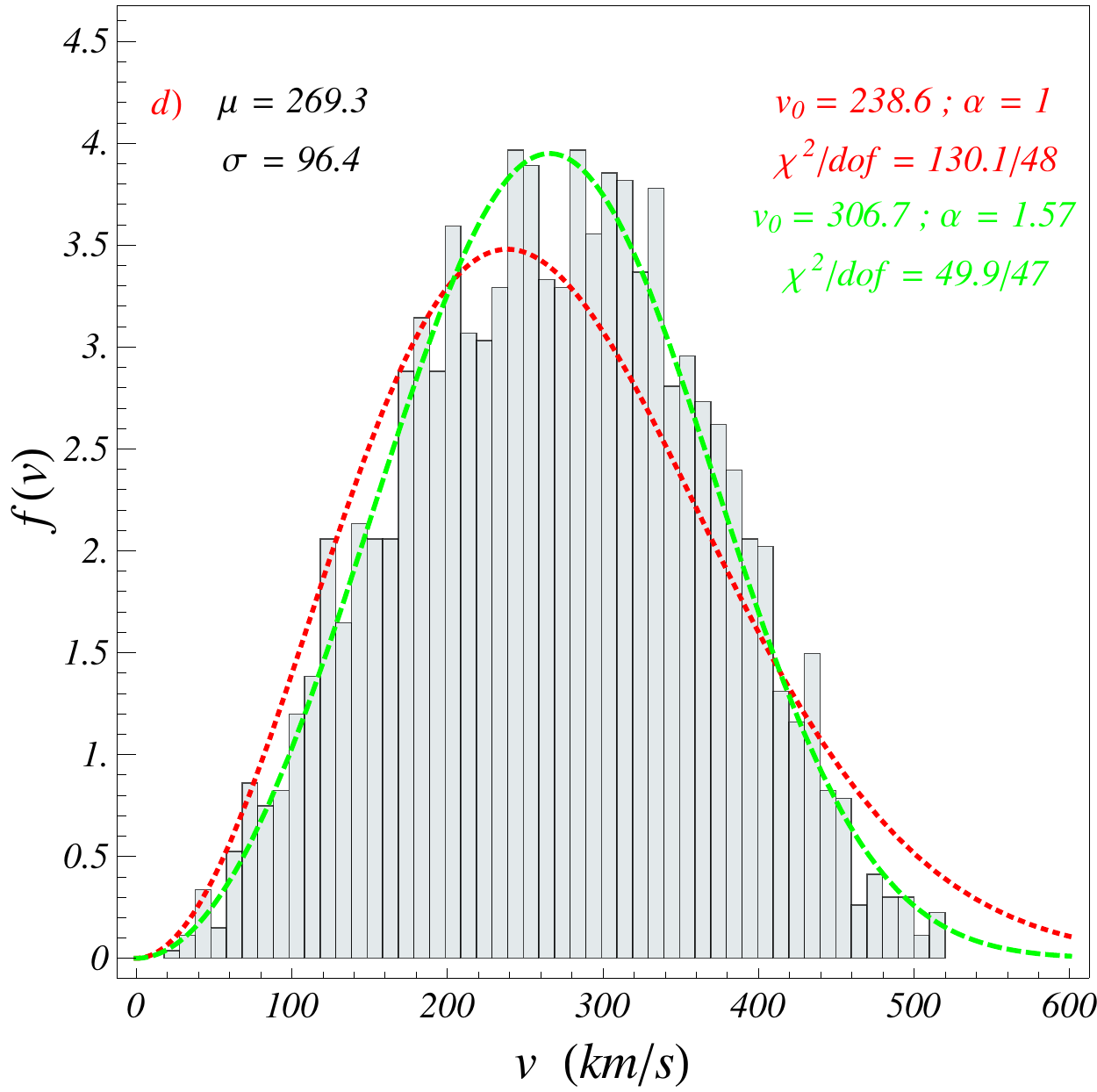}\\
\end{tabular}
\caption{\small \it 
Velocity distributions of dark matter particles ($N_{ring}=2,662$) in a ring $7<R<9$ kpc, $\vert z \vert <1$~kpc around the galactic plane.
\newline
a) Radial velocity $v_r$, with Gaussian (red) and generalized Gaussian (green) fits (cfr. Eq.~(\ref{eq:ggauss})). 
\newline
b) Tangential velocity $v_\phi$, with a double Gaussian fit. $f$ indicates the fraction of each component.
\newline
c) Velocity across the galactic plane $v_z$, with Gaussian (red) and generalized Gaussian (green) fits (cfr. Eq.~(\ref{eq:ggauss})). 
\newline
d) Velocity module, with Maxwellian (red) and a generalized Maxwellian (green) fit (cfr. Eq.~(\ref{eq:gmaxw})).
\newline
$\mu$, $\sigma$ (both in km/s) and $K$ stand for the mean, the standard deviation and the Kurtosis parameter of the distribution.
The goodness of fit is indicated by the value of the $\chi^2$ vs. the number of degrees of freedom (dof).
}
\label{fig:histovgalacticplane}
\end{center}
\end{figure}
In this section, we extract and analyze the velocity distributions given by our N-body simulation,
and compare them with the usual Gaussian/Maxwellian assumption.
As the goal of this study is to calculate direct detection predictions with a realistic DM halo,
the emphasis will be put on the local phase-space structure, in the solar neighborhood.

The Sun being located at a distance $R_0 \simeq 8$~kpc from the galactic center, we first select all DM particles 
in the spherical shell $7\ {\rm kpc}<R<9\ {\rm kpc}$. The number of particles in the selection is $N_{shell} = 16,545$.
Following Ref.~\cite{Vogelsberger:2008qb}, we calculate the velocity distributions along the principal axes of 
the velocity dispersion tensor $\sigma_{ij}^2$. They are shown on Fig.~\ref{fig:histovAquariuslike}.
The Gaussian fit for each component and the Maxwellian fit for the velocity module are shown in red.
A clear departure from these distributions is seen. 

In order to model and parameterize these deviations, one can introduce generalized Gaussian and Maxwellian distributions, which are convenient fitting functions.
The generalized Gaussian distribution with a mean velocity $\mu$, a velocity dispersion parameter $v_0$, and 
a Gaussianity parameter $\alpha$ is defined by
\be
{\rm f}(v) = \frac{1}{N(v_0,\alpha)} e^{-\left( (v-\mu)^2/v_0^2 \right)^\alpha} \quad ,
\label{eq:ggauss}
\ee
with a normalization factor $N(v_0,\alpha) = 2 v_0 \Gamma(1+1/2\alpha)$. For $\alpha=1$, the usual Gaussian distribution is recovered.
Similarly, the generalized Maxwellian distribution can be defined by
\be
{\rm f}(\vec{v}) = \frac{1}{N(v_0,\alpha)} e^{-\left( \vec{v}^2/v_0^2 \right)^\alpha} \quad ,
\label{eq:gmaxw}
\ee
with a normalization factor $N(v_0,\alpha) = 4\pi v_0^3 \Gamma(1+3/2\alpha)/3$. For $\alpha=1$, we get the usual Maxwellian distribution.

For the velocity components, a fit with a generalized Gaussian distribution (in green in Fig.~\ref{fig:histovAquariuslike}, panels a), b) and c))
shows that the velocity distribution given by the simulation is systematically more flat than a Gaussian distribution.
This property can be described by the so-called Kurtosis parameter $K$, which compares the fourth-order moment with the square of the variance.
A Kurtosis parameter $K<3$ corresponds to a distribution that is platykurtic, \ie more flat than a Gaussian distribution with the same standard deviation,
while $K>3$ corresponds to a distribution that is leptokurtic, \ie more peaked around the central value compared to the Gaussian one.
So velocity distributions extracted from this simulation are systematically platykurtic, in contrast with results obtained in DM only 
simulations~\cite{Diemand:2006ik,Vogelsberger:2008qb}.

The goodness-of-fit for each theoretical distribution is calculated using the Pearson's $\chi^2$ test.
The values of $\chi^2$ (calculated with a number of bins $n_{bin}=50$) printed in each panel of Fig.~\ref{fig:histovAquariuslike} 
show that the fit is considerably improved when using a generalized Gaussian distribution with $\alpha>1$ 
(the actual values of $v_0$ and $\alpha$ are estimated on the basis of the standard deviation $\sigma$ 
and the Kurtosis parameter $K$ extracted from the velocity distribution).
We also notice that along the principal direction ``3", the velocity dispersion and the deviation from the Gaussian distribution are smallest.
It appears that this direction is actually very close to the galactic pole axis. 

For the velocity module, a clear deviation from a Maxwellian distribution (in red) is apparent, with a sharper drop of the high velocity tail,
particularly crucial for the inelastic recoil scenario (see Sec.~\ref{sec:inel}).
There is no evidence of bumps in the velocity distribution, as seen in Ref.~\cite{Vogelsberger:2008qb}.  
A fit of the distribution with a generalized Maxwellian distribution with $\alpha>1$ leads to some improvement of the goodness-of-fit. 
However, no value of $\alpha$ enables to model the simulation distribution in a satisfactory way.
With two free parameters, only the mean value and the standard deviation can be accommodated, while the Kurtosis parameter is derived. 
It appears that the N-body velocity distribution is ``fatter" than the best-fit generalized Maxwellian ($v_0 = 301.6$~km/s and $\alpha=1.55$), 
with an observed Kurtosis parameter $K=2.39$ compared to $K=2.71$ for the fit. 
In panel d) of Fig.~\ref{fig:histovAquariuslike}, we show two possible fits with a generalized 
Maxwellian distribution, corresponding to $\alpha=1.5$ and $\alpha=1.95$.
The former gives a better global fit for the whole distribution and for the low velocity bins,
the latter gives a better fit of the high velocity tail.

In fact, the sharp drop of the high velocity tail has been recognized as a universal behavior of relaxed 
collisionless structures~\cite{Hansen:2005yj}.
The presence of long-range gravitational forces in dark matter structures indicates that these systems should be described
by non-extensive statistical mechanics. 
The generalization of the usual Boltzmann-Gibbs entropy to non-extensive systems by Tsallis~\cite{tsallis-88}
leads to distribution functions of the form
\be
{\rm f}(\vec{v}) = \frac{1}{N(v_0,q)} \left(1-(1-q) \frac{\vec{v}^2}{v_0^2} \right)^{q/(1-q)} \quad ,
\label{eq:tsallis}
\ee
where $N(v_0,q)$ is a normalization constant.
The Maxwell-Botzmann distribution is recovered by taking the limit $q \rightarrow 1$.
Equilibrated self-gravitating collisionless structures have been shown to exhibit Tsallis distributions
both analytically~\cite{lima-2001-86}, and numerically~\cite{Hansen:2004dg,Hansen:2005yj}.
For the particles in the spherical shell around 8~kpc in this simulation, the velocity module distribution is indeed 
very well fit by a Tsallis distribution with $v_0 = 267.2$~km/s and $q = 0.773$.
With these values, the Kurtosis parameter for the distribution is $K = 2.44$, very close to the observed value.

As argued in Ref.~\cite{Hansen:2004qs}, there should be some anisotropy in the velocity distributions
between the radial and the tangential directions, as particles moving in the radial direction see a change in the potential.
The velocity (dispersion) anisotropy is usually described by the parameter $\beta = 1-\sigma_t^2/\sigma_r^2$.
Also, the departure from a Maxwellian distribution, parameterized by $1-q$ for Tsallis distributions could show the same anisotropy. 
In this simulation, for the DM particles in the spherical shell $7<R<9$~kpc, 
we find a velocity anisotropy $\beta \simeq 0.12$, together with a slightly smaller
Kurtosis parameter in the radial direction $K = 2.41$, in general agreement with the conclusions drawn from
the universal relation between the density slope and the velocity anisotropy presented in Ref.~\cite{Hansen:2004qs}. To calculate the velocity anisotropy, the radial velocity dispersion $\sigma_r$ is directly extracted from the radial velocity distribution, while the tangential velocity dispersion $\sigma_t$ can be deduced from the mean $\mu$
and the standard deviation $\sigma$ of the distribution of the velocity module as 
$\sigma_r^2 + 2 \sigma_t^2 = \mu^2 + \sigma^2$.


%
\begin{figure}[t]
\begin{center}
\begin{tabular}{cc}
\includegraphics[width=0.45\textwidth]{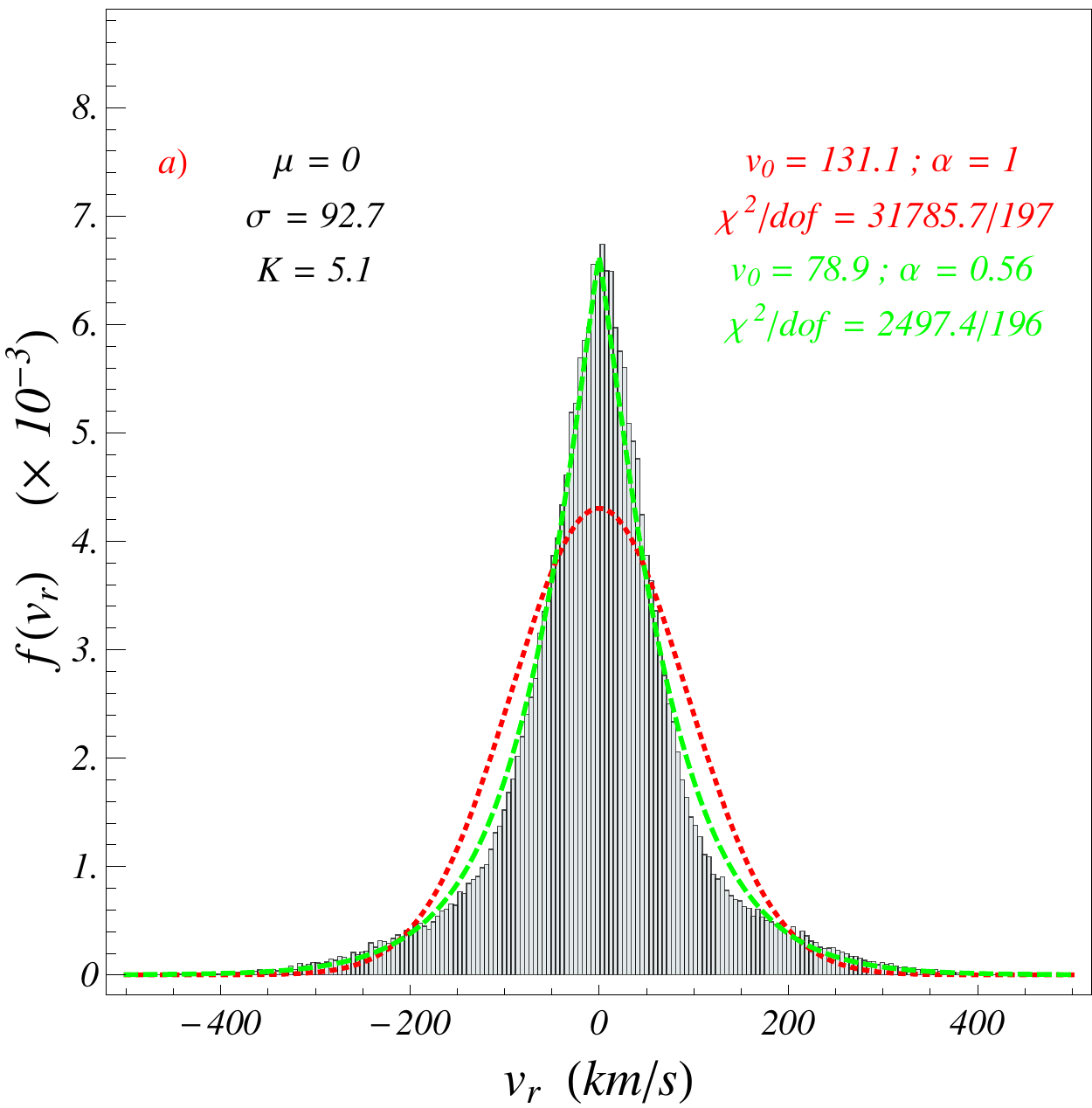}&
\includegraphics[width=0.45\textwidth]{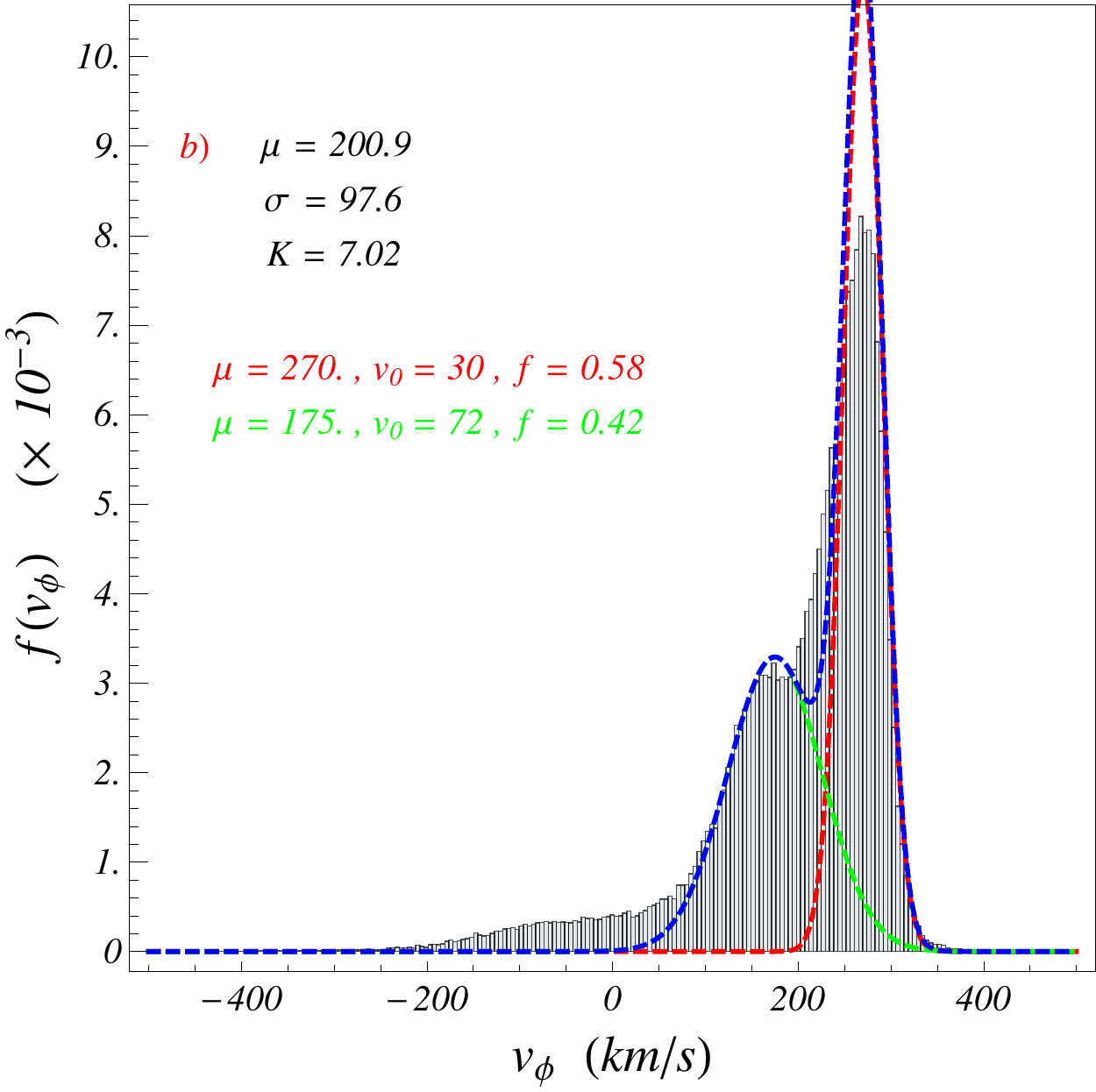}\\
\end{tabular}
\caption{\small \it 
Velocity distributions of star particles ($N_{star}=143,320$) in a ring $7<R<9$ kpc, $\vert z \vert <3$~kpc around the galactic plane.
\newline
a) Radial velocity $v_r$ with Gaussian (red) and generalized Gaussian (green) fits (cfr. Eq.~(\ref{eq:ggauss})).
\newline
b) Tangential velocity $v_\phi$ with a double Gaussian fit, with a thin (red) and a thick (green) disk components. 
$f$ indicates the fraction for each component.
\newline
$\mu$, $\sigma$ (both in km/s) and $K$ stand for the mean, the standard deviation and the Kurtosis parameter of the distribution.
The goodness of fit is indicated by the value of the $\chi^2$ vs. the number of degrees of freedom (dof).}
\label{fig:stardist}
\end{center}
\end{figure}
In dark matter-only simulations, one has a $4\pi$ uncertainty on the Sun position. 
This freedom is lifted, at least partially, in simulations with baryons where the galactic plane can be identified.
To determine it, we select {\it star } particles with $3<R<10$~kpc to avoid the galactic bulge. We have $N_{star}=1,935,104$ in this selection.
The galactic plane is then rotated onto the $xy$ plane by diagonalizing the position tensor. 
In this new reference frame, denoted by $\{ \vec{1}_x, \vec{1}_y, \vec{1}_z \}$, we select DM particles with
$7<R<9$~kpc and $|z|<1~{\rm kpc}$, corresponding to particles located in a ring around the galactic plane. 
The number of particles in this selection is $N_{ring}=2,662$.
Fig.~\ref{fig:histovgalacticplane} shows the corresponding velocity distributions in cylindrical coordinates $\{r,\phi,z\}$.
Anisotropy is manifest. There is a significant rotation along the galactic disk, with $\langle v_{\phi} \rangle=35.2$~km/s.
For the radial and $z$ velocity components, the mean of the distribution is compatible with zero.
The velocity dispersions and Kurtosis parameters also exhibit some anisotropy in the $z$ direction compared to the galactic plane.
For the distribution of the velocity module, a strong deviation from a Maxwellian distribution is again visible,
with a cut-off of the high velocity tail. 
However, despite a coarser resolution, we can observe that the distribution for ring particles differs significantly
from the Tsallis form that is expected for equilibrated systems. Therefore, we can already stress that predictions of
direct detection signals that use equilibrium distributions should be taken with some caution.

\subsection{Dark disk, halo rotation and local density}

\begin{figure}[t]
\begin{center}
\begin{tabular}{cc}
\includegraphics[width=0.45\textwidth]{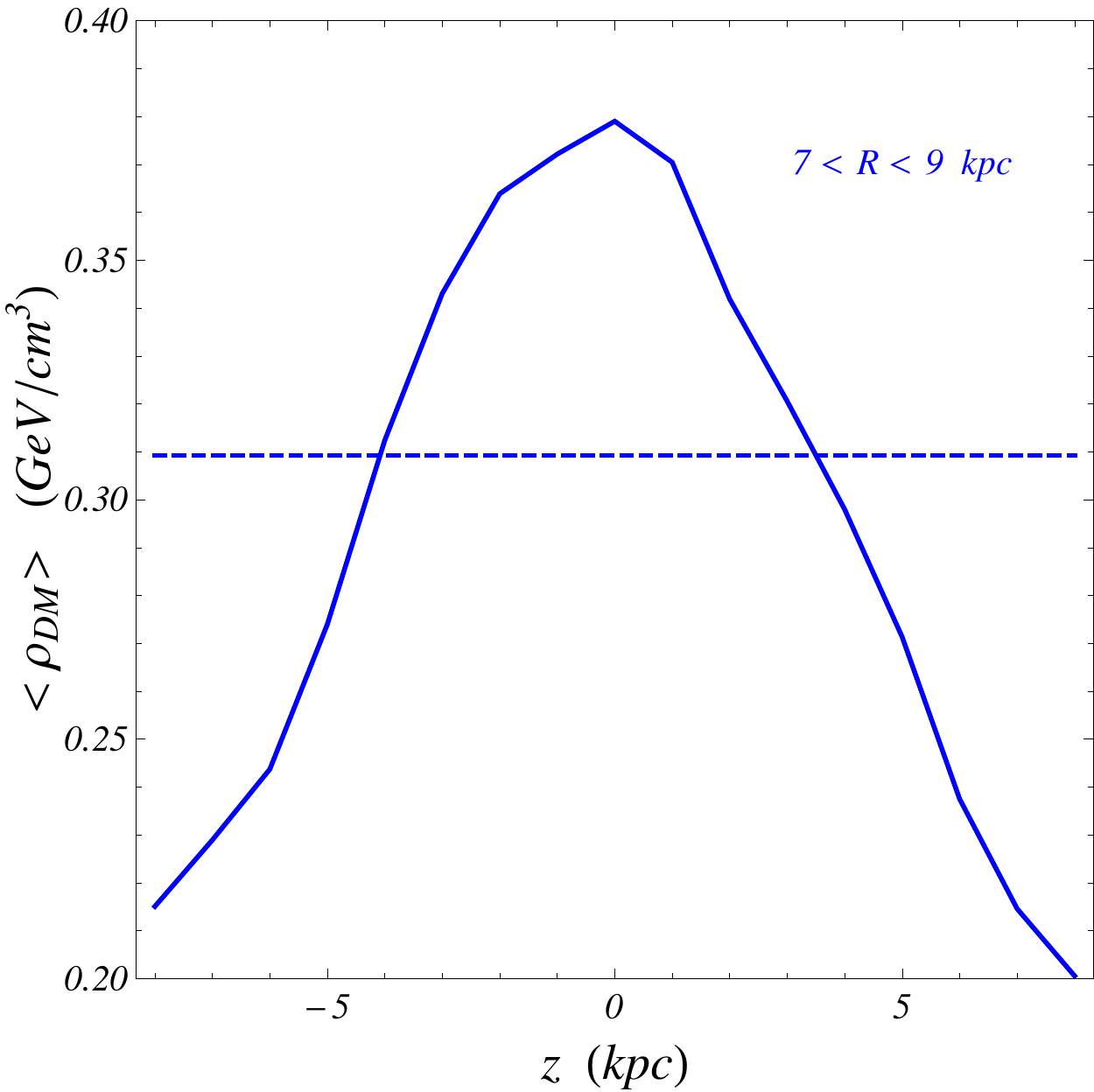}&
\includegraphics[width=0.45\textwidth]{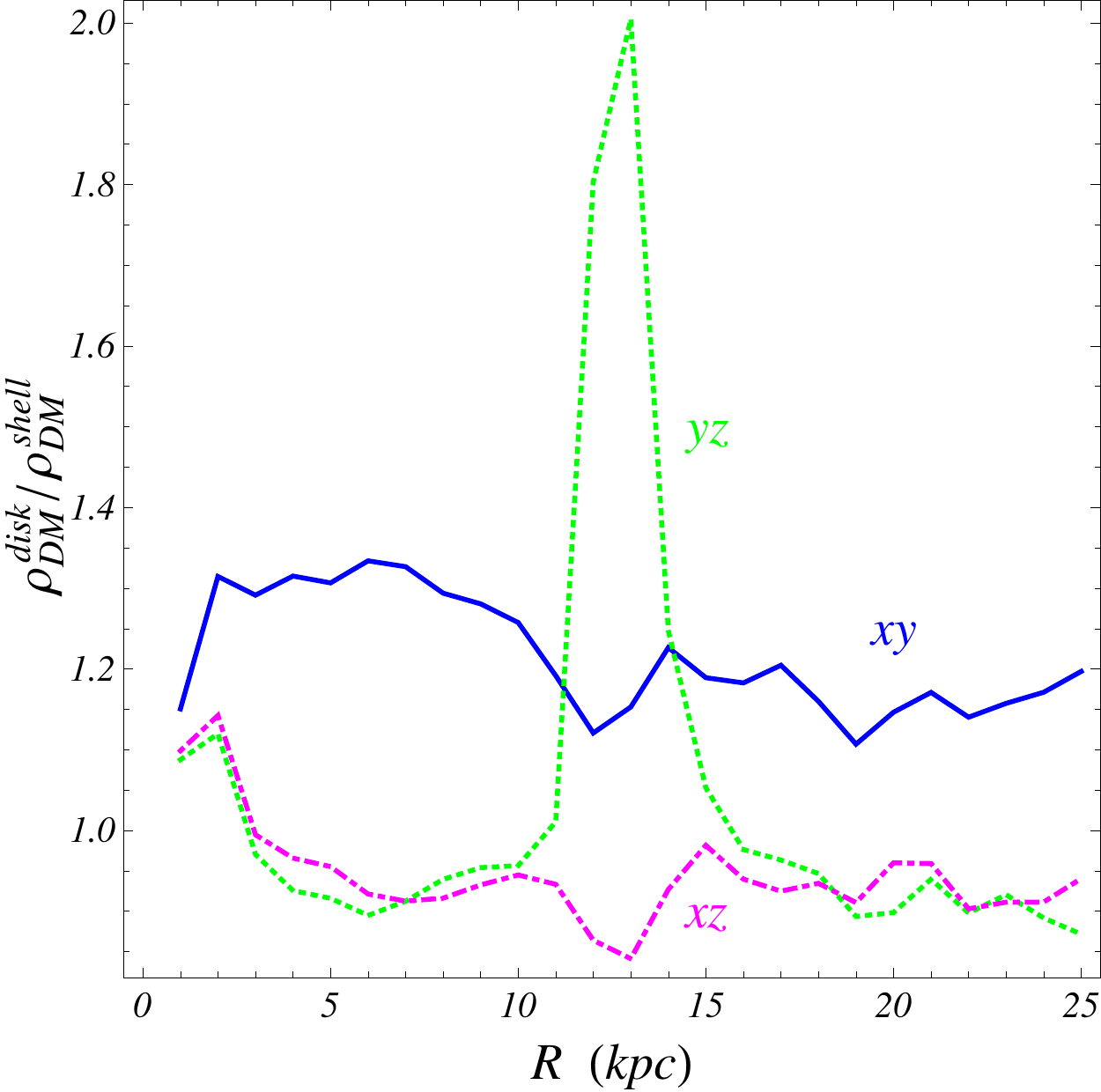}
\end{tabular}
\caption{\small \it 
(Left) Average density of DM particles with $7<R<9$~kpc as a function of the height from the galactic disk z (R is the spherical radius to the galactic center).
The dashed line gives the average value for the entire spherical shell. To select particles in $z$ slices, we used a thickness $\delta z=2$~kpc.
\newline
(Right) Ratio of ring to shell densities as a function of distance from the galactic center for different planes. 
The ratio fluctuates around 1.2 for the galactic plane (blue), while it drops to a value $\sim 0.9$ for other planes (green, magenta). 
For the plane $yz$, the sudden peak at $R \simeq 13$~kpc is due to the presence of a satellite halo, visible on Fig.~\ref{fig:rhomap}.b.
} 
\label{fig:darkdisk}
\end{center}
\end{figure}
\begin{figure}[t]
\begin{center}
\begin{tabular}{cc}
\includegraphics[width=0.45\textwidth]{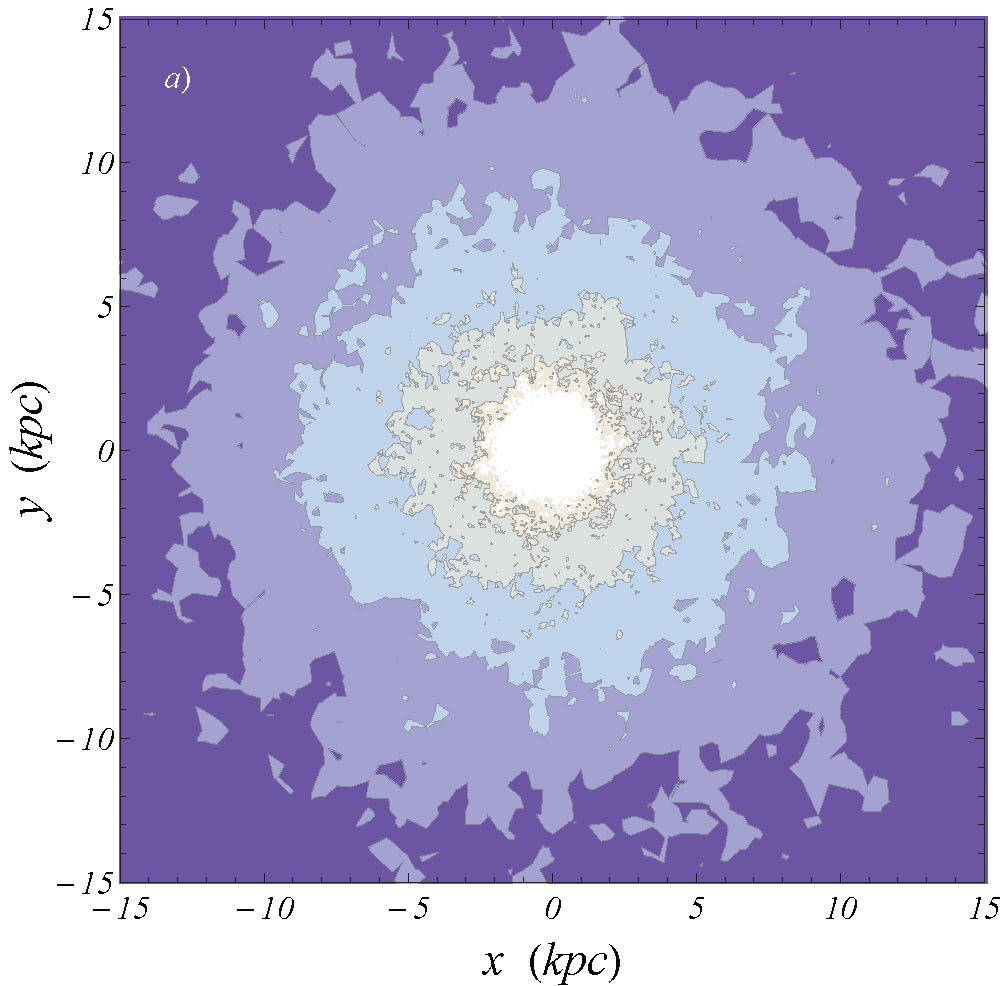}&
\includegraphics[width=0.45\textwidth]{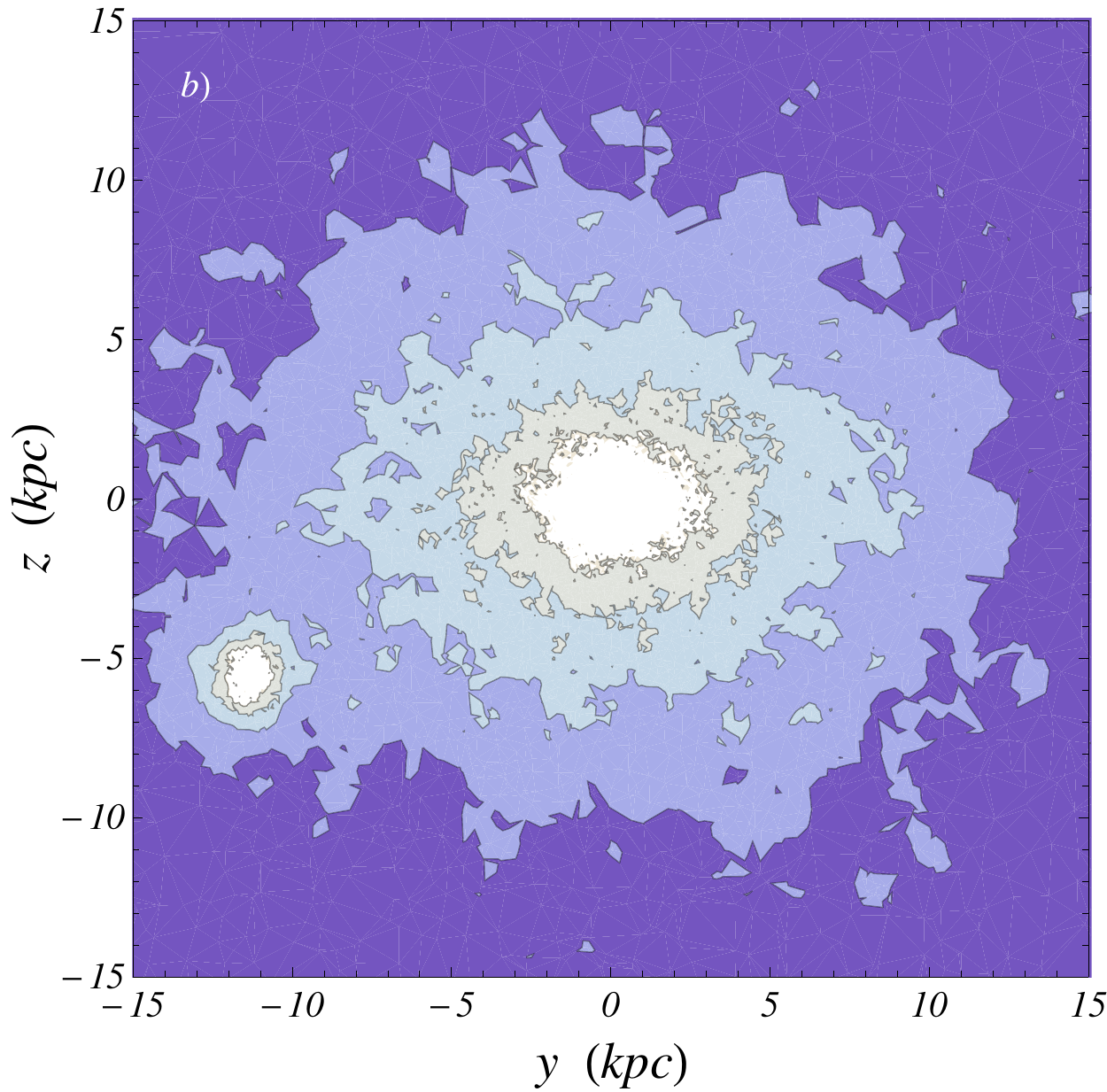}\\
\end{tabular}
\caption{\small \it 
Density maps of the dark matter halo in the planes a) $xy$ (galactic plane), b) $yz$.
\newline
Contours correspond to $\rho_{DM} = \{0.1, 0.3, 1.0, 3.0\}~{\rm GeV/cm^3}$.
}
\label{fig:rhomap}
\end{center}
\end{figure}
\begin{figure}[t]
\begin{center}
\begin{tabular}{cc}
\includegraphics[width=0.45\textwidth]{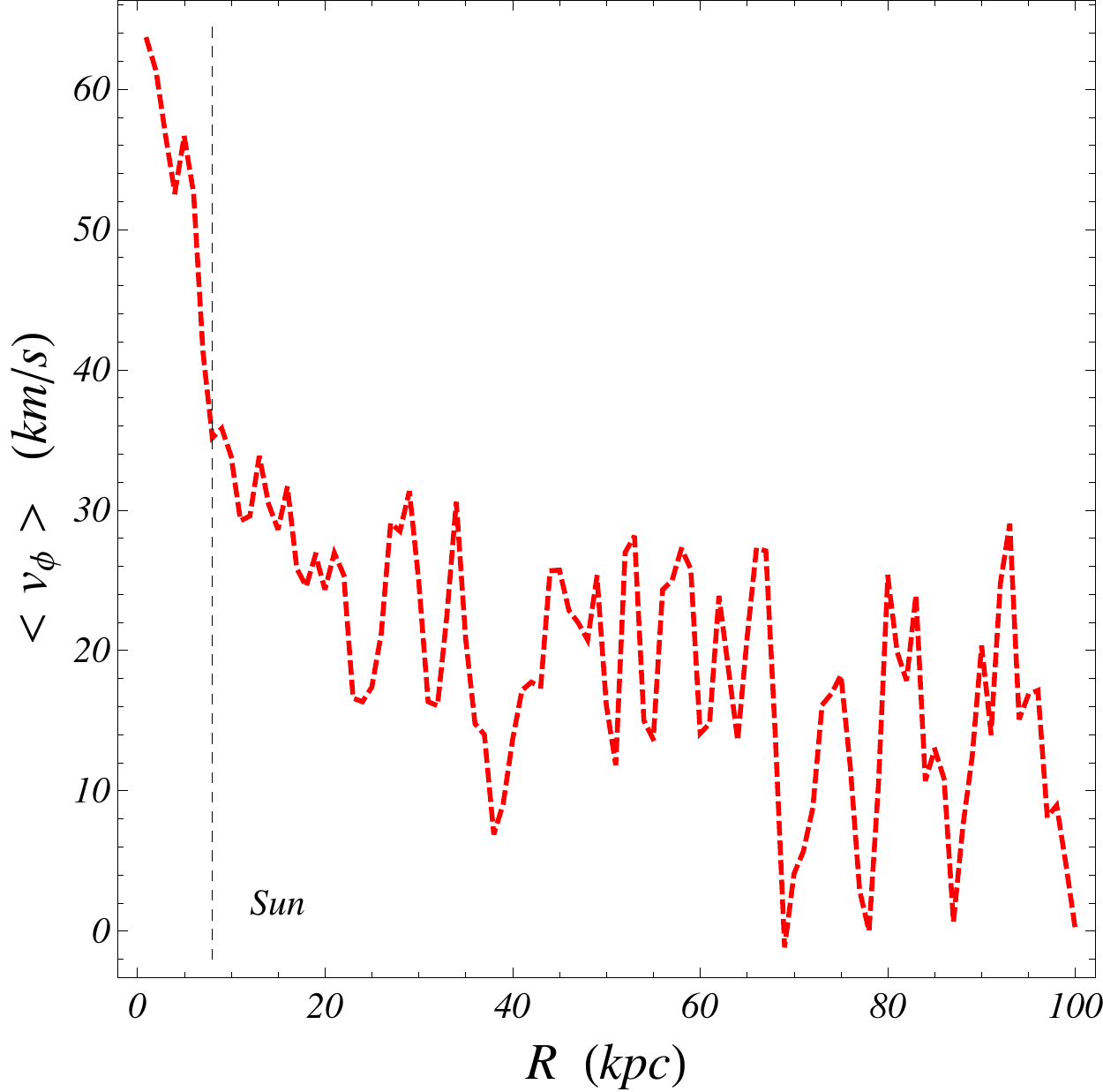}&
\includegraphics[width=0.45\textwidth]{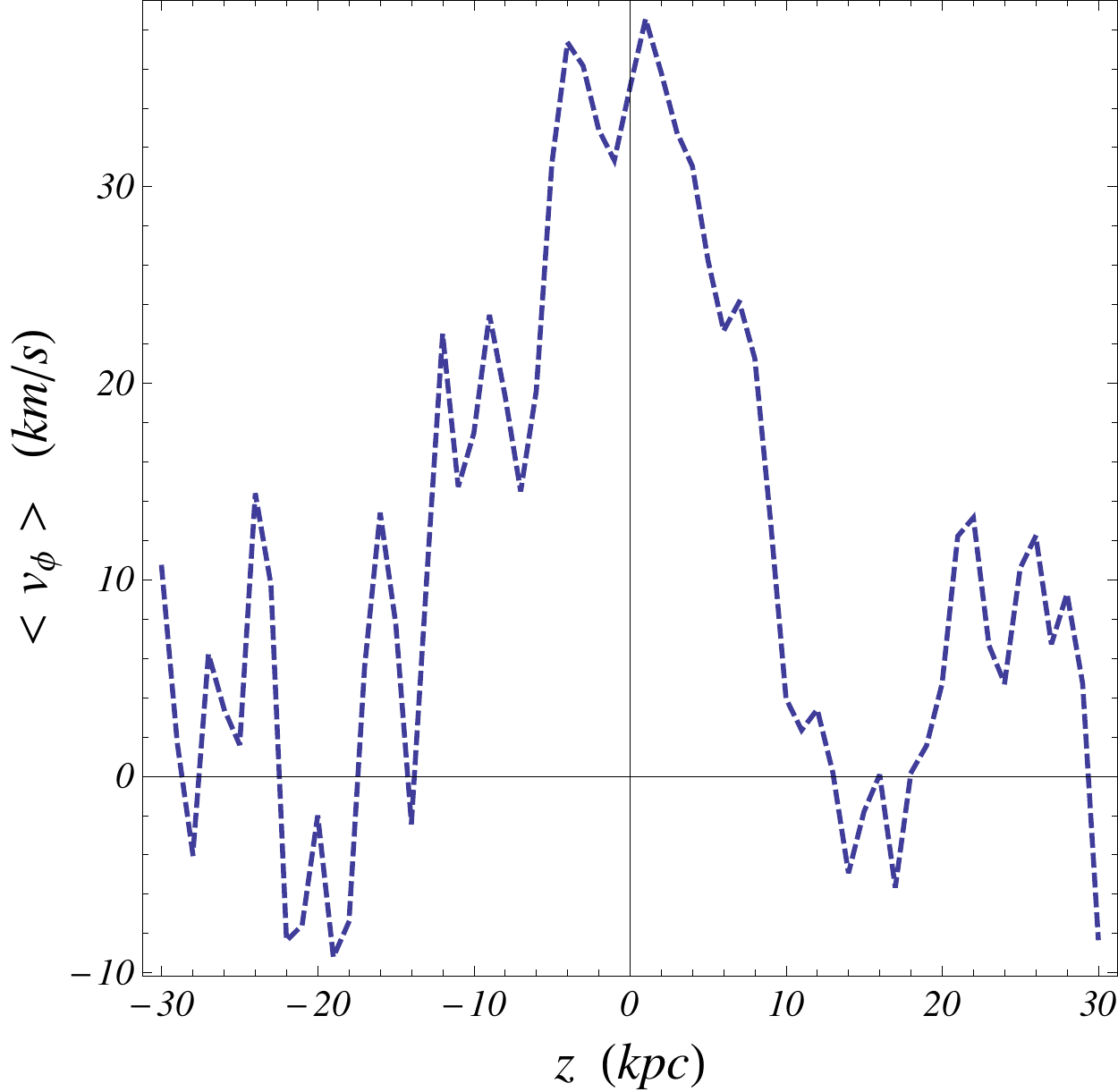}
\end{tabular}
\caption{\small \it 
Mean tangential velocity $\langle v_{\phi} \rangle$ as a function 
of $R$ for $z=0$ (left panel), and as a function of $z$ for $r=8$~kpc (right), where $r=\sqrt{x^2+y^2}$ is the polar radius.
}
\label{fig:vphi}
\end{center}
\end{figure}
As pointed out in Ref.~\cite{Read:2008fh}, the rotation of the DM halo observed in Fig.~\ref{fig:histovgalacticplane} is expected 
in $\Lambda$CDM cosmology with hierarchical structure formation. 
At relatively low redshift, $z<2$, merging satellite galaxies are preferentially dragged towards the already well-formed disk of a host galaxy, 
where they are disrupted by tidal forces.
The accreted material settles in a thick stellar disk and a thick dark matter disk, dubbed the \emph{dark disk}.
The vertical distribution of the stars of the host galaxy thickens due to the impact, but all new stars form in a thin 
disc~\cite{Walker:1995ef,Penarrubia:2006bu,Kazantzidis:2007hy}.
Observational evidence for a thick stellar disk in the Milky Way has been first pointed out by 
Gilmore and Reid (1983)~\cite{Gilmore:1983bv}.
Although the detailed morphology and some quantitative properties of the final galaxy depend on its merger history, 
the existence of a thick stellar disk and a co-rotating dark disk in the $\Lambda$CDM paradigm seems quite robust.
Using three cosmological hydrodynamics simulations, Read {\it et al.}~\cite{Read:2009iv} conclude that the accreted dark matter
can contribute $\sim 0.25 - 1.5$ times the non-rotating halo density at the solar position.
Moreover, the DM $v_\phi$ distribution is best-fit with a double Gaussian, one for the non-rotating halo component,
and one for the rotating accreted DM. The rotation lag of the rotating component is in the range
$0-150$~km/s. A large value of the rotation lag, corresponding to a small halo circular speed, is only found
for galaxies which had no significant merger after a redshift $z=2$.
The importance of the dark disk for the prediction of DM direct detection rates has been acknowledged by several 
authors~\cite{Bruch:2008rx,Read:2009iv}. It could lead to an increase by up to a factor of 3 in the $5-20$~keV recoil energy range.
The signal modulation can also be boosted and the modulation phase is shifted.
However, in Ref.~\cite{Purcell:2009yp}, the authors use high-resolution simulations of accretion events in order to bracket the
range of co-rotating accreted dark matter. They find that the Milky Way's merger history must have been
unusually quiescent compared to median $\Lambda$CDM expectations. 
As a result, the fraction of accreted dark matter near the Sun is less than 20\% of the host halo density,
and does not change the likelihood of detection significantly.

To investigate this topic further, we select star particles with $7<R<9$~kpc and $\vert z \vert<3$~kpc, we have $N_{star}=143,320$.
The distribution of $v_r$ and $v_{\phi}$ are shown on Fig.~\ref{fig:stardist}.
We observe that the dark matter and the star particles are indeed co-rotating in the solar neighborhood.
The mean tangential velocity is $\langle v_{\phi} \rangle=201$~km/s but tends towards
$\langle v_{\phi} \rangle=225$~km/s for stars closer to the galactic plane, 
which is consistent with Milky Way rotation curve data~\cite{Sofue:2008wt}.
Moreover, the $v_{\phi}$ distribution is clearly bimodal, a double Gaussian gives a reasonable fit.
The low velocity component ($\mu=175$~km/s) has a larger velocity dispersion parameter ($v_0=72$~km/s)
than the high velocity component ($\mu=270$~km/s, $v_0=30$~km/s).
To check whether these correspond to the thick and thin disks populations,
we take subsets of the previous selected star particles with $0<v_\phi<150$~km/s
and $v_\phi>275$~km/s respectively, and calculate the disc scale height as well as the total velocity dispersion. 
We obtain $z_d^{thick} \simeq 3$~kpc, $z_d^{thin} \simeq 0.67$~kpc, 
$\sigma_{tot}^{thick} \simeq 165.5$~km/s, $\sigma_{tot}^{thin} \simeq 65.5$~km/s.
These values are higher than those measured for the Milky Way:
from SDSS tomographic results, $z_{MW}^{thick} \simeq 1$~kpc, $z_{MW}^{thin} \simeq 0.35$~kpc~\cite{Juric:2005zr};
for the velocity dispersion, $\sigma_{MW}^{thick} \simeq 85$~km/s~\cite{Soubiran:2002sf}, 
$\sigma_{MW}^{thin} \simeq 40$~km/s~\cite{Nordstrom:2004wd}.
Although the thick and thin stellar disks could be revealed by applying severe cuts on $v_\phi$,
the inferred values of the disk scale height and the total velocity dispersion are not reliable,
they probably overestimate the correct values by up to a factor 2. 
Hence, we stay inconclusive regarding the consistency of stellar populations in this simulation with Milky Way observations. 

The existence of two DM components in the solar neighborhood cannot be assessed undoubtfully in this simulation
due to the limited local resolution. Nevertheless, as shown on panel b of Fig.~\ref{fig:histovgalacticplane}, 
a fit with a double Gaussian gives a better description of the $v_\phi$ distribution than with a single Gaussian.
From this fit, it appears that the mean velocity of the rotating component is comparable for DM and for stars.
The dark disk particles also have a smaller velocity dispersion. 
also, it appears that the dark disk constitutes around 25\% of the total local density in the solar neighborhood.
The value of this fraction can be further checked by calculating average densities from the simulation.
For DM particles in the spherical shell $7<R<9$~kpc, we find $\langle \rho_{DM} \rangle = 0.31~{\rm GeV/cm^3}$,
in agreement with what is found in DM only simulations and the commonly adopted value in this field, $\rho_{DM} = 0.3~{\rm GeV/cm^3}$.
However, when we restrict DM particles to a ring of thickness $\delta z$ around the galactic disk, the average density is higher.
We find $\langle \rho_{DM} \rangle = 0.37~{\rm GeV/cm^3}$ for $\delta z/2 = 1$~kpc and 
$\langle \rho_{DM} \rangle \simeq 0.39~{\rm GeV/cm^3}$ for $\delta z/2 \leq 0.1$~kpc.
The density increase near the galactic disk is indeed about 25\% of the total density in the solar neighborhood.

The extent of the density increase is shown on Fig.~\ref{fig:darkdisk}. 
On the left panel, DM particles with $7<R<9$~kpc are partitioned into slices of thickness $\delta z=2$~kpc parallel to the plane $xy$
(galactic plane), and the average density is shown as a function of $z$.
If the halo was close to spherical, there would be no bump around $z=0$.
On the right panel, we show the disk to shell density ratio for different planes.
For the galactic plane, the density increase due to the dark disk is typically between 15 and 30\% for $R \leq 25$~kpc.
For other planes, the ratio oscillates around 0.9, except when satellite halos are encountered, this is the case for
the plane $yz$ and $R \simeq 13$~kpc, as illustrated by the density maps of Fig.~\ref{fig:rhomap}. 
With the dark disk component, the entire DM halo appears with an oblate shape.
The local density increase due to the dark disk is relatively mild in this simulation,
compared to more extreme cases~\cite{Read:2009iv}.
Therefore, following the results of Ref.~\cite{Bruch:2008rx}, we do not expect a strong increase of the direct detection signal 
in this simulation compared to a standard Maxwellian halo.

Finally, we checked that the DM rotation is not limited to a thin galactic plane but is present in a rather thick dark disk.
In particular, Fig.~\ref{fig:vphi} shows the variation of the mean tangential velocity $v\phi$ as a function
of $R$ for $z=0$ (left panel), and as a function of $z$ for $r=8$~kpc (right), where $r=\sqrt{x^2+y^2}$ is the polar radius.
A significant rotation is found in the entire galactic plane, up to $R=100$~kpc. For $r=8$~kpc, the rotation remains important
up to heights of $\sim 15$~kpc from the galactic plane.

\section{Direct detection experiments and signals}
\label{sec:dde}

Direct detection experiments aim to measure the energy deposited during scatterings of
Weakly Interacting Massive Particles (WIMPs) with the detector material. 
Given the weak interaction rate and the low expected signal,
radiopurity of the material and control of other backgrounds are essential
although experimentally challenging.
In order to attenuate these difficulties, one possibility is to look for a typical signature of the signal 
that is hard to mimic with a background noise.
The DAMA/Libra experiment~\cite{Bernabei:2008yi} is the only one that has observed a positive signal 
with such a signature, namely the annual modulation of the event rate.
However, as the signal could not be confirmed by any other experiment, neither with a heavier or a lighter
target, the compatibility of the DAMA result and the exclusion limits set by these other experiments becomes
more and more controversial~\cite{Gelmini:2008vi,Hooper:2009zm}. 

Ways of reconciliation have been sought in the detector characteristics 
(quenching factors uncertainties and channelling effects~\cite{Bernabei:2007hw,Bottino:2007qg}), 
in particle physics uncertainties (nuclear form factors~\cite{Lewin:1995rx}, 
nucleon effective couplings~\cite{Andreas:2008xy})
or scenarios (elastic scattering~\cite{Gelmini:2004gm,Gondolo:2005hh,Bottino:2002ry,Bottino:2003iu,Bottino:2003cz}, 
inelastic scattering~\cite{TuckerSmith:2001hy,TuckerSmith:2004jv,Chang:2008gd,MarchRussell:2008dy,Cui:2009xq,Finkbeiner:2009ug}),
and in the Dark Matter halo velocity 
distribution~\cite{Gelmini:2000dm,Ling:2004aj,Fairbairn:2008gz,Smith:2006ym,Savage:2009mk,MarchRussell:2008dy}.

In this section, we give all the relevant steps necessary to compute the spin independent scattering signal
of Dark Matter on nuclei in direct detection experiments.
Reference values for all the useful parameters and quantities will be given, and discussed in light of the DAMA controversy.
In this paper, the accent will be put on the impact of a realistic velocity distribution 
(as given by our cosmological simulation with gas and stars)
on the direct detection signal and its modulation. 
In particular, we will focus on how the fit to DAMA  and the other exclusion limits 
change compared to a standard Maxwellian halo. Both the elastic and the inelastic cases will be covered.
For the influence of other parameters, the reader will be referred to the existing literature.

\subsection{Event rate formalism}

We consider collisions between dark matter particles from the Galactic halo and the nuclei of a given low background detector. 
The relevant characteristics of the halo are the local density of dark matter, taken to have the fiducial value 
$\rho_{DM}=0.3$ GeV/cm$^{3}$ at the Sun's location, and the local  distribution of velocities ${\rm f}(\vec v)$ with respect to the Earth.   

The differential event rate of nuclear recoils as a function of the recoil energy $E_R$ is conveniently factored as
\begin{equation}
\label{eq:diffrate}
\frac{d \mathcal{R}}{d E_{R}} = \frac{\rho_{DM}}{M_{DM}} \frac{d\sigma}{d E_{R}} \; \eta (E_R,t)
\end{equation}
where $M_{DM}$ is the WIMP mass, $d \sigma/d E_{R}$ encodes all the particle and nuclear physics factors, and
$\eta (E_R,t)$ is the mean inverse velocity of incoming DM particles that can deposit a recoil energy $E_R$. 
The time dependence of the velocity distribution is induced by the motion of the Earth around the Sun, which leads to a 
seasonal modulation of the event rate \cite{Drukier:1986tm,Freese:1987wu}. 

The total recoil rate per unit detector mass in a given energy bin $[E_1,E_2]$ is obtained by integrating Eq.~(\ref{eq:diffrate}),
\begin{equation}
\label{eq:totrate}
\mathcal{R}(t) = \int_{E_1}^{E_2} d E_R\  \epsilon(E_R)\  \left( \frac{d \mathcal{R}}{d E_{R}} \star G(E_R,\sigma(E_R)) \right) \quad ,
\end{equation}
where $\epsilon$ is the efficiency of the detector and the finite energy resolution of the experiment is taken into account by
convoluting the differential rate with a Gaussian distribution with spread $\sigma(E_R)$. 
For detectors made of several elements, the total rate is the average of the rates $\mathcal{R}_i(t)$ for each component $i$, 
weighted by its mass fraction $f_i$
\begin{equation}
\mathcal{R}(t) = \sum_{i} f_i \mathcal{R}_i(t)
\end{equation}
Finally the expected number of observed events per unit time is the product of the total rate times the detector mass $M_{det}$. 

The particle physics is enclosed in the term $d \sigma/d E_{R}$, which is generally parameterized as
\begin{equation}
\frac{d \sigma}{d E_{R}} = \frac{M_N \sigma^0_n}{2 \mu^2_n}\ \frac{\Big(Z f_p + (A-Z) f_n \Big)^2}{f_n^2} F^2(E_R) \quad ,
\label{eq:sigma}
\end{equation}
where $M_N$ is the nucleus mass, $\mu_n$ is the reduced WIMP/neutron mass, $\sigma^0_n$ is the zero momentum WIMP-neutron effective cross-section, 
$Z$ and $A$ are respectively the number of protons and the atomic number of the element, $f_p$ ($f_n$) 
are the WIMP effective coherent coupling to the proton (resp. neutron),
and $F^2(E_R)$ is the nuclear form factor.
In the sequel, we will consider the case of scalar interactions, 
most relevant for the elastic scattering scenario, for which $f_p \simeq f_n$ and 
the case of weak vector interactions through a $Z$ boson, most relevant for the inelastic scattering scenario, 
for which $f_p =4 \sin^2\theta_W-1$ and $f_n=1$.
The form factor $F^2(E_R)$ characterizes the loss of coherence for non zero momentum transfer.
We use the simple parameterization given by Helm~\cite{Helm:1956zz,Lewin:1995rx}, defined as
\begin{equation}
\label{eq:F_iodine}
F(E_R) = 3 e^{- q^2 s^2/2}\ \frac{J_1(qr)}{qr} \quad ,
\end{equation} 
with $q=\sqrt{2 M_N E_R}$, $s=1.0$~fm, $r=\{ (1.23 (m_N/m_n)^{1/3} - 0.6)^2 + 0.63 \pi^2 - 5 s^2 \}^{1/2}$~fm the effective nuclear radius,
$J_1$ is a spherical Bessel function of the first kind. 
This form factor is optimal for scattering on Iodine \cite{Lewin:1995rx}. 
For simplicity we use the same form factor for all targets (more accurate form factor are off by at most 
${\cal O}(20 \%)$, for some targets and at large recoil energies \cite{Duda:2006uk}).  

Finally, the velocity distribution appears in the quantity
\begin{equation}
\eta (E_R,t) =  \int_{v_{min}} d^3 \vec{v}\  \frac{ {\rm f}(\vec{{v}}(t))}{{v}}  \quad ,
\end{equation}
where $\vec{v}$ the WIMP velocity {\it wrt} the Earth and $v_{min}$ is the minimum velocity needed to provoke a recoil inside the detector. 
Two cases need to be considered. In the elastic scattering scenario, a dark matter particle is simply scattered off a nucleus.
In the inelastic scattering scenario, a dark matter particle $DM_1$ is supposed to scatter into a slightly heavier state $DM_2$,
with a mass splitting $\delta = M_{DM_2} - M_{DM_1} \sim 100$ keV. 
In this scenario, which has been first proposed in Ref.~\cite{TuckerSmith:2001hy} and confronted to the recent data  
in Refs.~\cite{TuckerSmith:2004jv,Chang:2008gd,MarchRussell:2008dy,Cui:2009xq,Finkbeiner:2009ug}, 
a much broader range of dark matter candidates may both fit DAMA and be consistent with the other experiments. 
The threshold velocity is given by
\be
v_{min} = \sqrt{\frac{1}{2 M_N E_R}} \Big(\frac{M_N E_R}{\mu}+\delta\Big) \quad ,
\label{thresvel}
\ee
with $\mu$ the WIMP-nucleus reduced mass. 
This formula encompasses both the elastic ($\delta$ = 0) and the inelastic ($\delta \neq 0$) scattering scenarios.

When the velocity distribution in the galactic frame ${\rm f_{gal}}(v)$ is isotropic, $\eta$ is given by
\be
\eta = \frac{2\pi}{v_\oplus} \int_{v_-}^{v_+} \big( {\rm F}(v_{esc}) - {\rm F}(v) \big) \, dv \quad ,
\label{myeta}
\ee
with $v_\pm = \min\{v_{esc},v_{min} \pm { v}_{\oplus}\}$, $v_{esc}$ is the galactic escape velocity,
$\vec{v}_\oplus(t)$ is the Earth's velocity in the galactic frame, which varies with the time of year $t$,
and ${\rm F}(v)=\int v \, {\rm f_{gal}}(v) \, dv$. In deriving this formula, we have used the realistic assumption 
$v_{esc} > v_\oplus$ at any time $t$.
Notice that ${\rm F}(v)$ is an even function of $v$, 
because ${\rm f_{gal}}$ only depends on the modulus $\vert \vec{v} \vert$ of the velocity.
For a standard Maxwellian distribution with a mean velocity $v_0$,
\be
{\rm f_{gal}}(\vec{v}) = \frac{1}{N(v_{esc})\pi^{3/2}v_0^3} \, e^{-v^2/v_0^2} \quad {\rm for}~v<v_{esc} \quad , 
\ee
where $N(v_{esc})$ is a normalization factor, Eq.~(\ref{myeta}) reads
\be
\eta = \frac{1}{2N(v_{esc})v_\oplus} \left\{ {\rm Erf}\Big( \frac{v_+}{v_0} \Big) - {\rm Erf}\Big( \frac{v_-}{v_0} \Big) 
- \frac{2(v_+-v_-)}{\sqrt{\pi} \, v_0} \, e^{-v_{esc}^2/v_0^2}  \right\} \quad ,
\ee
which agrees with the result of Ref.~\cite{Savage:2008er}. It might seem curious that Eq.~(\ref{myeta}) is expressed
as a definite integral between $v_-$ and $v_+$ while particles with any velocity above $v_{min}$ should contribute to $\eta$.
However, the function $F(v)$ is itself an integral on the velocity distribution. Also the dependence in $v_{esc}$ is now explicit.
In the rest of this paper, we will assume an escape velocity $v_{esc}=600$~km/s, 
somewhat on the upper part of the expected range, $498\  \mbox{\rm km/s}\,< v_{esc}< 608 \,\mbox{\rm km/s}$~\cite{Smith:2006ym}. 
Modifications of the velocity distributions and their impact on the fits, have been discussed in various works, 
see for instance Refs.~\cite{Fairbairn:2008gz,Savage:2009mk,MarchRussell:2008dy,Cui:2009xq}.

\begin{figure}[t]
\begin{center}
\includegraphics[width=0.6\textwidth]{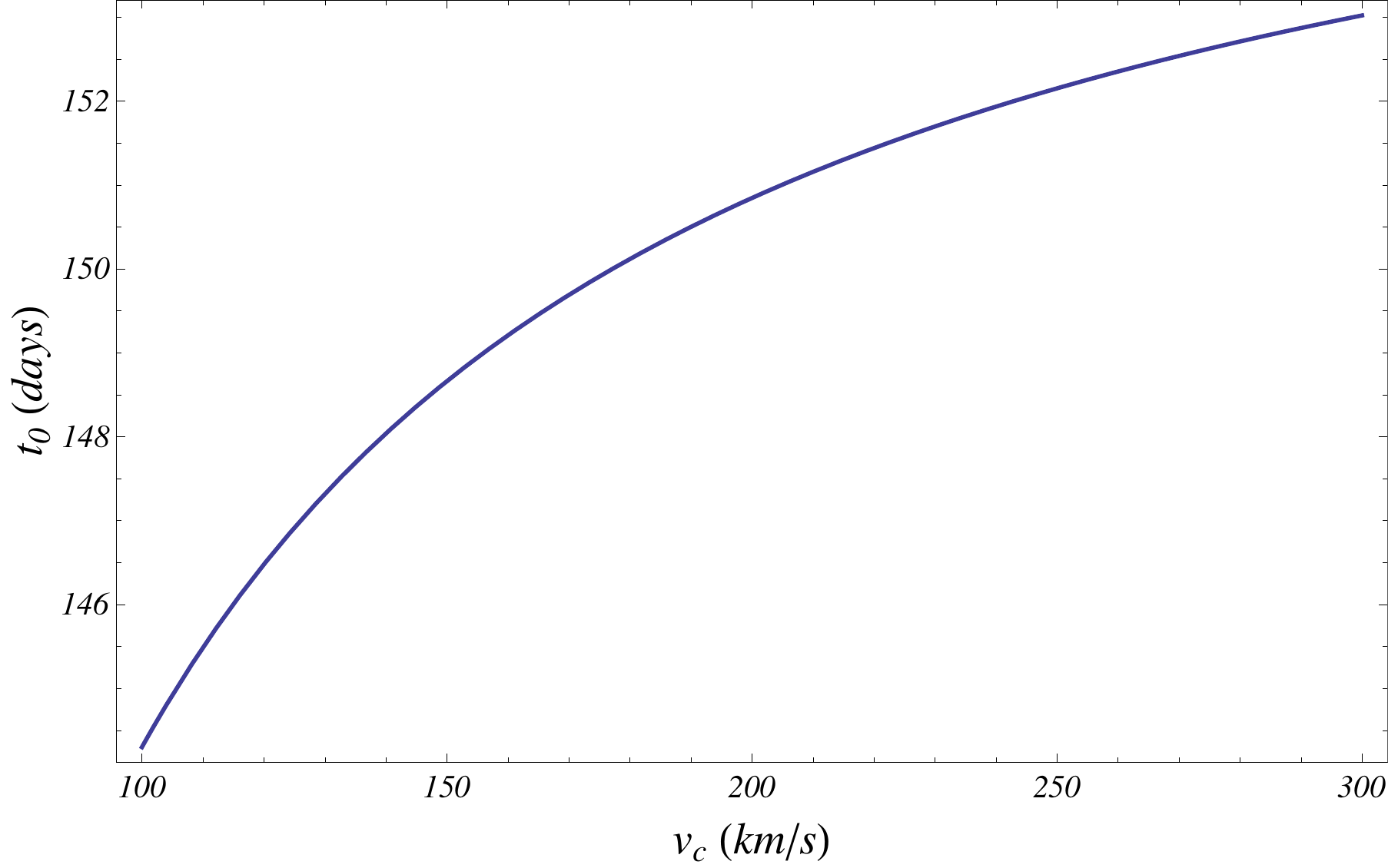}
\caption{\small \it
Variation of the phase $t_0$ corresponding to a maximal Earth velocity and event rate as a function
of the galactic circular velocity $v_c$.}
\label{fig:t0}
\end{center}
\end{figure}
The seasonal modulation of the event rate is caused by the variation of the Earth's velocity throughout the year.
Following Refs.~\cite{Gelmini:2000dm,Green:2003yh}, in the galactic reference frame such that $\vec{1}_X$ is pointing towards the galactic
center, $\vec{1}_Y$ is along the disk rotation, and $\vec{1}_Z$ is pointing towards the galactic north pole, we write
\bea
\vec{v}_\oplus(t) &=& \vec{v}_\odot + \vec{v}_{{\rm EO}}(t) \quad , \nonumber  \\ 
\vec v_\odot &=& \vec{v}_c + \vec{v}_{\odot,pec} = (0,220,0) + (10,5,7)~{\rm km/s} \quad , \nonumber \\ 
\vec{v}_{{\rm EO}}(t) &=& v_{\rm EO}(\vec{e}_1 \sin \omega(t-t_1) - \vec{e}_2 \cos \omega(t-t_1)) \quad ,
\eea
where the velocity $\vec v_\odot$ of the Sun with respect to the halo~\cite{Dehnen:1997cq} is decomposed as the sum of 
the galactic disk circular velocity $\vec{v}_c$ and a peculiar velocity $\vec{v}_{\odot,pec}$,
the Earth mean orbital velocity is $v_{\rm EO} = 29.8$~km/s, with $\omega = 2\pi/T$ and a period $T=1$~year,
$t_1 = 79.62$~days is the time of the vernal equinox, and $\vec{e}_1 = (-0.067, 0.4927, -0.8676)$, 
$\vec{e}_2 = (-0.9931, -0.1170, 0.01032)$ (the Earth's orbital eccentricity is neglected).
Therefore the Earth velocity follows a cosine function. As $v_{\rm EO} \ll v_\odot$, we have
\be
v_\oplus(t) \simeq v_\odot + v_{\rm EO} \sin \gamma \cos \omega (t-t_0) \quad ,
\ee
with $\gamma \simeq 29.3^\circ$ is the angle between the ecliptic plane perpendicular axis and $\vec v_\odot$.
The Earth velocity is maximal for $t=t_0$, with
\be
t_0 = t_1 + \frac{\pi}{2\omega} + 
\frac{1}{\omega} \arctan \frac{\vec{v}_\odot \cdot \vec{e}_2}{\vec{v}_\odot \cdot \vec{e}_1} 
\simeq 151.5~{\rm days} \quad .
\label{phase_t0}
\ee
With these values, the Earth velocity modulation amplitude amounts to $6.2\%$. 
As pointed out in Ref.~\cite{Reid:2009nj}, the circular velocity of the Sun may be larger than the value $v_c = 220~{\rm km/s}$ usually considered.
Also, if the DM halo is co-rotating with the galactic disk, the relative velocity relevant for calculating the direct detection signal is
effectively reduced. 
In Fig.~\ref{fig:t0}, the circular velocity is varied between 100~km/s and 300~km/s, the corresponding phase
$t_0$ increases with $v_c$, from $t_0 = 144$ to $t_0 = 153$~days.

For an isotropic velocity distribution, the differential rate also follows a temporal cosine function at first order in $v_{\rm EO}/v_\odot$.
Indeed, the only time dependent factor in Eq.~(\ref{eq:diffrate}) is $\eta$, for which we get, in the limit $v_{esc}>v_+$
\be
\eta(E_R,t) \simeq \eta_0(E_R) + \eta_1(E_R) \frac{v_{\rm EO}}{v_\odot} \sin \gamma \cos \omega (t-t_0) \quad ,
\label{eq:eta_t}
\ee
with 
\bea
\eta_0(E_R) &=& \eta(E_R,v_{\rm EO}=0) \quad ,  \nonumber \\
\eta_1(E_R) &=& \frac{2\pi}{v_\odot} \int_{v_-^0}^{v_+^0} F(v) \, dv
-2 \pi \Big( F(v_+^0) + F(v_-^0) \Big) \quad ,
\label{eta1}
\eea
with  $v_\pm^0 = v_{min}(E_R) \pm { v}_{\odot}$.
It should be stressed that the time of year $t=t_0$ does not necessarily correspond to a maximum of the event rate.
For low energy events, Eq.~(\ref{eta1}) shows that it is a minimum instead, as $\eta_1$ becomes negative.

In this paper, we also need to consider discrete velocity distributions as the output of a simulation gives a list
of particles with their position and velocity.
If we denote by $N_{part}$ the number of particles, and by $\vec{v}_i$ the velocity of the particle $i$ in the galactic
frame, the distribution can be written as
\be
{\rm f_{gal}}(\vec{v}) = \frac{1}{N_{part}} \sum_i \delta(\vec{v}-\vec{v}_i) \quad ,
\ee
where $\delta$ is the Dirac delta distribution in 3~dimensions.
Therefore, for $\eta$, we get in this case
\be
\eta = \frac{1}{N_{part}} \sum_i \frac{1}{w_i} H(w_i-v_{min}) \quad ,
\ee
with $w_i = |\vec{v}_i-\vec{v}_\oplus|$ and $H(x)$ is the Heaviside step function.
To compute the total rate, or the modulation amplitude in some energy bin, we need to integrate the differential event rate.
For discrete distributions, we take advantage of the step-like shape of the differential event rate for a given flow $\vec{v}_i$
to get accurate results.

\subsection{DAMA modulation data}

The former DAMA/NaI \cite{Bernabei:2000qi} and its follow-up DAMA/LIBRA \cite{Bernabei:2008yi} are experiments run at the LNGS
at Gran Sasso, Italy using NaI(Tl) crystals as target. 
They are designed to detect the dark matter recoil off nuclei through the model independent annual modulation signature 
which is due to the motion of the Earth around the Sun.  The experimental results obtained by
DAMA/LIBRA, with an exposure of 0.53 ton$\times$yr collected over 4 annual cycles, combined with the ones of DAMA/NaI, 
for an exposure of 0.29 ton$\times$yr collected over 7 annual cycles, corresponding to a total exposure of 0.82 ton$\times$yr, 
show a modulated signal with a confidence level of 8.2 $\sigma$~\cite{Bernabei:2008yi}. 

\begin{figure}[t]
\begin{center}
\includegraphics[width=0.6\textwidth]{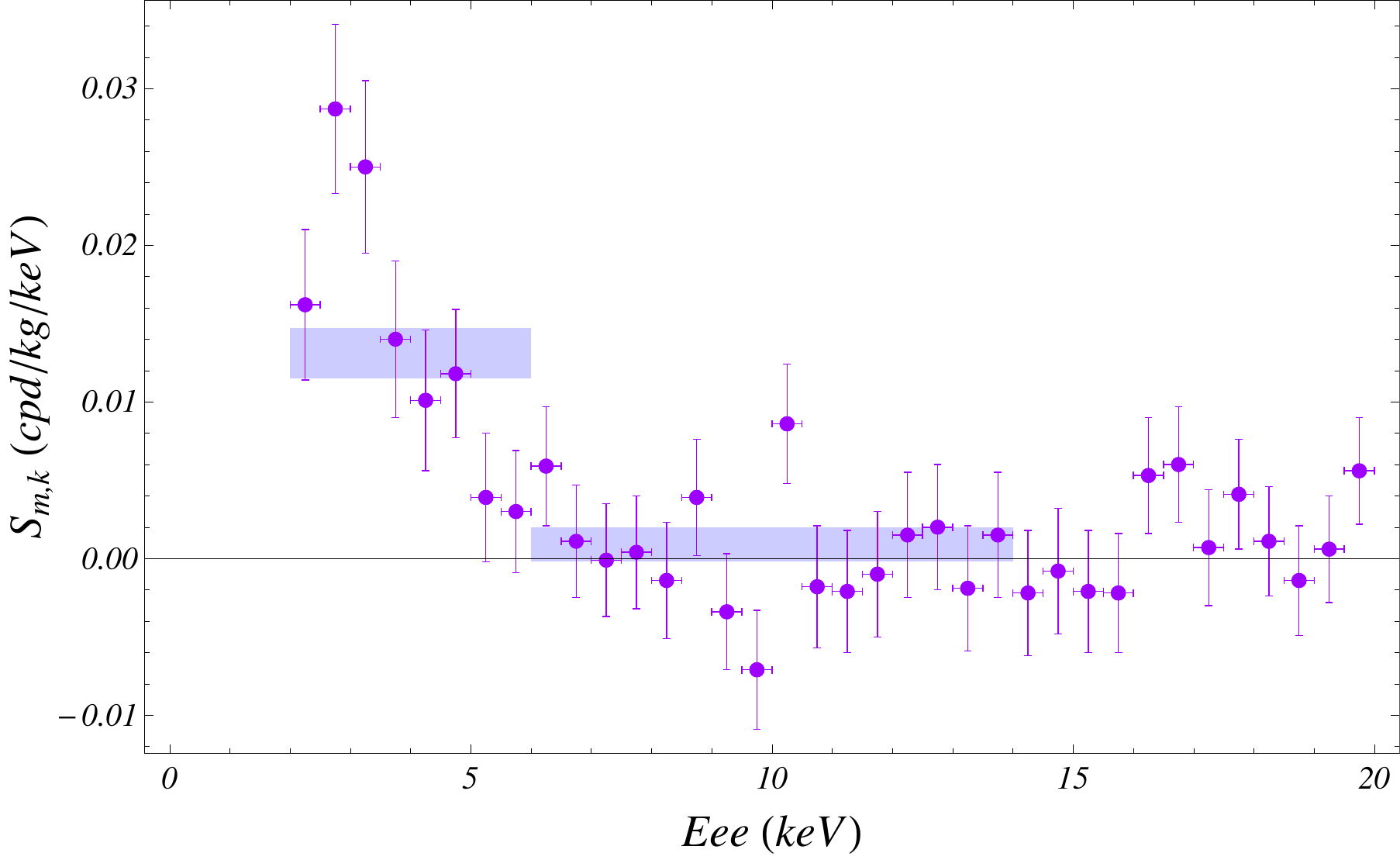}
\caption{\small \it
DAMA modulation amplitude data (in counts per day per kg  per keV) 
as a function of the recoil energy, expressed in electron-equivalent keV units.
The 36 bins, from 2 to 20 keV, show a modulation signal in the low energy part, from 2 to 6 keV.
The upper part is compatible with zero modulation, as shown in the two bins version of the data
(from 2 to 6 keV and from 6 to 14 keV).}
\label{fig:DAMAmod}
\end{center}
\end{figure}
The DAMA modulation data, reproduced on Fig.~\ref{fig:DAMAmod}, are obtained by adjusting (with a likelihood function) 
the observed event rate in each energy bin $k$ to a temporal cosine function 
\begin{equation}
 S_k = S_{0,k} + S_{m,k} \cos \omega(t-t_0)  \; , 
\end{equation}
where $S_{0,k}$ is the constant part of the signal, $S_{m,k}$ is the modulation amplitude.
The DAMA collaboration uses the value $t_0=152.5$~days (2nd of June) for the phase corresponding to a maximal Earth velocity
and a maximal event rate.
It should be stressed that the true value of $t_0$ depends in general upon the velocity of the Sun,
and upon the DM velocity distribution.
However, for an isotropic distribution, Eqs.~(\ref{phase_t0}-\ref{eq:eta_t}) show that the phase $t_0$ is completely
determined once the velocity of the Sun in the galaxy is specified. 
For the low energy region where the signal modulation is present, the experimental value of $t_0$ has been determined 
by the DAMA collaboration.
When the phase is allowed to vary freely, and fitting again the event rate in the energy interval $2 \leq Eee \leq 6$~keV with 
a cosine function, the best-fit value for $t_0$ is 
\be
t_0 = 144.0 \pm 7.5~{\rm days}~(1 \sigma) \quad .
\ee

The DAMA spectrum is given in  keVee (electron-equivalent keV). 
The observed energy released in scintillation light is  related to the nucleus recoil energy through a so-called 
quenching factor $q$, $E_{\mbox{\rm scint.}} = q \cdot E_{\mbox{\rm recoil}}$. This expresses the fact that a recoiling 
nucleus may loose energy by collisions with other nuclei, hence in the form of heat, or through collisions with electrons, 
which create scintillation light.  
The reference values for Iodine and Sodium are respectively $q_I = 0.09$ and $q_{Na} = 0.3$. 
However it has been pointed that the so-called channelled events may play a role \cite{Bernabei:2007hw,Bottino:2007qg}. 
This refers to events in which a recoiled nucleus moves along the axis of the NaI crystal, 
losing most of its energy by collisions with electrons, in which case the quenching factor may be larger, up to $q\approx 1$.  
Once channelling is taken into account, collisions of light WIMPs with Iodine become relevant, 
while recoils on Sodium become negligible in all 
scenarios~\cite{Bottino:2007qg,Bottino:2008mf,Petriello:2008jj,Savage:2008er,Fairbairn:2008gz,Chang:2008xa}.  
We  use the fraction $f$ of channelled events given in Ref.~\cite{Fairbairn:2008gz}, 
\begin{equation}
f_{\rm Na}(E_R) = {e^{-E_R/18}\over 1+ 0.75 E_R}\,, \qquad f_{\rm I} = {e^{-E_R/40}\over 1 + 0.65 E_R}\, \quad ,
\end{equation}
where the recoil energy $E_R$ is in keV. We have verified that the other parameterizations found in the literature 
\cite{Chang:2008gd,Petriello:2008jj,Savage:2008er} give identical results.
Finally, we use the energy resolution of the detector given by the collaboration~\cite{Bernabei:2008yi}
\begin{equation}
\frac{\sigma(E)}{E}= \frac{0.45}{\sqrt{E}}+0.0091 
\end{equation}
and a detector efficiency $\epsilon = 1$~\cite{Bernabei:2008yi,Savage:2008er}. 

To fit the observed energy spectrum of Fig.~\ref{fig:DAMAmod} with a theoretical model, we use a
goodness-of-fit (GOF) method with a $\chi^2$ metric, for the 12 first bins of data, with width $0.5$ keVee, 
from $2.0$ to $8.0$ keVee~\cite{Bernabei:2008yi}. 
\begin{equation}
\chi^2 = \sum_{i=1}^{n}\frac{(S_i-S_i^{obs})^2}{\sigma^2_i}
\end{equation}
where $S_i$ is the predicted value of the modulation amplitude in the bin number $i$,
$S_i^{obs}$ is the value reported by DAMA and $\sigma_i$ is the corresponding experimental uncertainty. 
As data above $8.0$~keVee are essentially compatible with no modulation signal, including them in the $\chi^2$ fit enlarges
the allowed regions in the parameter space.
For explicit numerical values of $S_i^{obs}$, we refer the reader to Table~III of Ref.~\cite{Savage:2008er}.  
As emphasized earlier, these values were obtained by the DAMA collaboration with the hypothesis $t_0 = 152.5$.
To be consistent with their analysis and to facilitate comparisons with previous analyses, 
we will also use this fixed value of $t_0$ for the determination
of the allowed regions in the parameter space, although the true phase corresponding to the maximal event rate
might be different for some energy bins.   
For two parameters ($M_{DM}$ and $\sigma$ in the elastic scenario, and $\delta$ and $M_{DM}$ in the inelastic scenario),
the $\chi^2$ metric has 10 degrees of freedom, therefore the 90\% (99\% and $99.9\%$) 
confidence level (CL) corresponds to $\chi^2 < 16.0$ (resp. $\chi^2<23.2$ and $\chi^2< 29.6$).

\subsection{Other experiments exclusion limits}

So far all the other direct detection experiments searching for dark matter are compatible with null results. 
In this section we briefly describe the experiments that lead to the most constraining limits on both the elastic and the inelastic scenarios.

In the elastic scenario, light nuclei, like Aluminum ($A=27$) and Silicon ($A=29$), are more sensitive to light WIMPs scattering $M_{DM} \sim $ multi-GeV, 
and provide the strongest upper bounds on the allowed parameter space favored by DAMA. In the inelastic scenario, 
which involves heavier candidates, experiments made of heavy nuclei, like Iodine ($A=127$), Xenon ($A=129$) 
and Tungsten ($A=184$) are the most constraining ones. 
Germanium ($A=73$) made detectors  fall in between. 

In computing the rate of Eq.~(\ref{eq:totrate}) for each experiment, 
we  uniformly assume that the small number of events seen, if any,  are signals from dark matter and we use the Poisson 
statistics to find the parameter space excluded at a given confidence level (CL). Integrating over the exposure time, 
the upper bounds are obtained by requiring that the total number of events $N_{tot}$  is compatible with the number of observed 
events at $99.9\%$ of CL, while Ref.~\cite{Cui:2009xq} uses $99\%$ CL. 
This means that we take less severe constraints, leading to slightly smaller excluded regions.
We have checked that the exclusion curves that we obtain reproduce fairly well the published 90\% CL upper bounds for each experiment. 
In the following we describe the main features of the experiments considered, their particular characteristics, and the $\chi^2$ function adopted.\\

\vspace{0.5cm}

\underline{CDMS}: \  The Cryogenic CDMS experiment at Soudan Underground Laboratory operates Ge and Si made solid-state detectors. 
For a heavy WIMP, the Ge data are more constraining: we have considered the ensemble of the three released runs, 
with respectively an exposure of 19.4 kg-day~\cite{Akerib:2004fq}, 34 kg-day~\cite{Akerib:2005kh} after cuts and a total 
exposure of 397.8 kg-days before cuts for the third run~\cite{Ahmed:2008eu}. 
The sensitivity to nuclear recoils is in the energy range between 10-100 keV while the efficiencies and the energy resolution 
of the detectors are given in~\cite{Gondolo:2005hh}. 
For the third run, the efficiency is parameterized as $\epsilon(E_R)=0.25+0.05(E_R-10)/5$ for $10\ \rm{keV}<E_R<15\ \rm{keV}$ and 
$\epsilon(E_R)=0.30$ for $E_R \geq 15\ \rm{keV}$. 
The searches on Si are more sensitive to light DM candidates. For the run released in Ref.~\cite{Akerib:2003px},
two 100~g Si ZIP detectors were used, and the data set corresponds to 65.8~live days.
The energy-dependent efficiency has been modeled in Ref.~\cite{Savage:2008er} as
$\epsilon(E_R) = 0.80 \times 0.95 (0.10+0.30 (E_R-5)/15)$ for $5\ \rm{keV}<E_R< 20\ \rm{keV}$ and 
$\epsilon(E_R)= 0.80 \times 0.95 (0.40+0.10 (E_R-20)/80)$ for $20\ \rm{keV}<E_R<100\ \rm{keV}$. 
Two candidate events with energies $E_R \simeq 55$~keV and $E_R \simeq 95$~keV were observed,
consistent with zero event once expected neutron background is taken into account. 
For the run released in Ref.~\cite{Akerib:2005kh}, the total exposure is 12~kg-day after cuts, and the energy range
has a higher minimum threshold $7\ \rm{keV}<E_R<100\ \rm{keV}$. 

For zero observed event and Poisson distributed data, the $\chi^2$ function reduces to
\be
\chi^2 = 2 N_{tot}
\ee
where $N_{tot}$ is the total predicted number of events in the entire range of energy, for given values $(p_1, p_2, \dots)$ of 
parameters $(P_1, P_2, \dots)$ of the tested model.
Therefore, for a $99.9\%$ CL limit and two parameters, the exclusion limit is obtained by requiring that
$N_{tot}$ is less than $6.9$.

\vspace{0.2cm}
 
\underline{XENON10}: \   XENON10 is a dual-phase Xenon chamber operating at LNGS,
with a total exposure of 316.4 kg-days and 10 candidate events in the recoil energy range 
between 4.5-26.9~keV~\cite{Angle:2007uj,Angle:2008we}.
For the data analysis, we follow Ref.~\cite{Fairbairn:2008gz}, using the 7 energy bins provided by the 
collaboration and a $\chi^2$ for Poisson distributed data
\be
\chi^2 = 2 \sum_i \left \{ N_i^{pred} + B_i - D_i + D_i \log \left( \frac{D_i}{N_i^{pred} + B_i} \right) \right\} \quad ,
\ee
with $N_i^{pred}$ is the predicted number of events in the $i$th bin, $B_i = (0.2, 0.3, 0.2, 0.8, 1.4, 1.4, 2.7)$ is the expected background,
$D_i = (1, 0, 0, 0, 3, 2, 4)$ is the number of detected events, and the log term is zero if $D_i = 0$. 
For a $99.9\%$ CL limit and two parameters, the value of the $\chi^2$ function must be less than 13.8.
As observed in Ref.~\cite{Fairbairn:2008gz}, using a constant nuclear recoil scintillation efficiency $\mathcal{L}_{eff}=0.19$ gives
a slightly underestimated energy threshold of 4.5~keV. Correcting for this bias leads to a slightly less severe exclusion limit.
In this paper however, we will keep the energy bins as given by the XENON collaboration.
Finally, we can notice that this experiment is also sensitive to inelastic dark matter in a similar way as the DAMA experiment, 
due to the close proximity of the target nucleus masses.

\vspace{0.2cm}

\underline{CRESST}: \ The CRESST-I and CRESST-II experiments at LNGS are mostly sensitive to light and heavy WIMPs respectively. 
In the first data release (CRESST-I)~\cite{Angloher:2002in}, the prototype detector module uses $\rm{Al}_2 \rm{O}_3$ crystals as target, 
with a total exposure of 1.51 kg-days covering the energy range between 0.6 - 20 keV.
The energy resolution is given by $\sigma(E) = \sqrt{ (0.519\  keV)^2+(0.0408\  E)^2}$ and 
the collaboration reports eleven observed events after cuts. 
Therefore, for Poisson distributed data, the number of predicted events cannot exceed 23 at 99.9\% CL.
For the second commissioned run of 2007 (CRESST-II)~\cite{Angloher:2004tr,Angloher:2008jj}, the target is changed to $\rm{CaWO}_4$.
The presence of heavy nuclei of Tungsten enhances the sensitivity to spin-independent inelastic scatterings. 
The energy range for the CRESST-II Zora and Verena detectors is 12-100 keV, with a total exposure of 47.9 kg-days
before cuts and an acceptance of 0.9 on Tungsten recoil.
The collaboration measured seven events, leading to a limit $N_{tot}<16$ at 99.9\% CL.\\

\vspace{0.5cm}

Other experiments are potentially relevant, like ZEPLIN, CoGeNT and TEXONO. 
Also, the total rate observed by DAMA provides a constraint that excludes part of the parameter space. 
However, to avoid the cluttering of the figures, we do not include these limits in this paper,
as they are weaker than the constraints presented.

\subsection{Results for the Elastic Scenario}
\label{sec:el}

\bfig[t]
\bc
\bt{cc}
\includegraphics[width=0.45\textwidth]{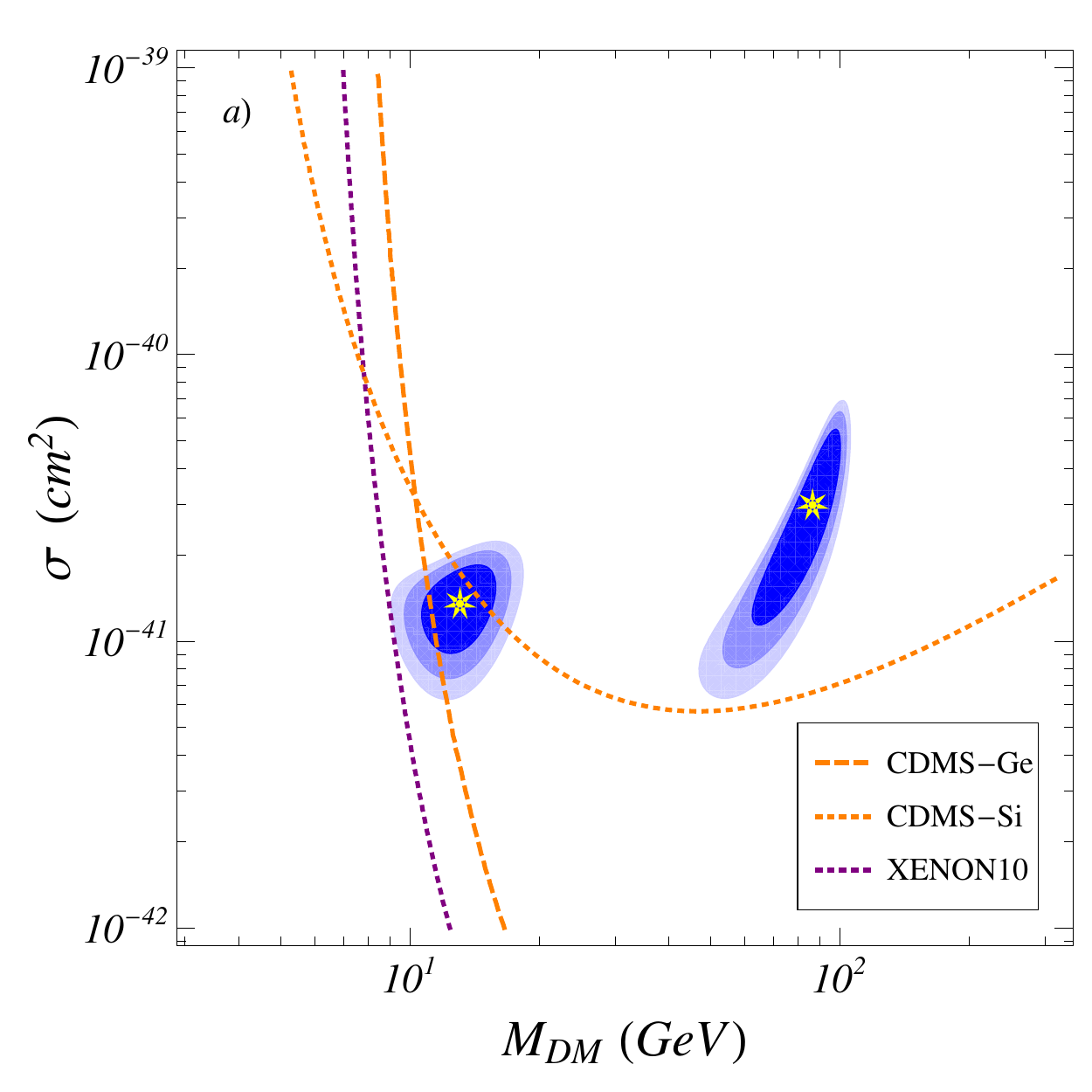} \quad &
\includegraphics[width=0.45\textwidth]{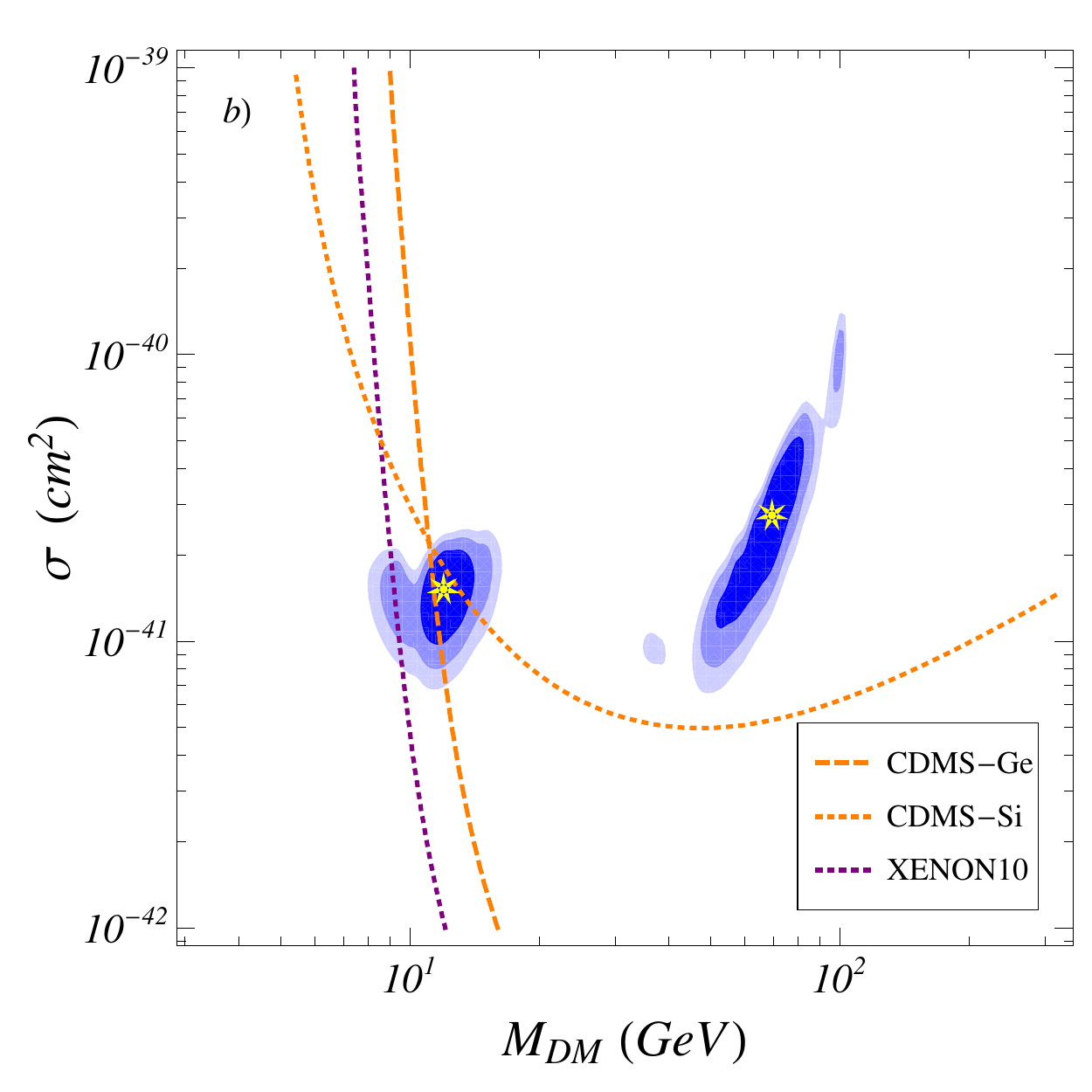} \quad \\
\includegraphics[width=0.45\textwidth]{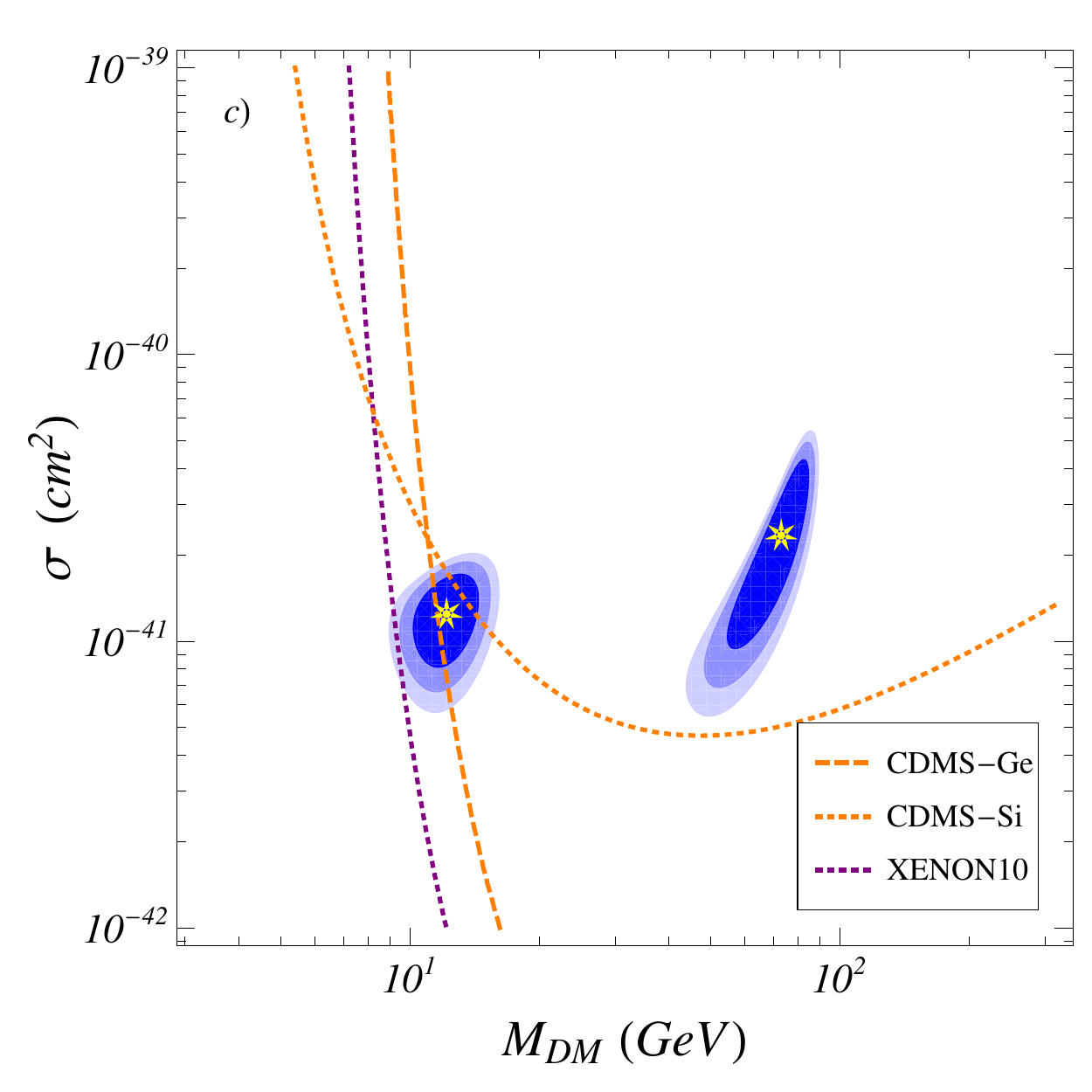} \quad &
\includegraphics[width=0.45\textwidth]{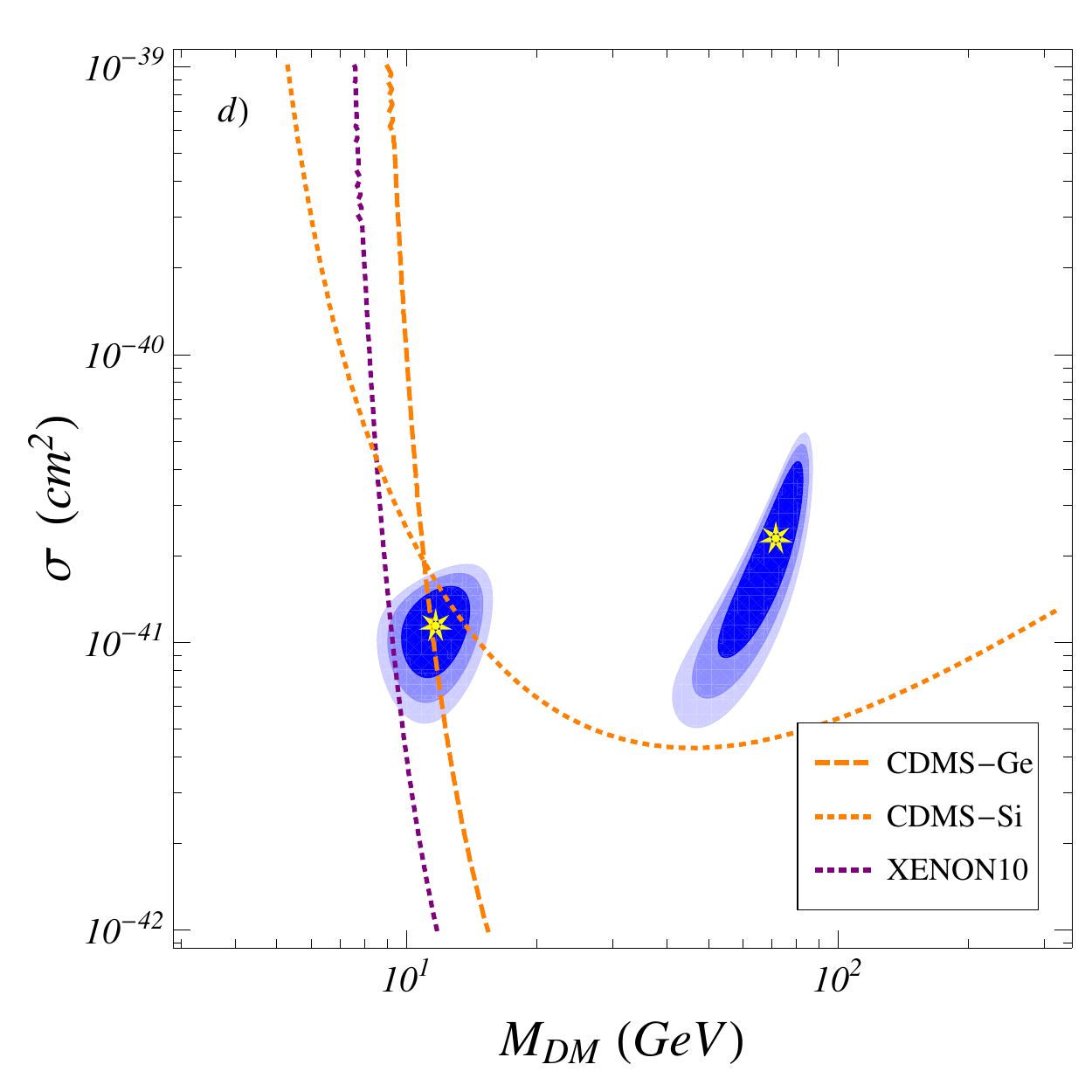} \quad
\et
\caption{\small \underline{Elastic Scenario}, allowed regions in the plane 
$M_{DM}-\sigma$ (DM mass vs. spin independent cross-section on nucleon at zero momentum transfer)
\it
\newline
a) Standard Maxwellian halo ($\rho_{DM}=0.3~{\rm GeV/cm^3}$, $v_c=220$~km/s, $v_0=220$~km/s, $v_{esc}=600$~km/s)
\newline
b) N-body simulation with baryons ($\rho_{DM}=0.37~{\rm GeV/cm^3}$, $v_c=220$~km/s)
\newline
c) Generalized Maxwellian distribution ($\rho_{DM}=0.37~{\rm GeV/cm^3}$, $v_c=190$~km/s, $v_0=301$~km/s, $\alpha=1.5$)
\newline
d) Tsallis distribution ($\rho_{DM}=0.37~{\rm GeV/cm^3}$, $v_c=190$~km/s, $v_0=267.2$~km/s, $q=0.773$)
\newline
For c) and d), the circular velocity $v_c$ has been reduced compared to the standard value in order to take into account a halo rotation.
DAMA contours correspond to $90$, $99$ and $99.9$\% CL. Stars indicate local best-fit points.
All other exclusion curves are at the $99.9$\% CL.}
\label{fig:elchi2}
\ec
\efig
\bfig[t]
\bc
\bt{cc}
\includegraphics[width=0.45\textwidth]{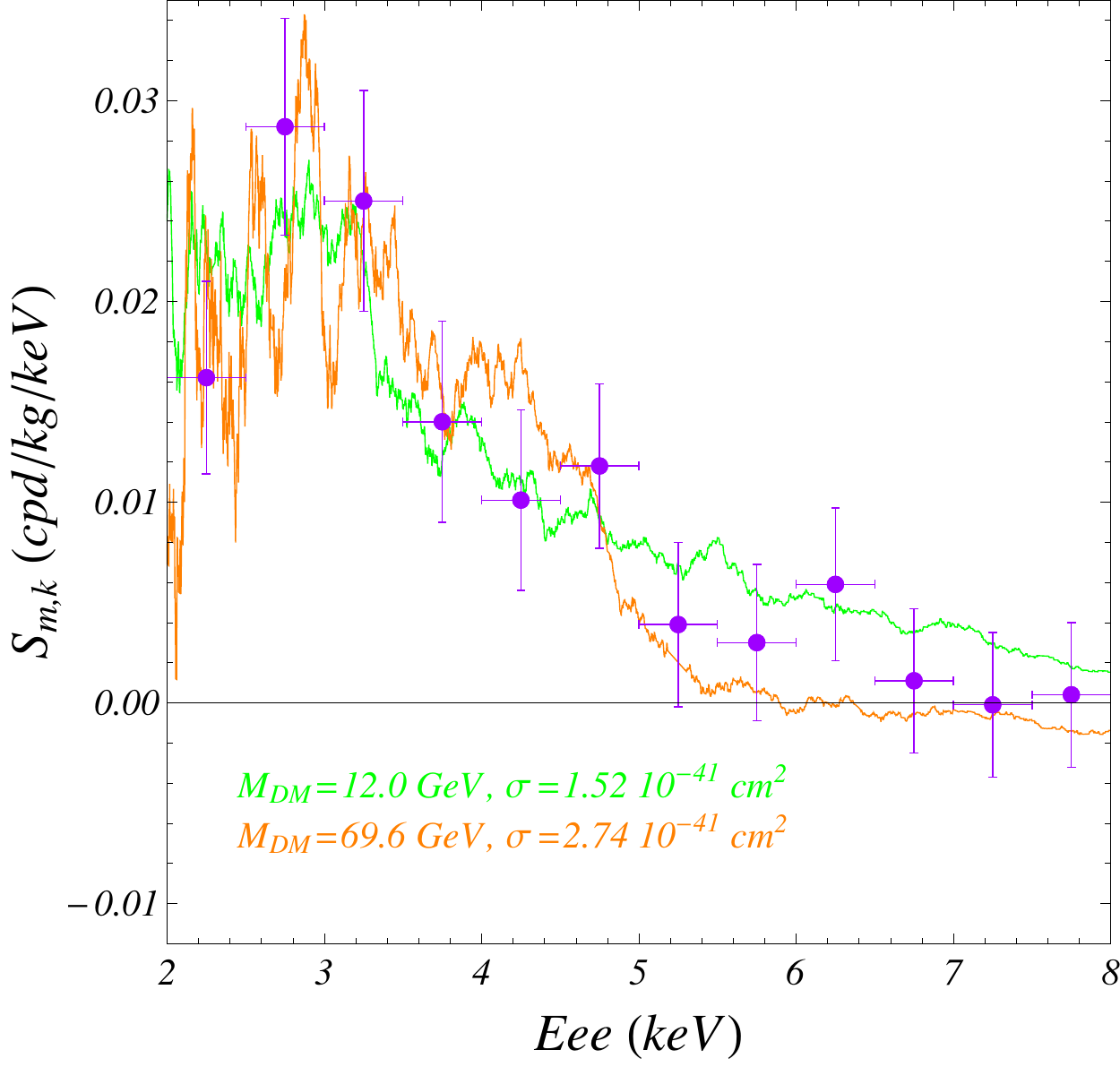} \quad &
\includegraphics[width=0.45\textwidth]{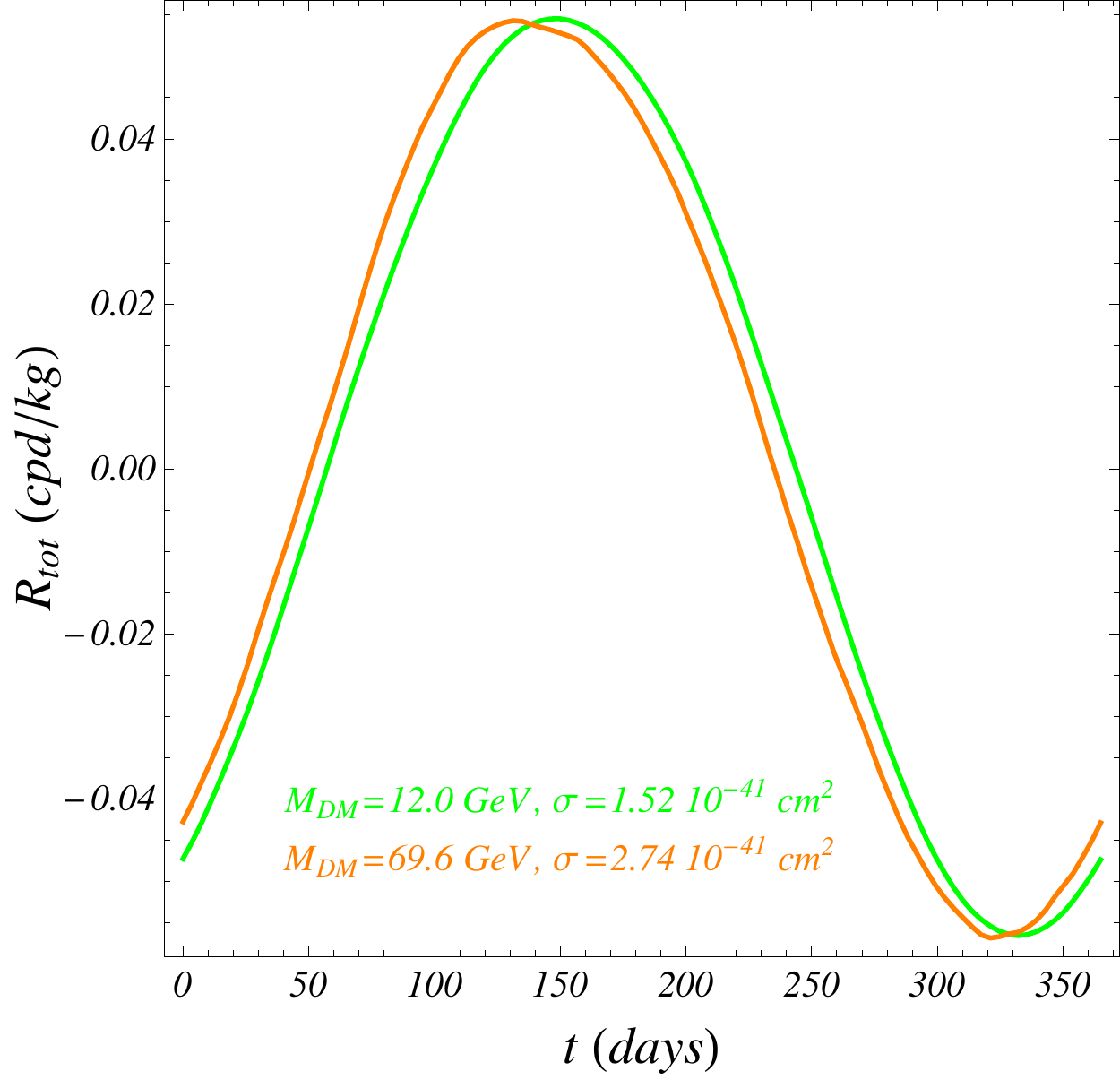} \quad
\et
\caption{\small \underline{Elastic Scenario} \it
Modulation amplitude as a function of recoil energy (left) or time (right) for the best-fit points of Fig.~\ref{fig:elchi2}.
In the case of the time variation, events with a recoil energy $2 \leq E_R \leq 6$~keV (in electron-equivalent units) were considered.}
\label{fig:elbest}
\ec
\efig
In the case of spin-independent elastic scatterings, the effective coherent couplings $f_p$ and $f_n$ of the WIMP to the proton and the neutron
have similar values. Therefore, the only relevant parameters are the WIMP mass $M_{DM}$ and the spin independent cross section on nucleon, 
here simply denoted as $\sigma$ (see Eq.~(\ref{eq:sigma})).
The main result for the elastic scenario is shown on the four panels of Fig.~\ref{fig:elchi2},
where the regions in the plane $M_{DM}$ - $\sigma$ favored by DAMA
are plotted against the exclusion limits set by CDMS and XENON10 (the excluded regions are on the right and above the curves).
From top left, the predictions for different halos: a) standard Maxwellian, b) this simulation, c) generalized Maxwellian with $\alpha=1.5$,
d) Tsallis with $q=0.773$.

In the elastic scenario, two regions are compatible with DAMA. For the region on the right, which is totally excluded by the other experiments,
the signal is due to quenched scatterings events on Iodine. The region on the left, due to channeled events on Iodine, is not totally excluded by CDMS.
The recent results of XENON10 set however a stronger limit, so that the elastic solution becomes very marginal. 
Here we have chosen to fit the DAMA data on 12 bins, which yields a smaller allowed region than with 36 bins.
For a standard Maxwellian halo, with $v_0=220$~km/s, it appears that the 99.9\% CL region of DAMA is totally excluded by XENON10 at 99.9\% CL. 

For the simulation halo, the direct detection prediction is made by analyzing the phase-space of the central and best resolved Milky-Way sized galactic DM halo.
The local structure is obtained by selecting the particles ($N_{ring}=2,662$) located in a ring ($7 \leq R \leq 9$~kpc and $\vert z \vert \leq 1$~kpc) around $R_0=8$~kpc
in the galactic plane. Despite the small number of flows in this selection, which causes some numerical noise and artifacts in the $\chi^2$ plots, 
a few observations can be made. The DAMA regions are slightly smaller than in the Maxwellian case, and the best-fit points move to the left.
Also, the region excluded by CDMS-Si slightly enlarges towards smaller cross sections.
Moreover, the channeling region extends outside the XENON exclusion limit, so that 
the compatibility between DAMA and the other experiments is improved with the realistic halo from the simulation.
While the best-fit point of the channeling region is still excluded, the 99.9\% and 99\% CL regions are not totally excluded anymore.

Several factors contribute to this improvement. 
The departure from a strict Maxwellian distribution has some impact, although moderate, as shown in panels c and d of Fig.~\ref{fig:elchi2}. 
We considered a generalized Maxwellian halo with the two values $\alpha=1.5$ and $\alpha=1.95$,
and a Tsallis halo with $q=0.773$. Their distributions are shown on Fig.~\ref{fig:histovAquariuslike}.
As in the simulation, a displacement of the best-fit points towards smaller masses compared to the standard Maxwellian case is noticeable.
On the contrary, if we take a Maxwellian halo with parameters tuned to fit as best as possible the simulation halo 
($\rho_{DM}=0.37$, $v_0=239~{\rm km/s}$, and smaller circular velocity $v_c=190~{\rm km/s}$ to reproduce the global average halo rotation velocity), 
the 99.9\% CL DAMA region is still totally excluded by XENON.
Other possible factors include the velocity dispersion anisotropy, shown on Fig.~\ref{fig:histovgalacticplane},
and a general anisotropy in the velocity field, induced by the co-rotating DM component, which has a small lag velocity
$v_{lag} \simeq 75$~km/s compared to the galactic disk. A detailed modeling and discussion of the impact of these features
will be presented in a subsequent paper~\cite{fsl}.

The panels of Fig.~\ref{fig:elbest} show the differential rate modulation for $2 \leq Eee \leq 6$~keV
and the time dependence of the total signal in this energy range for the two best-fit points of Fig.~\ref{fig:elchi2}.
Large fluctuations are seen in the modulation amplitude as a function of energy, they can be interpreted as numerical noise
due to the relatively small number of flows used to describe the local structure of the simulation halo around the Sun.
On the contrary, the integrated rate between 2 and 6~keV is a smooth, cosine-like function of time,
with a peak around $t \simeq 150$ as expected. The number of flows is however too small to reliably indicate any deviation from $t_0=152.5$.

\subsection{Results for the Inelastic Scenario}
\label{sec:inel}

\bfig[t]
\bc
\bt{cc}
\includegraphics[width=0.45\textwidth]{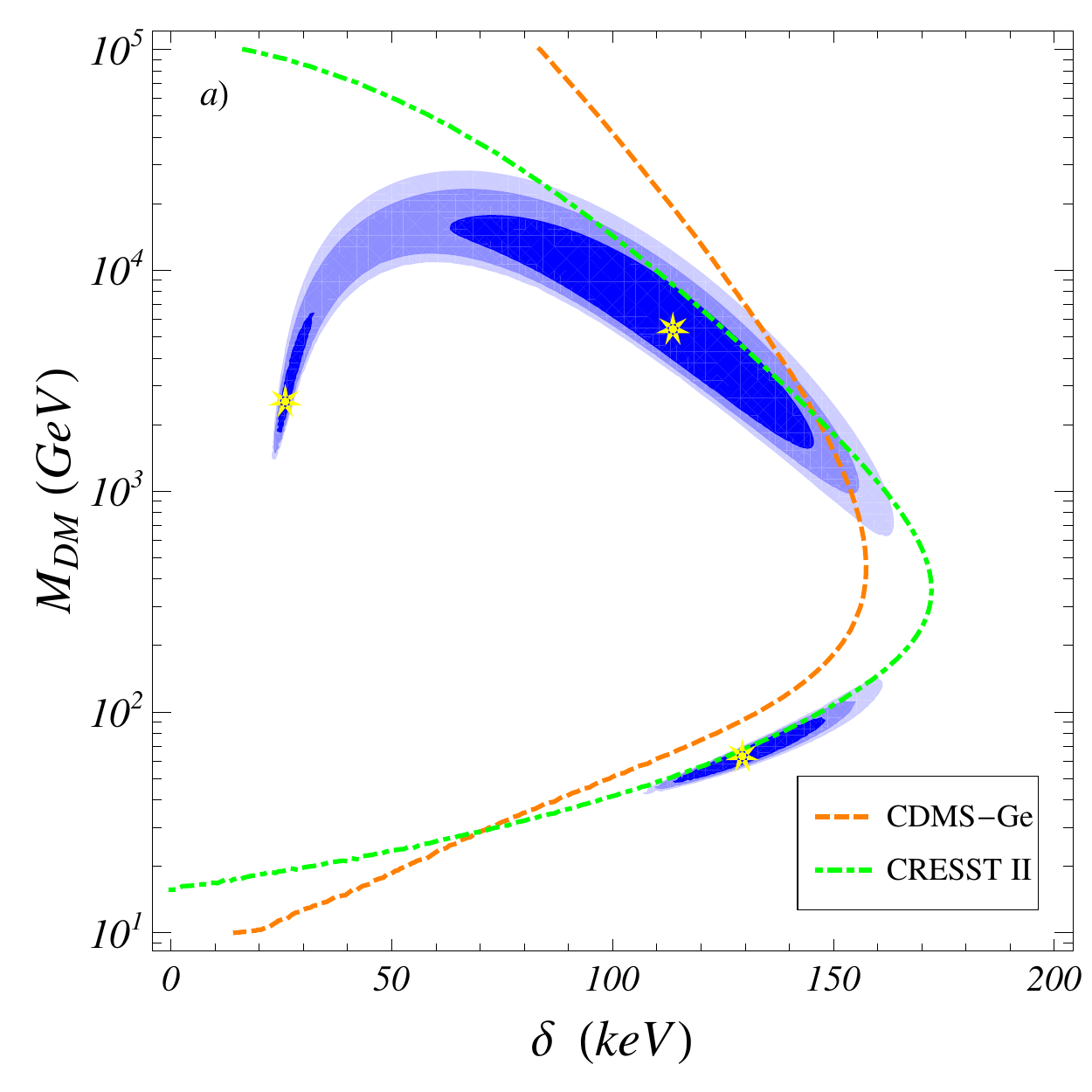} \quad &
\includegraphics[width=0.45\textwidth]{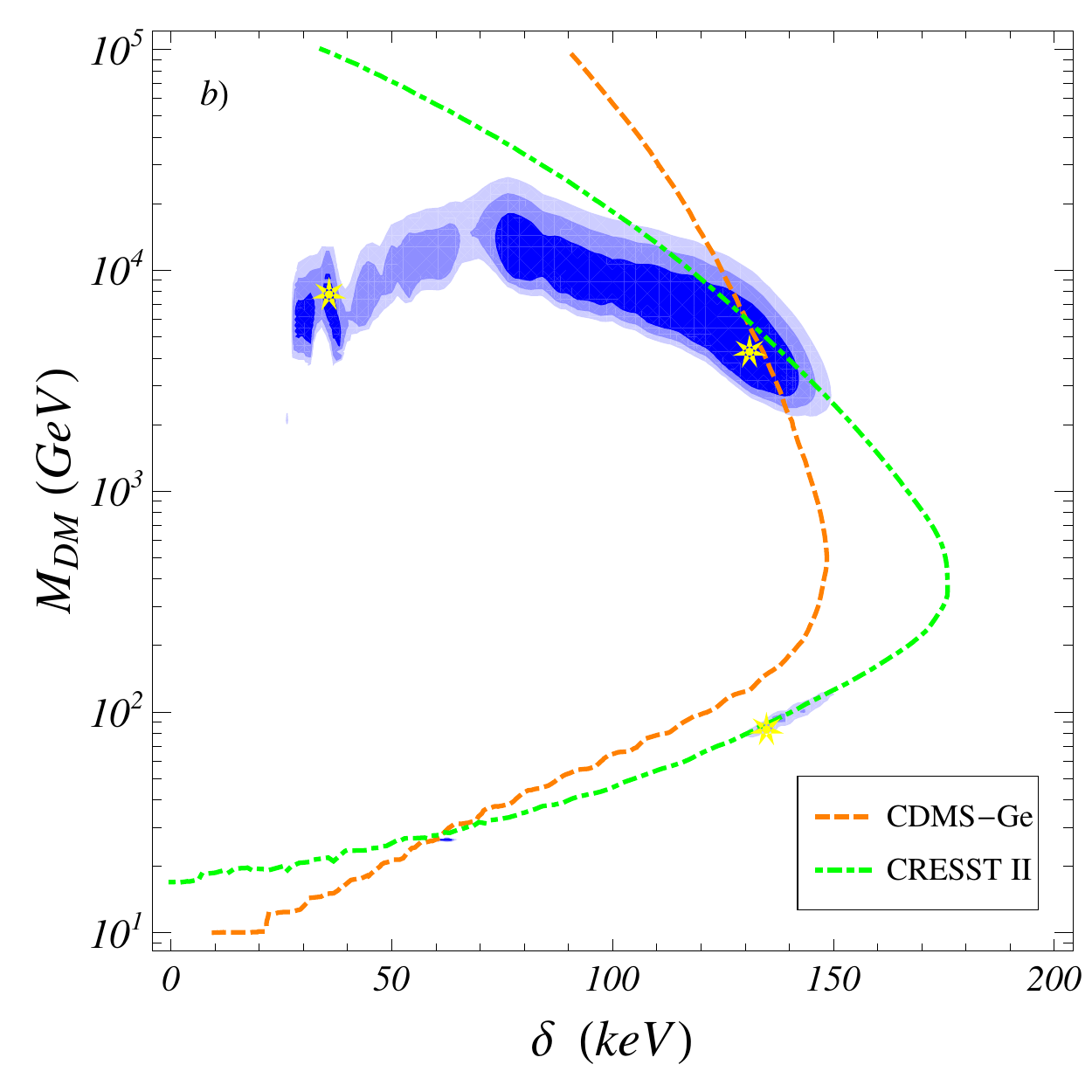} \quad \\
\includegraphics[width=0.45\textwidth]{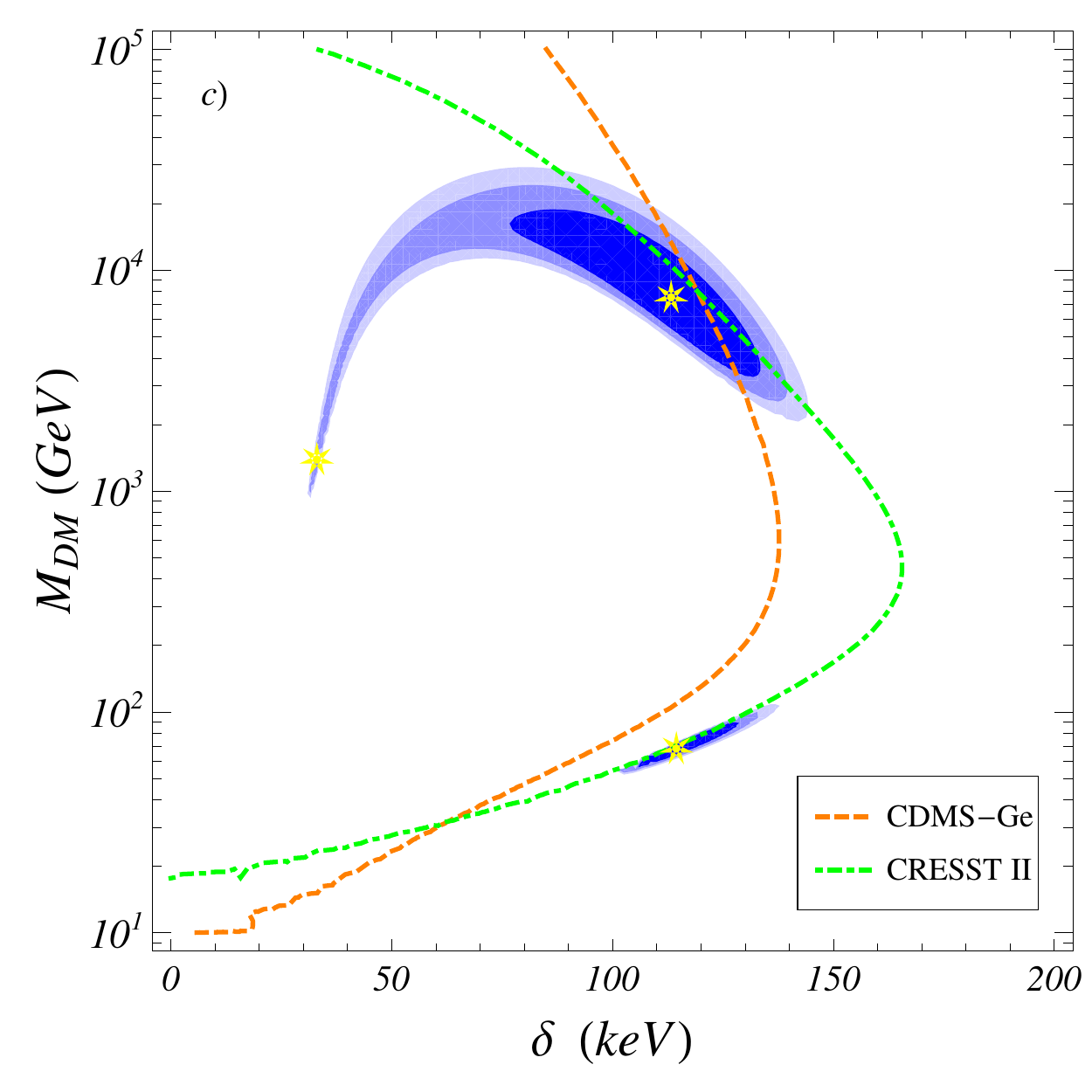} \quad &
\includegraphics[width=0.45\textwidth]{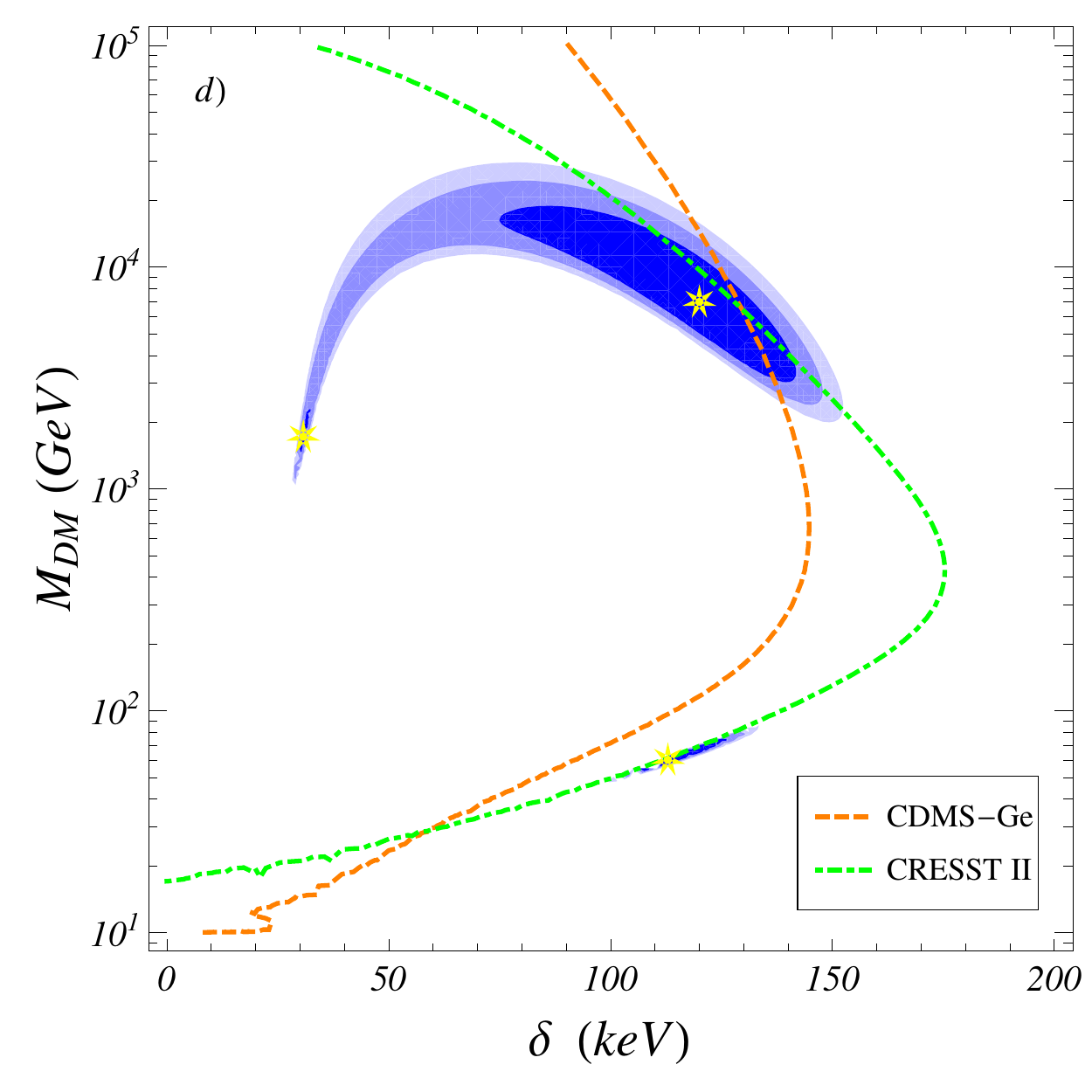} \quad
\et
\caption{\small \underline{Inelastic Scenario}, allowed regions in the plane 
$\delta-M_{DM}$ (DM mass splitting vs. DM mass)
\it
\newline
a) Standard Maxwellian halo ($\rho_{DM}=0.3~{\rm GeV/cm^3}$, $v_c=220$~km/s, $v_0=220$~km/s, $v_{esc}=600$~km/s)
\newline
b) N-body simulation with baryons ($\rho_{DM}=0.37~{\rm GeV/cm^3}$, $v_c=220$~km/s)
\newline
c) Generalized Maxwellian distribution ($\rho_{DM}=0.37~{\rm GeV/cm^3}$, $v_c=190$~km/s, $v_0=332$~km/s, $\alpha=1.95$)
\newline
d) Tsallis distribution ($\rho_{DM}=0.37~{\rm GeV/cm^3}$, $v_c=190$~km/s, $v_0=267.2$~km/s, $q=0.773$)
\newline
For c) and d), the circular velocity $v_c$ has been reduced compared to the standard value in order to take into account a halo rotation.
DAMA contours correspond to $90$, $99$ and $99.9$\% CL. Stars indicate local best-fit points.
All other exclusion curves are at the $99.9$\% CL.}
\label{fig:inelchi2}
\ec
\efig
\bfig[t]
\bc
\bt{cc}
\includegraphics[width=0.45\textwidth]{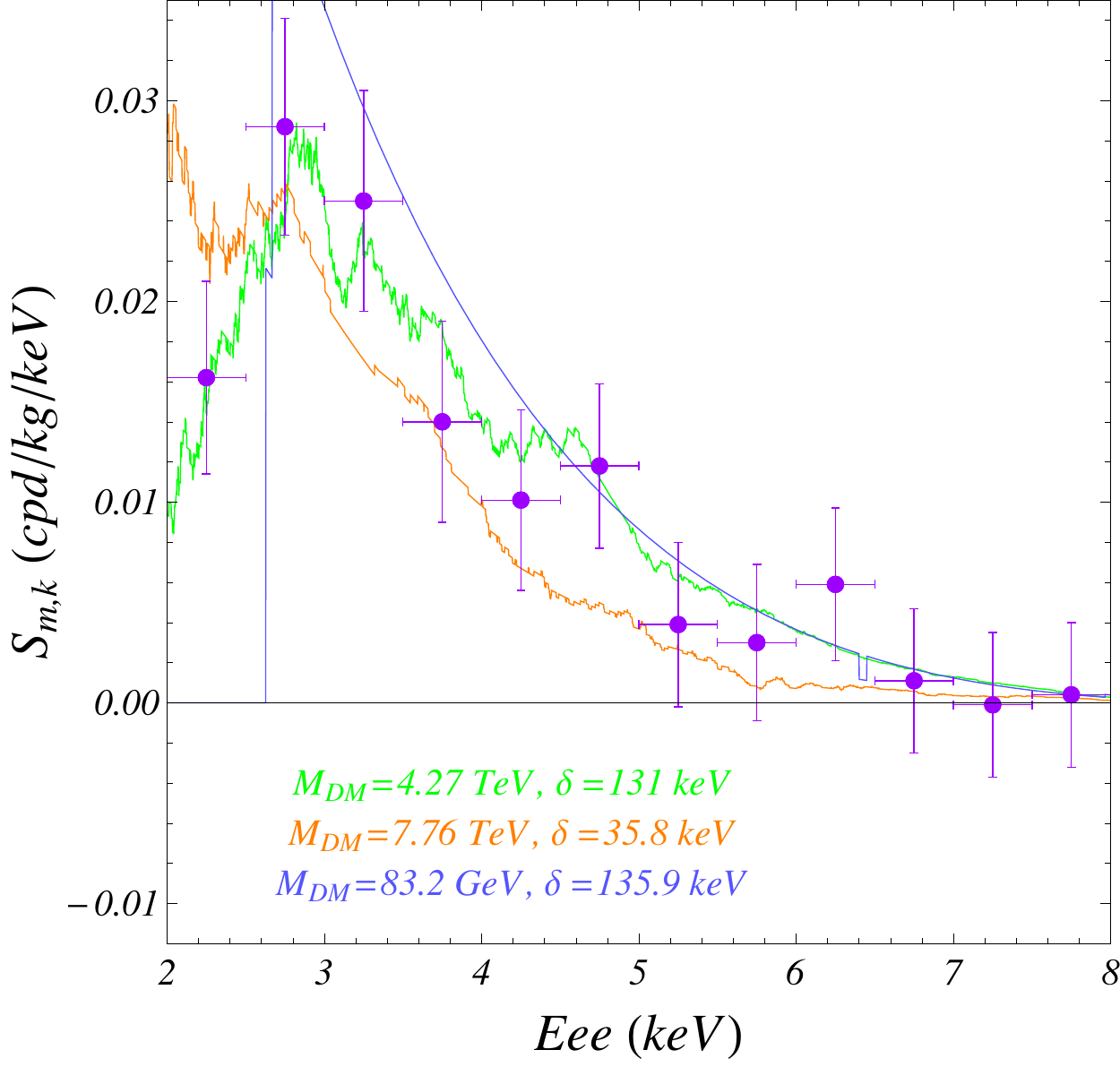} \quad &
\includegraphics[width=0.45\textwidth]{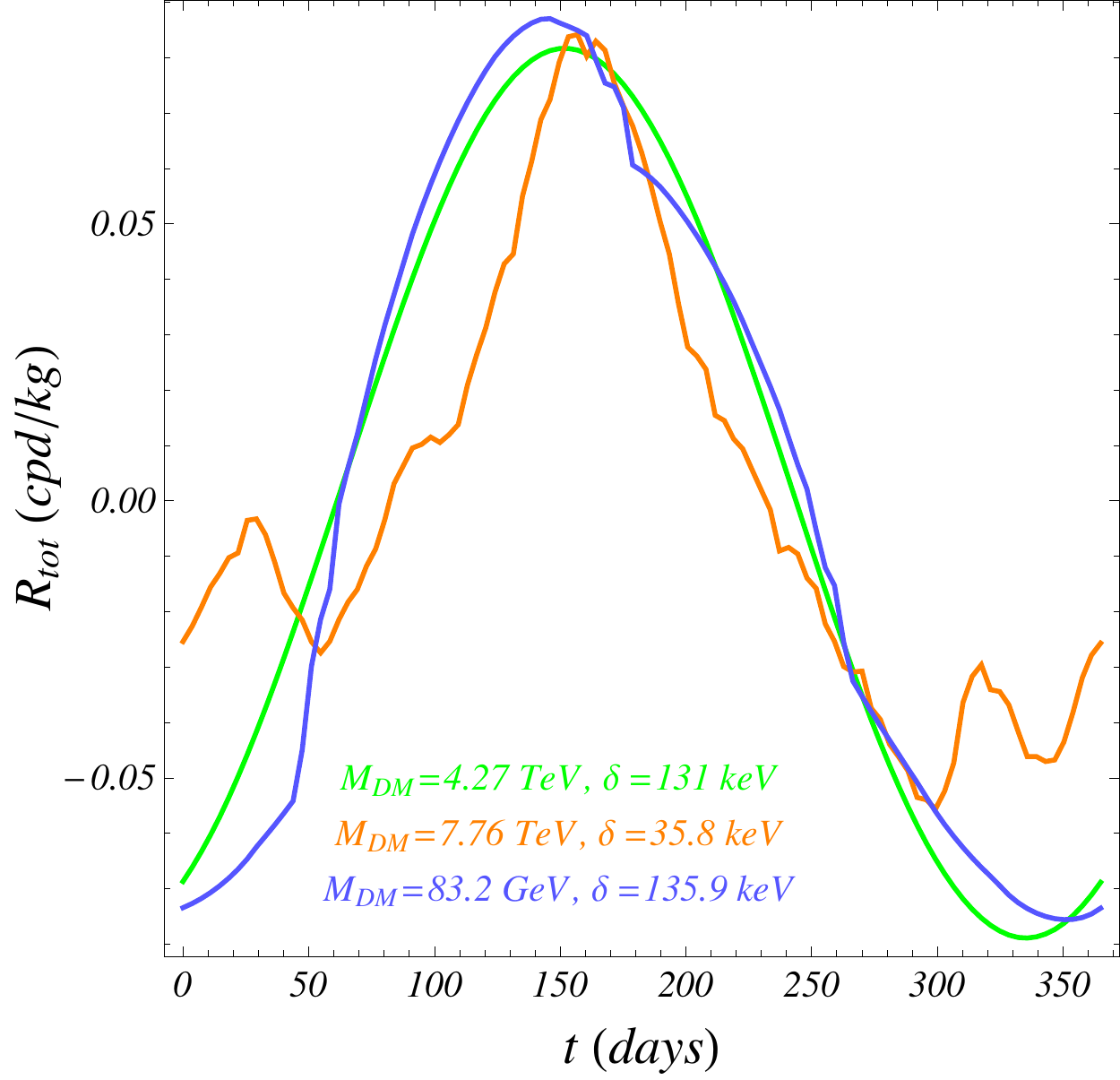} \quad
\et
\caption{\small \underline{Inelastic Scenario} \it
Modulation amplitude as a function of recoil energy (left) or time (right) for the best-fit points of Fig.~\ref{fig:inelchi2}.
In the case of the time variation, events with a recoil energy $2 \leq E_R \leq 6$~keV (in electron-equivalent units) were considered.}
\label{fig:inelbest}
\ec
\efig
The dominance of inelastic interactions over elastic ones is achieved in a natural way when the scattering is mediated by a massive gauge boson~\cite{Cui:2009xq}.
In the simplest scenario, the mediator is the weak $SU(2)_L$ boson $Z$, so that no new gauge boson needs to be introduced beyond the Standard Model.
The free parameters are therefore the mass $M_{DM}$ of the DM candidate, and its mass splitting $\delta$ with the next-to-lightest DM state.
In this paper, we will only consider this situation. More involved models, where the cross section differs from the weak cross section $\sigma_Z$
can be found in Ref.~\cite{Cui:2009xq}.

Results for the inelastic scenario are summarized on Fig.~\ref{fig:inelchi2}. 
The DAMA favored regions are shown together with the exclusion limits from CDMS and CRESST-II (the excluded region extends on the left of each curve). 
As in the elastic scenario, the four panels are the predictions for different halos: a) standard Maxwellian, b) this simulation, c) generalized Maxwellian with $\alpha=1.95$,
d) Tsallis with $q=0.773$.

In the inelastic scenario, for a standard Maxwellian halo ($\rho_{DM}=0.3~{\rm GeV/cm^3}$, $v_0=220~{\rm km/s}$, $v_{esc}=600~{\rm km/s}$), 
there are three 99\% CL regions compatible with DAMA, which are all mainly due to quenched scatterings on Iodine.
The region with $\delta \simeq 30$~keV is excluded by the constraints from both CDMS and CRESST.
It is actually excluded by DAMA itself, as it leads to a year averaged rate higher than observed.
The region with $\delta>50$~keV and $M_{DM}>1$~TeV is the largest in terms of parameter space. 
For a standard Maxwellian halo, this region is only marginally compatible with the exclusion limits of CDMS and CRESST.
Let us notice that in a minimal framework with only the two DM states, and their coupling to the $Z$,
the standard out-of-equilibrium freeze-out picture leads to a DM relic abundance higher than the value measured by WMAP~\cite{Hinshaw:2008kr},
because the (co)annihilation rates are suppressed for $M_{DM} \gg M_Z$.
For scalar candidates, the adjustable mass of the charged $SU(2)$ partner of the DM candidate enables to obtain the correct relic density
for a whole range of mass at the TeV scale~\cite{Hambye:2009pw}.
Finally, the DAMA region with $M_{DM}<100$~GeV is not excluded by other experiments. 
For these candidates, due to the proximity of the $Z$ pole, the relic abundance is lower than needed for WMAP unless some asymmetry
in the dark sector prevents DM density from collapsing during the freeze-out~\cite{Arina:2009um}.

The impact of the numerical noise from the finite number of flows in the simulation becomes more severe in the case of the inelastic scenario.
Indeed, for inelastic interactions, only particles with a velocity high enough can produce a recoil.
From Eq.~(\ref{thresvel}), it follows that the minimum velocity $v_*$ corresponding to the energy $E_*=\mu \delta/M_N$ is given by 
$v_*=\sqrt{2\delta/\mu}$. As a consequence, in the inelastic case, the signal is due to the high velocity tail of the distribution.
As panel b of Fig.~\ref{fig:inelchi2} shows, the DAMA region with $\delta \simeq 30$~keV is strongly affected by numerical noise.
In this region, only a small number of flows with a velocity around $v_* \simeq 235~{\rm km/s}$ contribute to the time dependent modulation.
The region with $M_{DM}<100$~GeV is also reduced, as it requires flows with a velocity higher than $v_* \simeq 700~{\rm km/s}$.
The third region is probably the most reliable. Its deformed shape, compared to a Maxwellian halo, as well as its relative position to the exclusion curves
are features that show that the velocity distribution of the simulation halo is not Maxwellian and not isotropic.
In particular, the compatibility between DAMA and the other experiments is strongly improved. 
DAMA solutions at 90\% CL are found, while only solutions
at 99\% CL were available in the case of the standard Maxwellian halo.
This improvement of the compatibility can also be attested by the different position of the best-fit point for each case.

Results for a generalized Maxwellian or a Tsallis halo are shown on panels c and d of Fig.~\ref{fig:inelchi2}.
The deviations from Maxwellianity cause the DAMA regions to shrink. 
Most importantly, in the region $\delta \simeq 120$~keV and $M_{DM} \simeq 1$~TeV,
the strong improvement of the fit between DAMA and the other experiments seen in the simulation is reproduced.
For this region, the relative position of the best-fit point and the exclusion curves of CDMS and CRESST-II
is comparable for the Tsallis and the generalized distributions, as both give a good fit of the high velocity 
tail of the velocity module distribution found in this simulation (see Fig.~\ref{fig:histovAquariuslike}).
However, a Tsallis distribution favors slightly larger mass splittings $\delta$, and can be seen to be
in better agreement with the simulation.

As in the elastic case, the differential rate modulation for $2 \leq Eee \leq 6$~keV
and the time dependence of the total signal in this energy range for the three best-fit points of Fig.~\ref{fig:inelchi2}
are shown on Fig.~\ref{fig:inelbest}.
The impact of numerical noise is clearly seen in the time modulation, which can strongly differ from a cosine-like behavior.

\section{ Summary \& Perspectives}
\label{sec:conclu}

In this paper, we have analyzed the phase-space structure of a galactic halo extracted from an advanced cosmological
N-body simulation which contains stars, gas, and DM, and used this information to make direct detection predictions.

Usually, such predictions are done with simplified astrophysical assumptions.
The local phase-space structure in the solar neighborhood is taken as a smooth, isotropic Maxwellian distribution.
However, it is known for quite some time that deviations should be expected in any realistic DM halo.
The action of long-range gravitational forces, as well as the presence of a non-virialized halo component in the form of cold streams 
from the continuous secondary galactic infall~\cite{Gelmini:2000dm,Ling:2004aj} 
already contribute to distort the simple picture of an isotropic Maxwellian halo.
Cosmological numerical simulations done in the $\Lambda$CDM paradigm have demonstrated the hierarchical character of structure formation,
leading to galactic structures with numerous sub-halos~\cite{Athanassoula:2008fn,Diemand:2009bm}.
Recent N-body simulations with very refined resolution, such as Aquarius~\cite{Springel:2008cc} 
or Via Lactea~\cite{Diemand:2008in} have even shown that non-Gaussianity and anisotropy 
(the velocities are rather in the radial direction) are present~\cite{Fairbairn:2008gz,MarchRussell:2008dy,Hansen:2005yj,Moore:2001vq}.
However, these simulations contain only DM particles, and cannot therefore provide a realistic description of a galaxy
in a satisfactory way, despite the large number of particles.
Dynamical interactions between the baryonic and the dark components of a galaxy are indeed expected to cause non negligible distortions on the
phase-space structure of both components~\cite{Athanassoula:2005xh}. 
Recently, simulations with baryons have shown that the presence of a galactic disk drags merging satellites in a 
preferential direction~\cite{Read:2009iv}, which leads to the formation of a thick stellar disc and a so-called dark disk. 

In this simulation, it appears that the DM halo contains a dark disk component that is co-rotating with the galactic disk. 
The local DM density in the solar neighborhood has a value $\rho_{DM} = 0.37$~to~$0.39~{\rm GeV/cm^3}$,
consistent with the recent determination of Ref.~\cite{Catena:2009mf}, but higher than the usually quoted value of $0.3~{\rm GeV/cm^3}$.
Despite large fluctuations, the average circular velocity differs significantly from zero throughout a thick region of the halo. 
At the Sun's location, a double Gaussian provides a reasonable fit of the distribution of the tangential velocity $v_\phi$. 
We infer that the rotating DM component has a mean lag velocity $v_{lag} \simeq 75$~km/s compared to the stellar disc,
and contributes to around 25\% of the total local density.
This rather small value, compared to results of other $\Lambda$CDM simulations~\cite{Read:2009iv},
rather points towards a quiescent merger history for the galactic halo considered in our simulation. 
For the other velocity components and for the velocity module, 
we also observe strong deviations from Gaussian and Maxwellian distributions, which we parameterized by introducing 
different generalizations of 
Gaussian and Maxwellian distributions, see Eqs.~(\ref{eq:ggauss}--\ref{eq:tsallis}).
For the velocity module of DM particles around $R_0=8$~kpc, we find that a Tsallis
distribution with a parameter $q=0.773$ gives an excellent fit.

The direct detection predictions are given as fits for the DAMA modulation signal and exclusion limits for null experiments.
Both the elastic and the inelastic scenarios have been considered with spin-independent cross-sections.
As opposed to previous works~\cite{Fairbairn:2008gz,MarchRussell:2008dy}, here the regions in the parameter space are obtained
directly from the velocity distributions extracted from the simulation data, without modeling.
They are then compared to the predictions of various analytical models.

As the co-rotating dark disk contribution to the local density is rather small, the absolute detection rates
are not significantly enhanced compared to the standard Maxwellian case.
Nevertheless, the compatibility of DAMA vs. null experiments shows some improvement. 
For the elastic scenario, the rotating DM component causes the channeling region to enlarge outside of the XENON constraint.
However, the overall goodness of fit of a global solution is still poor as most of the channeling region remains excluded.
For the inelastic scenario, the situation is much brighter, as two solutions with $\delta \simeq 130$~keV are permitted.
The solution with a heavy candidate ($M_{DM} \sim~{\rm TeV}$) becomes much more acceptable with the realistic halo provided by our simulation.
Moreover, such candidate can also achieve naturally the relic abundance needed for WMAP~\cite{Hambye:2009pw}, 
and is very little constrained by accelerator data~\cite{Barbieri:2006dq}.
If the DAMA modulation signal is taken seriously, all these hints concur to some interesting new physics beyond the Standard Model at the TeV scale.
The solution with a mass around $M_{DM} \simeq 80$~GeV is also acceptable from both the direct detection data and the particle physics point of view.
A correct relic abundance however requires some asymmetry in the dark sector~\cite{Arina:2009um} or some non thermal production mechanism.
The accelerator constraints could be more stringent, but have to be analyzed within a particular model~\cite{Lundstrom:2008ai}. 
These questions are beyond the scope of the present paper.

\section*{Acknowledgments}

We are grateful towards Jean-Charles Lambert for his help in handling GLNemo to produce the pictures of the galaxy for this paper. We would also like to thank Chiara Arina and Arman Khalatyan for helpful comments and discussions.
RT would like to thank St\'{e}phanie Courty for her help in selecting the simulated halo
and generating the initial conditions.
FSL work is supported by a belgian FNRS grant and the IAP. 
This work was granted access to the HPC resources of CINES under the allocation 2009-SAP2191 made by GENCI
(Grand Equipement National de Calcul Intensif).
This work was partly supported by grant ANR-06-BLAN-0172.

\bibliographystyle{unsrt}
\bibliography{simu-DD}

\end{document}